\documentclass[a4paper,10pt]{article}
\usepackage{jheppub}
\usepackage[utf8]{inputenc}
\usepackage{amsmath,amssymb,bm,slashed,braket}
\usepackage{dsfont}
\usepackage{amsfonts}
\usepackage{bm,bbm}
\usepackage{graphicx}
\usepackage{slashed} 
\usepackage[dvipsnames,table]{xcolor}
\usepackage[normalem]{ulem}
\usepackage{hyperref}
\hypersetup{colorlinks,citecolor= blue,linkcolor= blue, urlcolor=blue}
\usepackage{array,boldline,makecell,booktabs}
\usepackage{verbatim}
\usepackage{epsfig}
\usepackage{multirow}
\usepackage{hhline}
\usepackage{bbold}
\usepackage{float}
\usepackage[version=4]{mhchem}
\usepackage[capitalise]{cleveref}
\usepackage{orcidlink}
\usepackage{subcaption}
\usepackage{graphicx}
\usepackage{caption}
\usepackage{placeins} 
\allowdisplaybreaks[4]
\usepackage{tikz}
\usetikzlibrary{decorations.markings}

\newcommand{\calO}{\mathcal{O}}

\newcommand{\C}{ {\tt C} }
\newcommand{\tS}{ {\tt S} }
\newcommand{\tV}{ {\tt V} }
\newcommand{\tT}{ {\tt T} }
\newcommand{\tL}{ {\tt L} }
\newcommand{\tR}{ {\tt R} }
\newcommand{\N}{ {\tt N} }
\newcommand{\hc}{\text{H.c.}}

\newcommand{\nn}{ {\nonumber} }

\title{\Large Nucleon decays into three leptons: contact contributions}
\author[a,b]{Yi Liao\,\orcidlink{0000-0002-1009-5483},}
\emailAdd{liaoy@m.scnu.edu.cn}
\author[a,b]{Xiao-Dong Ma\,\orcidlink{0000-0001-7207-7793},}
\emailAdd{maxid@scnu.edu.cn}
\author[a,b]{and Xiang Zhao\,\orcidlink{0009-0008-6024-7722}}
\emailAdd{zhaox@m.scnu.edu.cn}
\affiliation[a]{State Key Laboratory of Nuclear Physics and
Technology, Institute of Quantum Matter, South China Normal
University, Guangzhou 510006, China}
\affiliation[b]{Guangdong Basic Research Center of Excellence for
Structure and Fundamental Interactions of Matter, Guangdong
Provincial Key Laboratory of Nuclear Science, Guangzhou
510006, China}

\abstract{

Baryon number violating (BNV) nucleon decays into three leptons provide a unique probe of BNV interactions beyond the conventional two-body modes involving a single lepton and a light meson.
In a previous work [Nucleon decays into three leptons: noncontact contributions, \href{https://arxiv.org/abs/2512.02692}{arXiv:2512.02692}.], two of us analyzed the noncontact contributions to these decays arising from dimension-6 (dim-6) operators within the low-energy effective field theory (LEFT), and found that they are severely suppressed due to stringent constraints on these dim-6 operators. 
In this work, we continue this endeavor by systematically investigating the contact contributions originating from dim-9 LEFT operators. 
We construct a complete basis of dim-9 operators relevant to these processes, and subsequently match them onto chiral perturbation theory to calculate their decay widths. By employing existing experimental data, we derive stringent constraints on the relevant operators. In addition, we present the analysis of an ultraviolet-complete model to demonstrate its connection with our theoretical framework, thereby facilitating further studies of these exotic nucleon decays in upcoming neutrino experiments with large fiducial masses.
}

\keywords{Baryon/Lepton Number Violation, SMEFT}

\makeatletter
\gdef\@fpheader{}
\makeatother
\begin{document} 

\maketitle
\setcounter{page}{2}

%%%%%%%%%%%%%%%%%%%%%%%%%%%%%%%%%%%%%%
\section{Introduction}
%%%%%%%%%%%%%%%%%%%%%%%%%%%%%%%%%%%%%%

Baryon number is violated in the Standard Model (SM) although baryon minus lepton number is conserved. When going beyond the SM the fate of baryon and lepton numbers is open and varies from one theory to another. This is especially obvious in the framework of effective field theories (EFTs) where effective operators of different dimension display different conserving or violating patterns of them. 
Experimentally, the most promising way to search for baryon number violation (BNV) is through nucleon decays. 
This search has a long history, covering both conventional two-body nucleon decay modes involving a lepton and a pseudoscalar or vector meson and the more exotic three-body decays into three leptons~\cite{ParticleDataGroup:2024cfk}. 
Although no positive signals have been observed so far, stringent limits have been established on their occurrence. Currently, experiments such as Super-Kamiokande (Super-K) and JUNO~\cite{JUNO:2015zny} continue these efforts, while the next generation of neutrino detectors---including Hyper-Kamiokande~\cite{Hyper-Kamiokande:2018ofw}, DUNE~\cite{DUNE:2020ypp}, 
and the two proposed detectors THEIA~\cite{Theia:2019non} and ESSnuSB~\cite{ESSnuSB:2023ogw}--- 
will further probe various nucleon decay channels with unprecedented sensitivity. 

Theoretically, since nucleon decays occur at the GeV energy scale while the underlying BNV mechanism takes place at a much higher scale above the electroweak scale, it is standard practice to study these processes systematically in the framework of EFTs. 
In past years, two-body nucleon decays have been thoroughly explored in the framework of the SM EFT (SMEFT) and low-energy EFT (LEFT) through dimension-6 (dim-6) and -7 operators~\cite{Abbott:1980zj,Claudson:1981gh,Beneito:2023xbk,Gargalionis:2024nij,Liao:2025sqt}. 
For nucleon triple-lepton decay modes, these operators can also contribute via the SM mediators such as photons, $W,~Z$ bosons, leptons, and light hadrons. 
Defining a generalized lepton flavor change in a decay as
$\Delta F_L\equiv |\Delta F_e|+|\Delta F_\mu|+|\Delta F_\tau|$~\cite{Chen:2025mjt}, where $\Delta F_i\, (i=e,\mu,\tau)$ denotes the $i$-flavor change,
dim-6 LEFT interactions (i.e., dim-6 and -7 SMEFT interactions) can only induce processes with $\Delta F_L=1$. 
Assuming that the $\Delta F_L=1$ nucleon triple-lepton decays are induced by dim-6 interactions, 
new stringent lower bounds on their partial lifetimes have been derived in~\cite{Chen:2025mjt} based on existing constraints from two-body nucleon decays.
It was found that the noncontact contributions from most dim-6 operators are negligibly small, resulting in partial lifetime limits well beyond the reach of future experiments.   

Going beyond dim-6 or -7 BNV interactions in SMEFT, dim-9 and higher SMEFT operators can induce $\Delta F_L=3$ nucleon triple-lepton decay modes that cannot be realized through dim-8 or lower interactions~\cite{Weinberg:1980bf,Hambye:2017qix,Heeck:2019kgr}. 
Moreover, these higher-dimensional operators may provide dominant contributions as well to $\Delta F_L=1$ nucleon triple-lepton decay modes for which the effects of lower-dimensional operators are suppressed. 
Below the electroweak scale, these higher-dimensional SMEFT operators are matched onto dim-9 LEFT interactions at leading order. 
In light of these considerations, 
the purpose of this work is to establish a robust framework for reliably calculating nucleon triple-lepton decay modes arising from contact contributions induced by dim-9 LEFT operators. 

To be as general as possible, we work within the LEFT framework and construct the relevant dim-9 operators responsible for nucleon triple-lepton decays. 
Using the general BNV triple-quark structures developed in~\cite{Liao:2025vlj}, all relevant operators are organized as irreducible representations of the QCD chiral group $\rm SU(3)_\tL\otimes SU(3)_\tR$ for the light $u$, $d$, and $s$ quarks. To calculate the nucleon decay widths, we match these operators to chiral perturbation theory (ChPT) by the spurion field method, thereby obtaining chiral interactions between a nucleon and three leptons. We then calculate all allowed decay modes from the obtained  chiral Lagrangian and express their decay widths as functions of the Wilson coefficients (WCs) of the dim-9 LEFT operators. By applying available experimental limits, we derive stringent constraints on the effective scales associated with these WCs. 
Our formalism will facilitate future experimental searches for these decay modes.
For a given new physics model, one can integrate out its heavy particles to obtain effective SMEFT operators and transform the latter into our chosen basis, enabling us to set direct constraints on model parameters.

The remainder of the paper is organized as follows. In the next section, we classify all possible nucleon triple-lepton decay modes and 
construct the complete set of dim-9 LEFT operators responsible for them. This is followed in \cref{sec:chpt} by deriving their hadronic counterparts in ChPT. 
The calculation of the decay widths including momentum distributions, and constraints from existing data are presented in \cref{sec:calc_nucleon_decay}. 
\cref{sec:UVmodel} is dedicated to demonstrating how to connect ultraviolet (UV)-complete models with our formalism and to establishing meaningful constraints on the model parameters.  
Our conclusions are summarized in \cref{sec:summary}. 
The complete amplitudes for all relevant decay modes are collected in \cref{app:decay_width_WCsmplitude}, while the expressions for decay widths, parametrized in terms of the LEFT WCs, are provided in \cref{app:decay_width_WCs}.

%%%%%%%%%%%%%%%%%%%%%%%%%%%%%%%%%%%%%%
\section{Dimension-9 LEFT operator basis for nucleon triple-lepton decays}
\label{sec:dim9basis}
%%%%%%%%%%%%%%%%%%%%%%%%%%%%%%%%%%%%%%

\begin{table}[ht]
\center
\resizebox{0.8\linewidth}{!}{
\renewcommand{\arraystretch}{1.}
\begin{tabular}{|c|c|c|c|c|}
\hline
Class & $\Delta L = +1$ & $\Delta L = - 1$ & $\Delta L = + 3$ & $\Delta L = - 3$ 
\\\hline
\multirow{4}*{\rotatebox[origin=c]{90}{Process}}
&$p \to \ell_x^+ \ell_y^+ \ell_z^-~~~~$ 
&$n \to \ell_x^-\ell_z^+  \nu_y\,(\diamond)$ 
&$p \to \bar\nu_x \bar\nu_y \ell_z^+\,(\dagger) $ 
&$n \to \nu_x \nu_y \nu_z\,(\star)$ 
\\\cline{2-5}
&$n \to \ell_z^- \ell_y^+ \bar\nu_x\,(\diamond)$ 
&$p \to \nu_x \nu_y \ell_z^+\,(\dagger)$ 
&$n \to \bar\nu_x \bar\nu_y \bar\nu_z\,(\star)$  & 
\\\cline{2-5}
&$p \to \nu_z \bar \nu_y \ell_x^+~~~~$ 
&$n \to \nu_x \nu_y \bar\nu_z\,(*)$  &  & 
\\\cline{2-5}
&$n \to \bar\nu_x \bar \nu_y \nu_z\,(*)$  &  &  & 
\\\hline
\end{tabular}}
\caption{Classification of all possible nucleon triple-lepton decay modes, based on the lepton number change $\Delta L$ in each process. 
The pair of processes that are related by the simultaneous interchange of a neutrino and an antineutrino ($\nu \leftrightarrow \bar{\nu}$) are indicated by the same symbol in the parentheses.
}
\label{tab:triple_lepton_modes}
\end{table}

We begin by classifying all possible nucleon decay modes involving  three leptons. For this purpose, we denote the proton and neutron by $p$ and $n$, the charged leptons by $\ell^\pm$, and neutrinos and antineutrinos by $\nu$ and $\bar\nu$, respectively.
In terms of the lepton number change $\Delta L$ defined as the initial- minus final-state lepton number,
all these processes are tabulated in \cref{tab:triple_lepton_modes}, where $x,y,z=e,\mu,\tau$ represent the three lepton flavors but for the tau flavor refer only to neutrinos and antineutrinos.
The processes in the classes with $\Delta L=\pm 1$ that satisfy the flavor conditions $z=x{\rm~or~}y$ correspond to $\Delta F_L=1$, while the remaining processes, together with those in the $\Delta L=\pm 3$ classes, have $\Delta F_L=3$.
Within the LEFT framework, the $\Delta F_L=3$ processes can only be induced at leading order by dim-9 operators. In contrast, the $\Delta F_L=1$ processes can be mediated both by dim-9 operators and by dim-6 operators through a SM mediator. 
However, these latter contributions from dim-6 BNV interactions are strongly suppressed due to stringent constraints imposed by nucleon two-body decays involving a pseudoscalar meson~\cite{Chen:2025mjt,Girmohanta:2019xya}. 
Therefore, in this work, we focus on the contact contributions arising from dim-9 LEFT interactions. In the remainder of this section, we will construct the relevant dim-9 operators for all the processes listed in \cref{tab:triple_lepton_modes}.

\begin{table}[ht]
\center
\resizebox{\linewidth}{!}{
\renewcommand{\arraystretch}{1.}
\begin{tabular}{|c|c|c|c|c|c|}
\hline   
&~~Notation~~& Operator &~~Chiral Irrep.~~& 
\# of operators
&~~Process~~
\\\hline%%%DeltaL=+1
\multirow{26}*{\rotatebox[origin=c]{90}{
\pmb{$\Delta L = + 1 $} } }
& $\calO_{\ell\ell\bar\ell ,xyz}^{\tV\tR\tV\tL\tS\tL}$
& $(\overline{\ell_{\tL x}^\C}\gamma_{\mu}\ell_{\tR y})
(\overline{\ell_{\tL z}}\gamma^\mu{\cal N}_{uud}^{\tL\tL}) $  
& ${\color{purple}\pmb{8}_\tL\otimes{\pmb{1}}_\tR }$
& $n_\ell^3 \,\langle 27\rangle$ 
&\multirow{26}*{\rotatebox[origin=c]{90}{
$p\to\ell_x^+ \ell_y^+ \ell_z^-$ } }
\\\cline{2-5}%
& $\calO_{\ell\ell\bar\ell ,xyz}^{\tV\tR\tV\tL\tS\tR}$
& $(\overline{\ell_{\tL x}^\C}\gamma_{\mu}\ell_{\tR y})
(\overline{\ell_{\tL z}}\gamma^\mu{\cal N}_{uud}^{\tL\tR}) $  
& ${\color{Purple}\pmb{3}_\tL\otimes{\bar{\pmb{3}}}_\tR }$
& $n_\ell^3 \,\langle 27 \rangle$ &
\\
& $\calO_{\ell\ell\bar\ell ,xyz}^{\tV\tR\tS\tR\tV\tL1}$
& $(\overline{\ell_{\tL x}^\C}\gamma_\mu\ell_{\tR y})
(\overline{\ell_{\tL z}}{\cal N}_{udu}^{\tR\tL,\mu}) $  
& ${\color{Green}\pmb{3}_\tL\otimes{{\pmb{6}}}_\tR }$
& $n_\ell^3 \,\langle 27 \rangle$ &
\\\cline{2-5}%
& $\calO_{\ell\ell\bar\ell ,xyz}^{\tV\tR\tS\tR\tV\tL2}$
& $(\overline{\ell_{\tL x}^\C}\gamma_{\mu}\ell_{\tR y})
(\overline{\ell_{\tL z}}{\cal N}_{uud}^{\tR\tL,\mu}) $ 
& ${\color{Green}\pmb{3}_\tL\otimes\pmb{6}_\tR }$
& $n_\ell^3\,\langle 27 \rangle$ &
\\\cline{2-5}%%%
& $\calO_{\ell\ell\bar\ell ,xyz}^{\tS\tL\tS\tR\tS\tR}$
& $(\overline{\ell_{\tL x}^\C}\ell_{\tL y})
(\overline{\ell_{\tL z}} {\cal N}_{uud}^{\tR\tR}) $  
& ${\color{purple}\pmb{1}_\tL\otimes\pmb{8}_\tR}$
& $\frac{1}{2}n_\ell^2 (n_\ell+1)\,\langle 18 \rangle$ &
\\\cline{2-5}%
& $\calO_{\ell\ell\bar\ell ,xyz}^{\tS\tL\tS\tL\tS\tL}$
& $(\overline{\ell_{\tL x}^\C}\ell_{\tL y})
(\overline{\ell_{\tR z}} {\cal N}_{uud}^{\tL\tL}) $  
& ${\color{purple}\pmb{8}_\tL\otimes\pmb{1}_\tR}$
& $\frac{1}{2}n_\ell^2 (n_\ell+1)\,\langle 18 \rangle$ &
\\
& $\calO_{\ell\ell\bar\ell ,xyz}^{\tT\tL\tT\tL\tS\tL}$
& $(\overline{\ell_{\tL x}^\C}\sigma_{\mu\nu}\ell_{\tL y})
(\overline{\ell_{\tR z}} \sigma^{\mu\nu}{\cal N}_{uud}^{\tL\tL}) $  
& ${\color{purple}\pmb{8}_\tL\otimes\pmb{1}_\tR}$
& $\frac{1}{2}n_\ell^2 (n_\ell-1)\,\langle 9 \rangle$ &
\\
& $\calO_{\ell\ell\bar\ell ,xyz}^{\tT\tL\tS\tL\tT\tL}$
& $(\overline{\ell_{\tL x}^\C}\sigma_{\mu\nu}\ell_{\tL y})
(\overline{\ell_{\tR z}}{\cal N}_{uud}^{\tL\tL,\mu\nu}) $  
& ${\color{Blue}\pmb{10}_\tL\otimes\pmb{1}_\tR}$
& $\frac{1}{2}n_\ell^2 (n_\ell-1)\,\langle 9 \rangle$ &
\\\cline{2-5}%
& $\calO_{\ell\ell\bar\ell ,xyz}^{\tS\tL\tS\tR\tS\tL}$
& $(\overline{\ell_{\tL x}^\C} \ell_{\tL y})
(\overline{\ell_{\tL z}}{\cal N}_{uud}^{\tR\tL}) $  
&${\color{Purple}\bar{\pmb{3}}_\tL\otimes\pmb{3}_\tR }$
& $\frac{1}{2}n_\ell^2 (n_\ell+1)\,\langle 18 \rangle$ &
\\
& $\calO_{\ell\ell\bar\ell ,xyz}^{\tT\tL\tV\tL\tV\tR1}$
& $(\overline{\ell_{\tL x}^\C}\sigma_{\mu\nu} \ell_{\tL y})
(\overline{\ell_{\tL z}}\gamma^\mu{\cal N}_{udu}^{\tL\tR,\nu}) $  
& ${\color{Green}\pmb{6}_\tL \otimes\pmb{3}_\tR }$
& $\frac{1}{2}n_\ell^2 (n_\ell-1)\,\langle 9\rangle$ &
\\\cline{2-5}%
& $\calO_{\ell\ell\bar\ell ,xyz}^{\tT\tL\tV\tL\tV\tR2}$
& $(\overline{\ell_{\tL x}^\C}\sigma_{\mu\nu} \ell_{\tL y})
(\overline{\ell_{\tL z}}\gamma^\mu {\cal N}_{uud}^{\tL\tR,\nu}) $  
& ${\color{Green}\pmb{6}_\tL \otimes\pmb{3}_\tR }$
& $\frac{1}{2}n_\ell^2 (n_\ell-1)\,\langle 9 \rangle$ &
\\\cline{2-5}%
& $\calO_{\ell\ell\bar\ell ,xyz}^{\tS\tL\tS\tL\tS\tR}$
& $(\overline{\ell_{\tL x}^\C}\ell_{\tL y})
(\overline{\ell_{\tR z}}{\cal N}_{uud}^{\tL\tR}) $  
& ${\color{Purple}\pmb{3}_\tL\otimes\bar{\pmb{3}}_\tR }$
& $\frac{1}{2}n_\ell^2 (n_\ell+1)\,\langle 18 \rangle$ &
\\
& $\calO_{\ell\ell\bar\ell ,xyz}^{\tT\tL\tT\tL\tS\tR}$
& $(\overline{\ell_{\tL x}^\C}\sigma_{\mu\nu}\ell_{\tL y})
(\overline{\ell_{\tR z}}\sigma^{\mu\nu}{\cal N}_{uud}^{\tL\tR}) $  
& ${\color{Purple}\pmb{3}_\tL\otimes\bar{\pmb{3}}_\tR }$
& $\frac{1}{2}n_\ell^2 (n_\ell-1)\,\langle 9 \rangle$ &
\\\clineB{2-5}{2}%%%%
& $\calO_{\ell\ell\bar\ell ,xyz}^{\tV\tL\tV\tR\tS\tR}$
& $(\overline{\ell_{\tR x}^\C}\gamma_{\mu}\ell_{\tL y})
(\overline{\ell_{\tR z}}\gamma^\mu{\cal N}_{uud}^{\tR\tR}) $  
& ${\color{purple}\pmb{1}_\tL\otimes\pmb{8}_\tR }$
& $n_\ell^3\,\langle 27 \rangle$ &
\\\cline{2-5}%
& $\calO_{\ell\ell\bar\ell ,xyz}^{\tV\tL\tV\tR\tS\tL}$
& $(\overline{\ell_{\tR x}^\C}\gamma_{\mu}\ell_{\tL y})
(\overline{\ell_{\tR z}}\gamma^\mu{\cal N}_{uud}^{\tR\tL}) $  
& ${\color{Purple}{\bar{\pmb{3}}}_\tL\otimes\pmb{3}_\tR }$
& $n_\ell^3\,\langle 27 \rangle$ &
\\
& $\calO_{\ell\ell\bar\ell ,xyz}^{\tV\tL\tS\tL\tV\tR1}$
& $(\overline{\ell_{\tR x}^\C}\gamma_\mu\ell_{\tL y})
(\overline{\ell_{\tR z}}{\cal N}_{udu}^{\tL\tR,\mu}) $  
& ${\color{Green}\pmb{6}_\tL\otimes\pmb{3}_\tR }$
& $n_\ell^3\,\langle 27 \rangle$ &
\\\cline{2-5}%
& $\calO_{\ell\ell\bar\ell ,xyz}^{\tV\tL\tS\tL\tV\tR2}$
& $(\overline{\ell_{\tR x}^\C}\gamma_{\mu}\ell_{\tL y})
(\overline{\ell_{\tR z}}{\cal N}_{uud}^{\tL\tR,\mu}) $ 
& ${\color{Green}\pmb{6}_\tL\otimes{\pmb{3}}_\tR }$
& $n_\ell^3 \,\langle 27 \rangle$ &
\\\cline{2-5}%%%%
& $\calO_{\ell\ell\bar\ell ,xyz}^{\tS\tR\tS\tL\tS\tL}$
& $(\overline{\ell_{\tR x}^\C}\ell_{\tR y})
(\overline{\ell_{\tR z}} {\cal N}_{uud}^{\tL\tL}) $  
& ${\color{purple}\pmb{8}_\tL\otimes\pmb{1}_\tR}$
& $\frac{1}{2}n_\ell^2 (n_\ell+1)\,\langle 18 \rangle$ 
&
\\\cline{2-5}%
& $\calO_{\ell\ell\bar\ell ,xyz}^{\tS\tR\tS\tR\tS\tR}$
& $(\overline{\ell_{\tR x}^\C}\ell_{\tR y})
(\overline{\ell_{\tL z}} {\cal N}_{uud}^{\tR\tR}) $  
& ${\color{purple}\pmb{1}_\tL\otimes\pmb{8}_\tR}$
& $\frac{1}{2}n_\ell^2 (n_\ell+1)\,\langle 18 \rangle$ &
\\
& $\calO_{\ell\ell\bar\ell ,xyz}^{\tT\tR\tT\tR\tS\tR}$
& $(\overline{\ell_{\tR x}^\C}\sigma_{\mu\nu}\ell_{\tR y})
(\overline{\ell_{\tL z}} \sigma^{\mu\nu}{\cal N}_{uud}^{\tR\tR}) $  
& ${\color{purple}\pmb{1}_\tL\otimes\pmb{8}_\tR}$
& $\frac{1}{2}n_\ell^2 (n_\ell-1)\,\langle 9 \rangle$ &
\\
& $\calO_{\ell\ell\bar\ell ,xyz}^{\tT\tR\tS\tR\tT\tR}$
& $(\overline{\ell_{\tR x}^\C}\sigma_{\mu\nu}\ell_{\tR y})
(\overline{\ell_{\tL z}}{\cal N}_{uud}^{\tR\tR,\mu\nu}) $  
& ${\color{Blue}\pmb{1}_\tL\otimes\pmb{10}_\tR}$
& $\frac{1}{2}n_\ell^2 (n_\ell-1)\,\langle 9 \rangle$ &
\\\cline{2-5}%
& $\calO_{\ell\ell\bar\ell ,xyz}^{\tS\tR\tS\tL\tS\tR}$
& $(\overline{\ell_{\tR x}^\C} \ell_{\tR y})
(\overline{\ell_{\tR z}}{\cal N}_{uud}^{\tL\tR}) $  
&${\color{Purple}\pmb{3}_\tL\otimes\bar{\pmb{3}}_\tR }$
& $\frac{1}{2}n_\ell^2 (n_\ell+1)\,\langle 18 \rangle$ &
\\
& $\calO_{\ell\ell\bar\ell ,xyz}^{\tT\tR\tV\tR\tV\tL1}$
& $(\overline{\ell_{\tR x}^\C}\sigma_{\mu\nu} \ell_{\tR y})
(\overline{\ell_{\tR z}}\gamma^\mu{\cal N}_{udu}^{\tR\tL,\nu}) $  
& ${\color{Green}\pmb{3}_\tL\otimes\pmb{6}_\tR  }$
& $\frac{1}{2}n_\ell^2 (n_\ell-1)\,\langle 9 \rangle$ &
\\\cline{2-5}%
& $\calO_{\ell\ell\bar\ell ,xyz}^{\tT\tR\tV\tR\tV\tL2}$
& $(\overline{\ell_{\tR x}^\C}\sigma_{\mu\nu} \ell_{\tR y})
(\overline{\ell_{\tR z}}\gamma^\mu {\cal N}_{uud}^{\tR\tL,\nu}) $  
& ${\color{Green}\pmb{3}_\tL\otimes\pmb{6}_\tR }$
& $\frac{1}{2}n_\ell^2 (n_\ell-1)\,\langle 9 \rangle$ &
\\\cline{2-5}%
& $\calO_{\ell\ell\bar\ell ,xyz}^{\tS\tR\tS\tR\tS\tL}$
& $(\overline{\ell_{\tR x}^\C}\ell_{\tR y})
(\overline{\ell_{\tL z}}{\cal N}_{uud}^{\tR\tL}) $  
& ${\color{Purple}{\bar{\pmb{3}}}_\tL\otimes\pmb{3}_\tR }$
& $\frac{1}{2}n_\ell^2 (n_\ell+1)\,\langle 18 \rangle$ &
\\
& $\calO_{\ell\ell\bar\ell ,xyz}^{\tT\tR\tT\tR\tS\tL}$
& $(\overline{\ell_{\tR x}^\C}\sigma_{\mu\nu}\ell_{\tR y})
(\overline{\ell_{\tL z}}\sigma^{\mu\nu}{\cal N}_{uud}^{\tR\tL}) $  
&~~~${\color{Purple}{\bar{\pmb{3}}}_\tL\otimes\pmb{3}_\tR }~~~$
&~~~$\frac{1}{2}n_\ell^2 (n_\ell-1)\,\langle 9 \rangle$~~~&
\\\hline
\end{tabular}
}
\caption{
The dim-9 LEFT BNV and $\Delta L=+1$ operators involving three leptons and $u,d$ quarks responsible for nucleon decays.  
The chiral irrep. column lists the irreducible representations under the chiral group $\rm SU(3)_\tL\otimes SU(3)_\tR$.
The \# column counts the number of independent operators for general $n_{\ell}$ charged leptons and $n_\nu$ neutrinos, and the numbers in the angle brackets are for $n_\ell=n_\nu=3$.  }
\label{tab:BNVdim9opeLp1A}
\end{table}

\begin{table}[ht]
\center
\resizebox{\linewidth}{!}{
\renewcommand{\arraystretch}{1.05}
\begin{tabular}{|c|c|c|c|c|c|}
\hline   
&~~Notation~~& Operator &~~Chiral Irrep.~~& 
\# of operators
&~~Process~~
\\\hline%%DeltaL=+1
\multirow{25}*{\rotatebox[origin=c]{90}{
\pmb{$\Delta L = + 1 $} } }
& $\calO_{\bar\ell \ell\nu,xyz}^{\tV\tR\tV\tR\tS\tR}$
& $(\overline{\ell_{\tR x}}\gamma_\mu \ell_{\tR y})
(\overline{\nu_{\tL z}^\C}\gamma^\mu {\cal N}_{dud}^{\tR\tR}) $  
& ${\color{purple}\pmb{1}_\tL\otimes\pmb{8}_\tR}$
& $n_\ell^2 n_\nu\,\langle 27 \rangle$ 
&\multirow{17}*{\rotatebox[origin=c]{90}{
$n\to\ell_x^- \ell_y^+ \bar\nu_z$ } }
\\\cline{2-5}
& $\calO_{\bar\ell \ell\nu,xyz}^{\tV\tL\tV\tR\tS\tR}$
& $(\overline{\ell_{\tL x}}\gamma_\mu \ell_{\tL y})
(\overline{\nu_{\tL z}^\C}\gamma^\mu {\cal N}_{dud}^{\tR\tR}) $  
& ${\color{purple}\pmb{1}_\tL\otimes\pmb{8}_\tR}$
& $n_\ell^2 n_\nu\,\langle 27 \rangle$ &
\\\cline{2-5}%
& $\calO_{\bar\ell \ell\nu,xyz}^{\tV\tR\tV\tR\tS\tL}$
& $(\overline{\ell_{\tR x}}\gamma_\mu \ell_{\tR y})
(\overline{\nu_{\tL z}^\C}\gamma^\mu {\cal N}_{dud}^{\tR\tL}) $  
& ${\color{Purple}\bar{\pmb{3}}_\tL\otimes\pmb{3}_\tR   }$
& $n_\ell^2 n_\nu\,\langle 27 \rangle$ &
\\
& $\calO_{\bar\ell \ell\nu,xyz}^{\tV\tR\tS\tL\tV\tR1}$
& $(\overline{\ell_{\tR x}}\gamma_\mu \ell_{\tR y})
(\overline{\nu_{\tL z}^\C}{\cal N}_{udd}^{\tL\tR,\mu}) $  
& ${\color{Green}\pmb{6}_\tL \otimes\pmb{3}_\tR  }$
& $n_\ell^2 n_\nu\,\langle 27 \rangle$ &
\\\cline{2-5}%
& $\calO_{\bar\ell \ell\nu,xyz}^{\tV\tR\tS\tL\tV\tR2}$
& $(\overline{\ell_{\tR x}}\gamma_\mu \ell_{\tR y})
(\overline{\nu_{\tL z}^\C}{\cal N}_{ddu}^{\tL\tR,\mu}) $  
& ${\color{Green}\pmb{6}_\tL \otimes\pmb{3}_\tR }$
& $n_\ell^2 n_\nu\,\langle 27 \rangle$ &
\\\cline{2-5}%
& $\calO_{\bar\ell \ell\nu,xyz}^{\tV\tL\tV\tR\tS\tL}$
& $(\overline{\ell_{\tL x}}\gamma_\mu \ell_{\tL y})
(\overline{\nu_{\tL z}^\C}\gamma^\mu {\cal N}_{dud}^{\tR\tL}) $  
& ${\color{Purple}\bar{\pmb{3}}_\tL\otimes\pmb{3}_\tR }$
& $n_\ell^2 n_\nu\,\langle 27 \rangle$ &
\\
& $\calO_{\bar\ell \ell\nu,xyz}^{\tV\tL\tS\tL\tV\tR1}$
& $(\overline{\ell_{\tL x}}\gamma_\mu \ell_{\tL y})
(\overline{\nu_{\tL z}^\C}{\cal N}_{udd}^{\tL\tR,\mu}) $  
& ${\color{Green}\pmb{6}_\tL \otimes\pmb{3}_\tR }$
& $n_\ell^2 n_\nu\,\langle 27 \rangle$ &
\\\cline{2-5}%
& $\calO_{\bar\ell \ell\nu,xyz}^{\tV\tL\tS\tL\tV\tR2}$
& $(\overline{\ell_{\tL x}}\gamma_\mu \ell_{\tL y})
(\overline{\nu_{\tL z}^\C}{\cal N}_{ddu}^{\tL\tR,\mu}) $  
& ${\color{Green}\pmb{6}_\tL \otimes\pmb{3}_\tR }$
& $n_\ell^2 n_\nu\,\langle 27 \rangle$ &
\\\cline{2-5}%
& $\calO_{\bar\ell \ell\nu,xyz}^{\tS\tL\tS\tL\tS\tL}$
& $(\overline{\ell_{\tR x}}\ell_{\tL y})
(\overline{\nu_{\tL z}^\C}{\cal N}_{dud}^{\tL\tL}) $  
& ${\color{purple}\pmb{8}_\tL \otimes\pmb{1}_\tR   }$
& $n_\ell^2 n_\nu\,\langle 27 \rangle$ &
\\
& $\calO_{\bar\ell \ell\nu,xyz}^{\tT\tL\tT\tL\tS\tL}$
& $(\overline{\ell_{\tR x}}\sigma_{\mu\nu}\ell_{\tL y})
(\overline{\nu_{\tL z}^\C}\sigma^{\mu\nu} {\cal N}_{dud}^{\tL\tL}) $  
& ${\color{purple}\pmb{8}_\tL \otimes\pmb{1}_\tR }$
& $n_\ell^2 n_\nu\,\langle 27 \rangle$ &
\\
& $\calO_{\bar\ell \ell\nu,xyz}^{\tT\tL\tS\tL\tT\tL}$
& $(\overline{\ell_{\tR x}}\sigma_{\mu\nu}\ell_{\tL y})
(\overline{\nu_{\tL z}^\C}{\cal N}_{udd}^{\tL\tL,\mu\nu}) $  
& ${\color{Blue} \pmb{10}_\tL \otimes\pmb{1}_\tR  }$
& $n_\ell^2 n_\nu\,\langle 27 \rangle$ &
\\\cline{2-5}%
& $\calO_{\bar\ell \ell\nu,xyz}^{\tS\tR\tS\tL\tS\tL}$
& $(\overline{\ell_{\tL x}}\ell_{\tR y})
(\overline{\nu_{\tL z}^\C}{\cal N}_{dud}^{\tL\tL}) $  
& ${\color{purple}\pmb{8}_\tL \otimes\pmb{1}_\tR }$
& $n_\ell^2 n_\nu\,\langle 27 \rangle$ &
\\\cline{2-5}%
& $\calO_{\bar\ell \ell\nu,xyz}^{\tS\tL\tS\tL\tS\tR}$
& $(\overline{\ell_{\tR x}}\ell_{\tL y})
(\overline{\nu_{\tL z}^\C}{\cal N}_{dud}^{\tL\tR}) $  
& ${\color{Purple}\pmb{3}_\tL \otimes\bar{\pmb{3}}_\tR   }$
& $n_\ell^2 n_\nu\,\langle 27 \rangle$ &
\\
& $\calO_{\bar\ell \ell\nu,xyz}^{\tT\tL\tT\tL\tS\tR}$
& $(\overline{\ell_{\tR x}}\sigma_{\mu\nu}\ell_{\tL y})
(\overline{\nu_{\tL z}^\C}\sigma^{\mu\nu}{\cal N}_{dud}^{\tL\tR}) $  
& ${\color{Purple}\pmb{3}_\tL \otimes\bar{\pmb{3}}_\tR }$
& $n_\ell^2 n_\nu\,\langle 27 \rangle$ &
\\\cline{2-5}%
& $\calO_{\bar\ell \ell\nu,xyz}^{\tS\tR\tS\tL\tS\tR}$
& $(\overline{\ell_{\tL x}}\ell_{\tR y})
(\overline{\nu_{\tL z}^\C}{\cal N}_{dud}^{\tL\tR}) $  
& ${\color{Purple}\pmb{3}_\tL \otimes\bar{\pmb{3}}_\tR }$
& $n_\ell^2 n_\nu\,\langle 27 \rangle$ &
\\
& $\calO_{\bar\ell \ell\nu,xyz}^{\tT\tR\tV\tR\tV\tL1}$ 
& $(\overline{\ell_{\tL x}}\sigma_{\mu\nu}\ell_{\tR y})
(\overline{\nu_{\tL z}^\C}\gamma^\mu{\cal N}_{udd}^{\tR\tL,\nu}) $  
& ${\color{Green}\pmb{3}_\tL\otimes\pmb{6}_\tR    }$
& $n_\ell^2 n_\nu\,\langle 27 \rangle$ &
\\\cline{2-5}%
& $\calO_{\bar\ell \ell\nu,xyz}^{\tT\tR\tV\tR\tV\tL2}$ 
& $(\overline{\ell_{\tL x}}\sigma_{\mu\nu} \ell_{\tR y})
(\overline{\nu_{\tL z}^\C}\gamma^\mu{\cal N}_{ddu}^{\tR\tL,\nu}) $  
& ${\color{Green}\pmb{3}_\tL\otimes\pmb{6}_\tR }$
& $n_\ell^2 n_\nu\,\langle 27 \rangle$ &
\\\clineB{2-6}{2}%%lvv 
& $\calO_{\bar\nu \nu \ell,xyz}^{\tV\tL\tV\tR\tS\tR}$
& $(\overline{\nu_{\tL x}}\gamma_\mu \nu_{\tL y})
(\overline{\ell_{\tL z}^\C}\gamma^\mu {\cal N}_{uud}^{\tR\tR}) $  
& ${\color{purple}\pmb{1}_\tL\otimes\pmb{8}_\tR}$
& $n_\nu^2n_\ell \,\langle 27 \rangle$ &\multirow{8}*{\rotatebox[origin=c]{90}{
$p\to\nu_x \bar\nu_y \ell_z^+$ } }
\\\cline{2-5}%
& $\calO_{\bar\nu \nu \ell,xyz}^{\tV\tL\tV\tR\tS\tL}$
& $(\overline{\nu_{\tL x}}\gamma_\mu \nu_{\tL y})
(\overline{\ell_{\tL z}^\C}\gamma^\mu {\cal N}_{uud}^{\tR\tL}) $  
& ${\color{Purple}\bar{\pmb{3}}_\tL\otimes\pmb{3}_\tR }$
& $n_\nu^2n_\ell \,\langle 27 \rangle$ &
\\
& $\calO_{\bar\nu \nu \ell,xyz}^{\tV\tL\tS\tL\tV\tR1}$
& $(\overline{\nu_{\tL x}}\gamma_\mu \nu_{\tL y})
(\overline{\ell_{\tL z}^\C}{\cal N}_{udu}^{\tL\tR,\mu}) $  
& ${\color{Green}\pmb{6}_\tL \otimes\pmb{3}_\tR }$
& $n_\nu^2n_\ell \,\langle 27 \rangle$ &
\\\cline{2-5}%
& $\calO_{\bar\nu \nu \ell,xyz}^{\tV\tL\tS\tL\tV\tR2}$
& $(\overline{\nu_{\tL x}}\gamma_\mu \nu_{\tL y})
(\overline{\ell_{\tL z}^\C}{\cal N}_{uud}^{\tL\tR,\mu}) $  
& ${\color{Green}\pmb{6}_\tL \otimes\pmb{3}_\tR }$
& $n_\nu^2n_\ell \,\langle 27 \rangle$ &
\\\cline{2-5}%
& $\calO_{\bar\nu \nu \ell,xyz}^{\tV\tL\tV\tL\tS\tL}$
& $(\overline{\nu_{\tL x}}\gamma_\mu\nu_{\tL y})
(\overline{\ell_{\tR z}^\C}\gamma^\mu{\cal N}_{uud}^{\tL\tL}) $  
& ${\color{purple}\pmb{8}_\tL \otimes\pmb{1}_\tR }$
& $n_\nu^2n_\ell \,\langle 27 \rangle$ &
\\\cline{2-5}%
& $\calO_{\bar\nu \nu \ell,xyz}^{\tV\tL\tV\tL\tS\tR}$
& $(\overline{\nu_{\tL x}}\gamma_\mu\nu_{\tL y})
(\overline{\ell_{\tR z}^\C}\gamma^\mu{\cal N}_{uud}^{\tL\tR}) $  
& ${\color{Purple}\pmb{3}_\tL \otimes\bar{\pmb{3}}_\tR }$
& $n_\nu^2n_\ell \,\langle 27 \rangle$ &
\\
& $\calO_{\bar\nu \nu \ell,xyz}^{\tV\tL\tS\tR\tV\tL1}$ 
& $(\overline{\nu_{\tL x}}\gamma_{\mu}\nu_{\tL y})
(\overline{\ell_{\tR z}^\C}{\cal N}_{udu}^{\tR\tL,\mu}) $  
& ${\color{Green}\pmb{3}_\tL\otimes\pmb{6}_\tR    }$
& $n_\nu^2n_\ell \,\langle 27 \rangle$ &
\\\cline{2-5}%
& $\calO_{\bar\nu \nu \ell,xyz}^{\tV\tL\tS\tR\tV\tL2}$ 
& $(\overline{\nu_{\tL x}}\gamma_\mu \nu_{\tL y})
(\overline{\ell_{\tR z}^\C}{\cal N}_{uud}^{\tR\tL,\mu}) $  
& ${\color{Green}\pmb{3}_\tL\otimes\pmb{6}_\tR }$
& $n_\nu^2n_\ell \,\langle 27 \rangle$ &
\\\clineB{2-6}{2}%%vvv
& $\calO_{\nu\nu\bar\nu,xyz}^{\tS\tL\tS\tR\tS\tR}$
& $(\overline{\nu_{\tL x}^\C}\nu_{\tL y})(\overline{\nu_{\tL z}}{\cal N}_{dud}^{\tR\tR})$
& ${\color{purple} \pmb{1}_\tL \otimes \pmb{8}_\tR}$
& $\frac{1}{2}n_\nu^2(n_\nu+1)\,\langle 18 \rangle$
&\multirow{4}*{
\rotatebox[origin=c]{90}{
\shortstack{$n\to \bar\nu_x\bar\nu_y \nu_z $}
}}
\\\cline{2-5}%
& $\calO_{\nu\nu\bar\nu,xyz}^{\tS\tL\tS\tR\tS\tL}$
& $(\overline{\nu_{\tL x}^\C}\nu_{\tL y})(\overline{\nu_{\tL z}}{\cal N}_{dud}^{\tR\tL})$
&~~${\color{Purple}{  \bar{\pmb{3}}_\tL \otimes \pmb{3}}_\tR}$~~
& $\frac{1}{2}n_\nu^2(n_\nu+1)\,\langle 18 \rangle$ &
\\%%%%
& $\calO_{\nu\nu\bar\nu,xyz}^{\tT\tL\tV\tL\tV\tR1}$
&~~$(\overline{\nu_{\tL x}^\C}\sigma_{\mu\nu}\nu_{\tL y})(\overline{\nu_{\tL z}}\gamma^\mu {\cal N}_{udd}^{\tL\tR,\nu})$~~
& ${\color{Green} \pmb{6}_\tL \otimes \pmb{3}_\tR}$
& $\frac{1}{2}n_\nu^2(n_\nu-1)\,\langle 9 \rangle$ &
\\\cline{2-5}%
& $\calO_{\nu\nu\bar\nu,xyz}^{\tT\tL\tV\tL\tV\tR2}$
& $(\overline{\nu_{\tL x}^\C}\sigma_{\mu\nu}\nu_{\tL y})(\overline{\nu_{\tL z}}\gamma^\mu {\cal N}_{ddu}^{\tL\tR,\nu})$
&~~~${\color{Green} \pmb{6}_\tL \otimes  \pmb{3}_\tR}$~~~
&~~~~$\frac{1}{2}n_\nu^2(n_\nu-1)\,\langle 9 \rangle$~~~~&
\\\hline
\end{tabular}
}
\caption{
Continuation of \cref{tab:BNVdim9opeLp1A}.
}
\label{tab:BNVdim9opeLp1B}
\end{table}

\begin{table}[ht]
\center
\resizebox{\linewidth}{!}{
\renewcommand{\arraystretch}{1.05}
\begin{tabular}{|c|c|c|c|c|c|}
\hline   
&~~Notation~~& Operator &~~Chiral Irrep.~~& 
\# of operators
&~~Process~~
\\\hline%%DeltaL=+1
\multirow{30}*{\rotatebox[origin=c]{90}{
\pmb{$\Delta L = - 1 $} } }
& $\calO_{\bar\ell \ell\bar\nu,xyz}^{\tV\tL\tV\tL\tS\tL}$
& $(\overline{\ell_{\tL x}}\gamma_\mu \ell_{\tL y})
(\overline{\nu_{\tL z}}\gamma^\mu {\cal N}_{dud}^{\tL\tL}) $  
& ${\color{purple}\pmb{8}_\tL \otimes \pmb{1}_\tR}$
& $n_\ell^2 n_\nu\,\langle 27 \rangle$ 
&\multirow{17}*{\rotatebox[origin=c]{90}{
$n\to\ell_x^- \ell_y^+ \nu_z$ } }
\\\cline{2-5}
& $\calO_{\bar\ell \ell\bar\nu,xyz}^{\tV\tR\tV\tL\tS\tL}$
& $(\overline{\ell_{\tR x}}\gamma_\mu \ell_{\tR y})
(\overline{\nu_{\tL z}}\gamma^\mu {\cal N}_{dud}^{\tL\tL}) $  
& ${\color{purple}\pmb{8}_\tL \otimes \pmb{1}_\tR}$
& $n_\ell^2 n_\nu\,\langle 27 \rangle$ &
\\\cline{2-5}%
& $\calO_{\bar\ell \ell\bar\nu,xyz}^{\tV\tL\tV\tL\tS\tR}$
& $(\overline{\ell_{\tL x}}\gamma_\mu \ell_{\tL y})
(\overline{\nu_{\tL z}}\gamma^\mu {\cal N}_{dud}^{\tL\tR}) $  
& ${\color{Purple}\pmb{3}_\tL \otimes \bar{\pmb{3}}_\tR }$
& $n_\ell^2 n_\nu\,\langle 27 \rangle$ &
\\
& $\calO_{\bar\ell \ell\bar\nu,xyz}^{\tV\tL\tS\tR\tV\tL1}$
& $(\overline{\ell_{\tL x}}\gamma_\mu \ell_{\tL y})
(\overline{\nu_{\tL z}}{\cal N}_{udd}^{\tR\tL,\mu}) $  
& ${\color{Green}\pmb{3}_\tL \otimes \pmb{6}_\tR }$
& $n_\ell^2 n_\nu\,\langle 27 \rangle$ &
\\\cline{2-5}%
& $\calO_{\bar\ell \ell\bar\nu,xyz}^{\tV\tL\tS\tR\tV\tL2}$
& $(\overline{\ell_{\tL x}}\gamma_\mu \ell_{\tL y})
(\overline{\nu_{\tL z}}{\cal N}_{ddu}^{\tR\tL,\mu}) $  
& ${\color{Green}\pmb{3}_\tL\otimes \pmb{6}_\tR }$
& $n_\ell^2 n_\nu\,\langle 27 \rangle$ &
\\\cline{2-5}%
& $\calO_{\bar\ell \ell\bar\nu,xyz}^{\tV\tR\tV\tL\tS\tR}$
& $(\overline{\ell_{\tR x}}\gamma_\mu \ell_{\tR y})
(\overline{\nu_{\tL z}}\gamma^\mu {\cal N}_{dud}^{\tL\tR}) $  
& ${\color{Purple}\pmb{3}_\tL \otimes \bar{\pmb{3}}_\tR }$
& $n_\ell^2 n_\nu\,\langle 27 \rangle$ &
\\
& $\calO_{\bar\ell \ell\bar\nu,xyz}^{\tV\tR\tS\tR\tV\tL1}$
& $(\overline{\ell_{\tR x}}\gamma_\mu \ell_{\tR y})
(\overline{\nu_{\tL z}}{\cal N}_{udd}^{\tR\tL,\mu}) $  
& ${\color{Green}\pmb{3}_\tL \otimes \pmb{6}_\tR }$
& $n_\ell^2 n_\nu\,\langle 27 \rangle$ &
\\\cline{2-5}%
& $\calO_{\bar\ell \ell\bar\nu,xyz}^{\tV\tR\tS\tR\tV\tL2}$
& $(\overline{\ell_{\tR x}}\gamma_\mu \ell_{\tR y})
(\overline{\nu_{\tL z}}{\cal N}_{ddu}^{\tR\tL,\mu}) $  
& ${\color{Green}\pmb{3}_\tL\otimes \pmb{6}_\tR }$
& $n_\ell^2 n_\nu\,\langle 27 \rangle$ &
\\\cline{2-5}%
& $\calO_{\bar\ell \ell\bar\nu,xyz}^{\tS\tR\tS\tR\tS\tR}$
& $(\overline{\ell_{\tL x}}\ell_{\tR y})
(\overline{\nu_{\tL z}}{\cal N}_{dud}^{\tR\tR}) $  
& ${\color{purple}\pmb{1}_\tL \otimes  \pmb{8}_\tR }$
& $n_\ell^2 n_\nu\,\langle 27 \rangle$ &
\\
& $\calO_{\bar\ell \ell\bar\nu,xyz}^{\tT\tR\tT\tR\tS\tR}$
& $(\overline{\ell_{\tL x}}\sigma_{\mu\nu}\ell_{\tR y})
(\overline{\nu_{\tL z}}\sigma^{\mu\nu} {\cal N}_{dud}^{\tR\tR}) $  
& ${\color{purple}\pmb{1}_\tL \otimes  \pmb{8}_\tR }$
& $n_\ell^2 n_\nu\,\langle 27 \rangle$ &
\\
& $\calO_{\bar\ell \ell\bar\nu,xyz}^{\tT\tR\tS\tR\tT\tR}$
& $(\overline{\ell_{\tL x}}\sigma_{\mu\nu}\ell_{\tR y})
(\overline{\nu_{\tL z}}{\cal N}_{udd}^{\tR\tR,\mu\nu}) $  
& ${\color{Blue} \pmb{1}_\tL\otimes \pmb{10}_\tR }$
& $n_\ell^2 n_\nu\,\langle 27 \rangle$ &
\\\cline{2-5}%
& $\calO_{\bar\ell \ell\bar\nu,xyz}^{\tS\tL\tS\tR\tS\tR}$
& $(\overline{\ell_{\tR x}}\ell_{\tL y})
(\overline{\nu_{\tL z}}{\cal N}_{dud}^{\tR\tR}) $  
& ${\color{purple}\pmb{1}_\tL \otimes  \pmb{8}_\tR }$
& $n_\ell^2 n_\nu\,\langle 27 \rangle$ &
\\\cline{2-5}%
& $\calO_{\bar\ell \ell\bar\nu,xyz}^{\tS\tR\tS\tR\tS\tL}$
& $(\overline{\ell_{\tL x}}\ell_{\tR y})
(\overline{\nu_{\tL z}}{\cal N}_{dud}^{\tR\tL}) $  
& ${\color{Purple}\bar{\pmb{3}}_\tL \otimes  \pmb{3}_\tR }$
& $n_\ell^2 n_\nu\,\langle 27 \rangle$ &
\\
& $\calO_{\bar\ell \ell\bar\nu,xyz}^{\tT\tR\tT\tR\tS\tL}$
& $(\overline{\ell_{\tL x}}\sigma_{\mu\nu}\ell_{\tR y})
(\overline{\nu_{\tL z}}\sigma^{\mu\nu}{\cal N}_{dud}^{\tR\tL}) $  
& ${\color{Purple}\bar{\pmb{3}}_\tL \otimes  \pmb{3}_\tR }$
& $n_\ell^2 n_\nu\,\langle 27 \rangle$ &
\\\cline{2-5}%
& $\calO_{\bar\ell \ell\bar\nu,xyz}^{\tS\tL\tS\tR\tS\tL}$
& $(\overline{\ell_{\tR x}}\ell_{\tL y})
(\overline{\nu_{\tL z}}{\cal N}_{dud}^{\tR\tL}) $  
& ${\color{Purple}\bar{\pmb{3}}_\tL \otimes  \pmb{3}_\tR }$
& $n_\ell^2 n_\nu\,\langle 27 \rangle$ &
\\
& $\calO_{\bar\ell \ell\bar\nu,xyz}^{\tT\tL\tV\tL\tV\tR1}$ 
& $(\overline{\ell_{\tR x}}\sigma_{\mu\nu}\ell_{\tL y})
(\overline{\nu_{\tL z}}\gamma^\mu{\cal N}_{udd}^{\tL\tR,\nu}) $  
& ${\color{Green}\pmb{6}_\tL \otimes  \pmb{3}_\tR }$
& $n_\ell^2 n_\nu\,\langle 27 \rangle$ &
\\\cline{2-5}%
& $\calO_{\bar\ell \ell\bar\nu,xyz}^{\tT\tL\tV\tL\tV\tR2}$ 
& $(\overline{\ell_{\tR x}}\sigma_{\mu\nu} \ell_{\tL y})
(\overline{\nu_{\tL z}}\gamma^\mu{\cal N}_{ddu}^{\tL\tR,\nu}) $  
& ${\color{Green}\pmb{6}_\tL\otimes \pmb{3}_\tR }$
& $n_\ell^2 n_\nu\,\langle 27 \rangle$ &
\\\clineB{2-6}{2}%%lvv 
& $\calO_{\bar\nu\bar\nu\ell,xyz}^{\tS\tR\tS\tL\tS\tL}$
& $(\overline{\nu_{\tL x}}\nu_{\tL y}^\C)(\overline{\ell_{\tL z}^\C}{\cal N}_{uud}^{\tL\tL})$
& ${\color{purple} \pmb{8}_\tL\otimes \pmb{1}_\tR }$
& $\frac{1}{2}n_\ell n_\nu(n_\nu+1)\,\langle 18 \rangle$ 
&\multirow{9}*{\rotatebox[origin=c]{90}{
$p\to \nu_x\nu_y\ell_z^+$ } }
\\\cline{2-5}% 
& $\calO_{\bar\nu\bar\nu\ell,xyz}^{\tS\tR\tS\tR\tS\tR}$
&$(\overline{\nu_{\tL x}}\nu_{\tL y}^\C)(\overline{\ell_{\tR z}^\C}{\cal N}_{uud}^{\tR\tR})$
& ${\color{purple}\pmb{1}_\tL \otimes \pmb{8}_\tR}$
& $\frac{1}{2}n_\ell n_\nu(n_\nu+1)\,\langle 18 \rangle$  &
\\
& $\calO_{\bar\nu\bar\nu\ell,xyz}^{\tT\tR\tT\tR\tS\tR}$
&$(\overline{\nu_{\tL x}}\sigma_{\mu\nu}\nu_{\tL y}^\C)(\overline{\ell_{\tR z}^\C}\sigma^{\mu\nu} {\cal N}_{uud}^{\tR\tR})$
& ${\color{purple}\pmb{1}_\tL \otimes \pmb{8}_\tR}$
& $\frac{1}{2}n_\ell n_\nu(n_\nu-1)\,\langle 9 \rangle$  &
\\
& $\calO_{\bar\nu\bar\nu\ell,xyz}^{\tT\tR\tS\tR\tT\tR}$ 
&$(\overline{\nu_{\tL x}}\sigma_{\mu\nu}\nu_{\tL y}^\C)(\overline{\ell_{\tR z}^\C}{\cal N}_{uud}^{\tR\tR,\mu\nu})$
& ${\color{Blue} \pmb{1}_\tL\otimes \pmb{10}_\tR }$
& $\frac{1}{2}n_\ell n_\nu(n_\nu-1)\,\langle 9 \rangle$ &
\\\cline{2-5}% 
& $\calO_{\bar\nu\bar\nu\ell,xyz}^{\tS\tR\tS\tL\tS\tR}$
& $(\overline{\nu_{\tL x}}\nu_{\tL y}^\C)(\overline{\ell_{\tL z}^\C}{\cal N}_{uud}^{\tL\tR})$
& ${\color{Purple}\pmb{3}_\tL\otimes \bar{\pmb{3}}_\tR }$
& $\frac{1}{2}n_\ell n_\nu(n_\nu+1)\,\langle 18 \rangle$ &
\\
& $\calO_{\bar\nu\bar\nu\ell,xyz}^{\tT\tR\tV\tR\tV\tL1}$
& $(\overline{\nu_{\tL x}}\sigma_{\mu\nu}\nu_{\tL y}^\C)(\overline{\ell_{\tL z}^\C}\gamma^\mu {\cal N}_{udu}^{\tR\tL,\nu})$
& ${\color{Green} \pmb{3}_\tL\otimes \pmb{6}_\tR}$
& $\frac{1}{2}n_\ell n_\nu(n_\nu-1)\,\langle 9 \rangle$ &
\\\cline{2-5}%
& $\calO_{\bar\nu\bar\nu\ell,xyz}^{\tT\tR\tV\tR\tV\tL2}$
& $(\overline{\nu_{\tL x}}\sigma_{\mu\nu}\nu_{\tL y}^\C)(\overline{\ell_{\tL z}^\C}\gamma^\mu {\cal N}_{uud}^{\tR\tL,\nu})$
& ${\color{Green}\pmb{3}_\tL\otimes \pmb{6}_\tR }$
& $\frac{1}{2}n_\ell n_\nu(n_\nu-1)\,\langle 9 \rangle$ &
\\\cline{2-5}% 
& $\calO_{\bar\nu\bar\nu\ell,xyz}^{\tS\tR\tS\tR\tS\tL}$
& $(\overline{\nu_{\tL x}}\nu_{\tL y}^\C)(\overline{\ell_{\tR z}^\C} {\cal N}_{uud}^{\tR\tL})$
& ${\color{Purple}\bar{\pmb{3}}_\tL \otimes  \pmb{3}_\tR }$
& $\frac{1}{2}n_\ell n_\nu(n_\nu+1)\,\langle 18 \rangle$ &
\\
& $\calO_{\bar\nu\bar\nu\ell,xyz}^{\tT\tR\tT\tR\tS\tL}$ 
& $(\overline{\nu_{\tL x}}\sigma_{\mu\nu}\nu_{\tL y}^\C)(\overline{\ell_{\tR z}^\C} \sigma^{\mu\nu} {\cal N}_{uud}^{\tR\tL})$
& ${\color{Purple}\bar{\pmb{3}}_\tL \otimes  \pmb{3}_\tR }$
& $\frac{1}{2}n_\ell n_\nu(n_\nu-1)\,\langle 9 \rangle$ &
\\\clineB{2-6}{2}%%vvv
& $\calO_{\bar\nu\bar\nu\nu,xyz}^{\tS\tR\tS\tL\tS\tL}$
& $(\overline{\nu_{\tL x}}\nu_{\tL y}^\C)(\overline{\nu_{\tL z}^\C}{\cal N}_{dud}^{\tL\tL})$
& ${\color{purple}\pmb{8}_\tL \otimes  \pmb{1}_\tR}$
& $\frac{1}{2}n_\nu^2(n_\nu+1)\,\langle 18 \rangle$
&\multirow{4}*{
\rotatebox[origin=c]{90}{
\shortstack{$n\to \nu_x\nu_y \bar\nu_z $}
}}
\\\cline{2-5}%
& $\calO_{\bar\nu\bar\nu\nu,xyz}^{\tS\tR\tS\tL\tS\tR}$
& $(\overline{\nu_{\tL x}}\nu_{\tL y}^\C)(\overline{\nu_{\tL z}^\C}{\cal N}_{dud}^{\tL\tR})$
&~~${\color{Purple}{\pmb{3}}_\tL \otimes \bar{\pmb{3}}_\tR}$~~
& $\frac{1}{2}n_\nu^2(n_\nu+1)\,\langle 18 \rangle$ &
\\%%%%
& $\calO_{\bar\nu\bar\nu\nu,xyz}^{\tT\tR\tV\tR\tV\tL1}$
&~~$(\overline{\nu_{\tL x}}\sigma_{\mu\nu}\nu_{\tL y}^\C)(\overline{\nu_{\tL z}^\C}\gamma^\mu {\cal N}_{udd}^{\tR\tL,\nu})$~~
& ${\color{Green}\pmb{3}_\tL \otimes \pmb{6}_\tR}$
& $\frac{1}{2}n_\nu^2(n_\nu-1)\,\langle 9 \rangle$ &
\\\cline{2-5}%
& $\calO_{\bar\nu\bar\nu\nu,xyz}^{\tT\tR\tV\tR\tV\tL2}$
& $(\overline{\nu_{\tL x}}\sigma_{\mu\nu}\nu_{\tL y}^\C)(\overline{\nu_{\tL z}^\C}\gamma^\mu {\cal N}_{ddu}^{\tR\tL,\nu})$
&~~~~${\color{Green}\pmb{3}_\tL \otimes \pmb{6}_\tR}$~~~~
&~~~~$\frac{1}{2}n_\nu^2(n_\nu-1)\,\langle 9 \rangle$~~~~&
\\\hline
\end{tabular}
}
\caption{Same as \cref{tab:BNVdim9opeLp1A,tab:BNVdim9opeLp1B} but for $\Delta L=-1$ sector.}
\label{tab:BNVdim9opeLm1}
\end{table}

\begin{table}[ht]
\center
\resizebox{\linewidth}{!}{
\renewcommand{\arraystretch}{1.}
\begin{tabular}{|c|c|c|c|c|c|}
\hline   
&~~Notation~~& Operator &~~Chiral Irrep.~~& 
\# of operators
&~~Process~~
\\\hline%%DeltaL=-3
\multirow{10}*{\rotatebox[origin=c]{90}{
\pmb{$\Delta L = + 3 $} } } 
& $\calO_{\nu\nu\ell,xyz}^{\tS\tL\tS\tR\tS\tR}$
& $(\overline{\nu_{\tL x}^\C}\nu_{\tL y})(\overline{\ell_{\tR z}^\C}{\cal N}_{uud}^{\tR\tR})$
& ${\color{purple} \pmb{1}_\tL\otimes \pmb{8}_\tR }$
& $\frac{1}{2}n_\ell n_\nu(n_\nu+1)\,\langle 18 \rangle$ 
& \multirow{9}*{\rotatebox[origin=c]{90}{
$p\to \bar\nu_x\bar \nu_y \ell_z^+$ } }
\\\cline{2-5}%  
& $\calO_{\nu\nu\ell,xyz}^{\tS\tL\tS\tL\tS\tL}$ 
&$(\overline{\nu_{\tL x}^\C}\nu_{\tL y})(\overline{\ell_{\tL z}^\C}{\cal N}_{uud}^{\tL\tL})$
& ${\color{purple}\pmb{8}_\tL \otimes \pmb{1}_\tR}$
& $\frac{1}{2}n_\ell n_\nu(n_\nu+1)\,\langle 18 \rangle$  &
\\
& $\calO_{\nu\nu\ell,xyz}^{\tT\tL\tT\tL\tS\tL}$
&~~$(\overline{\nu_{\tL x}^\C}\sigma_{\mu\nu}\nu_{\tL y})(\overline{\ell_{\tL z}^\C}\sigma^{\mu\nu} {\cal N}_{uud}^{\tL\tL})$~~
& ${\color{purple}\pmb{8}_\tL \otimes \pmb{1}_\tR}$
& $\frac{1}{2}n_\ell n_\nu(n_\nu-1)\,\langle 9 \rangle$ 
&
\\
& $\calO_{\nu\nu\ell,xyz}^{\tT\tL\tS\tL\tT\tL}$ 
&$(\overline{\nu_{\tL x}^\C}\sigma_{\mu\nu}\nu_{\tL y})(\overline{\ell_{\tL z}^\C}{\cal N}_{uud}^{\tL\tL,\mu\nu})$
& ${\color{Blue} \pmb{10}_\tL\otimes \pmb{1}_\tR }$
& $\frac{1}{2}n_\ell n_\nu(n_\nu-1)\,\langle 9 \rangle$ &
\\\cline{2-5}% 
&$\calO_{\nu\nu\ell,xyz}^{\tS\tL\tS\tR\tS\tL}$ 
& $(\overline{\nu_{\tL x}^\C}\nu_{\tL y})(\overline{\ell_{\tR z}^\C}{\cal N}_{uud}^{\tR\tL})$
& ${\color{Purple}\bar{\pmb{3}}_\tL\otimes \pmb{3}_\tR }$
& $\frac{1}{2}n_\ell n_\nu(n_\nu+1)\,\langle 18 \rangle$ &
\\
& $\calO_{\nu\nu\ell,xyz}^{\tT\tL\tV\tL\tV\tR1}$ 
& $(\overline{\nu_{\tL x}^\C}\sigma_{\mu\nu}\nu_{\tL y})(\overline{\ell_{\tR z}^\C}\gamma^\mu {\cal N}_{udu}^{\tL\tR,\nu})$
& ${\color{Green} \pmb{6}_\tL\otimes \pmb{3}_\tR}$
& $\frac{1}{2}n_\ell n_\nu(n_\nu-1)\,\langle 9 \rangle$ &
\\\cline{2-5}%
& $\calO_{\nu\nu\ell,xyz}^{\tT\tL\tV\tL\tV\tR2}$ 
& $(\overline{\nu_{\tL x}^\C}\sigma_{\mu\nu}\nu_{\tL y})(\overline{\ell_{\tR z}^\C}\gamma^\mu {\cal N}_{uud}^{\tL\tR,\nu})$
& ${\color{Green}\pmb{6}_\tL\otimes \pmb{3}_\tR }$
& $\frac{1}{2}n_\ell n_\nu(n_\nu-1)\,\langle 9 \rangle$ &
\\\cline{2-5}% 
& $\calO_{\nu\nu\ell,xyz}^{\tS\tL\tS\tL\tS\tR}$ 
& $(\overline{\nu_{\tL x}^\C}\nu_{\tL y})(\overline{\ell_{\tL z}^\C} {\cal N}_{uud}^{\tL\tR})$
& ${\color{Purple}\pmb{3}_\tL \otimes  \bar{\pmb{3}}_\tR }$
& $\frac{1}{2}n_\ell n_\nu(n_\nu+1)\,\langle 18 \rangle$ &
\\
& $\calO_{\nu\nu\ell,xyz}^{\tT\tL\tT\tL\tS\tR}$ 
& $(\overline{\nu_{\tL x}^\C}\sigma_{\mu\nu}\nu_{\tL y})(\overline{\ell_{\tL z}^\C} \sigma^{\mu\nu} {\cal N}_{uud}^{\tL\tR})$
& ${\color{Purple}\pmb{3}_\tL \otimes  \bar{\pmb{3}}_\tR }$
& $\frac{1}{2}n_\ell n_\nu(n_\nu-1)\,\langle 9 \rangle$ &
\\\clineB{2-6}{2}%%
& $\calO_{\nu\nu\nu,xyz}^{\tS\tL\tS\tL\tS\tL}$
& $(\overline{\nu_{\tL x}^\C}\nu_{\tL [y})(\overline{\nu_{\tL z]}^\C}{\cal N}_{dud}^{\tL\tL})$
& ${\color{purple}\pmb{8}_\tL \otimes  \pmb{1}_\tR}$
& ${1\over 3} n_\nu (n_\nu^2-1)\,\langle 8 \rangle$ 
&\multirow{3}*{\rotatebox[origin=c]{90}{$n\to\bar\nu\bar\nu\bar\nu$ } } 
\\
& $\calO_{\nu\nu\nu,xyz}^{\tT\tL\tS\tL\tT\tL}$
& $(\overline{\nu_{\tL [x}^\C}\sigma_{\mu\nu}\nu_{\tL y})(\overline{\nu_{\tL z]}^\C}{\cal N}_{udd}^{\tL\tL,\mu\nu})$ 
& ${\color{Blue}\pmb{10}_\tL \otimes \pmb{1}_\tR}$
& ${1\over 6} n_\nu (n_\nu-2)(n_\nu-1)\,\langle 1 \rangle$ &
\\\cline{2-5}%
& $\calO_{\nu\nu\nu,xyz}^{\tS\tL\tS\tL\tS\tR}$
& $(\overline{\nu_{\tL x}^\C}\nu_{\tL [y})(\overline{\nu_{\tL z]}^\C}{\cal N}_{dud}^{\tL\tR})$
& ${\color{Purple}\pmb{3}_\tL \otimes \bar{\pmb{3}}_\tR}$
& ${1\over 3} n_\nu (n_\nu^2-1)\,\langle 8 \rangle$ &
\\\cline{0-1}\clineB{2-6}{2}%%
\multirow{3}*{\rotatebox[origin=c]{90}{
\pmb{$\Delta L = - 3 $} } } 
& $\calO_{\bar\nu\bar\nu\bar\nu,xyz}^{\tS\tR\tS\tR\tS\tR}$
& $(\overline{\nu_{\tL x}}\nu_{\tL [y}^\C)(\overline{\nu_{\tL z]}}{\cal N}_{dud}^{\tR\tR})$
& ${\color{purple}\pmb{1}_\tL \otimes  \pmb{8}_\tR}$
& ${1\over 3} n_\nu (n_\nu^2-1)\,\langle 8 \rangle$ 
&\multirow{3}*{\rotatebox[origin=c]{90}{
$n\to\nu\nu\nu$ } } 
\\
& $\calO_{\bar\nu\bar\nu\bar\nu,xyz}^{\tT\tR\tS\tR\tT\tR}$
& $(\overline{\nu_{\tL [x}}\sigma_{\mu\nu}\nu_{\tL y}^\C)(\overline{\nu_{\tL z]}}{\cal N}_{udd}^{\tR\tR,\mu\nu})$ 
& ${\color{Blue}\pmb{1}_\tL \otimes \pmb{10}_\tR}$
&~~${1\over 6} n_\nu (n_\nu-2)(n_\nu-1)\,\langle 1 \rangle$~~&
\\\cline{2-5}%
& $\calO_{\bar\nu\bar\nu\bar\nu,xyz}^{\tS\tR\tS\tR\tS\tL}$
& $(\overline{\nu_{\tL x}}\nu_{\tL [y}^\C)(\overline{\nu_{\tL z]}}{\cal N}_{dud}^{\tR\tL})$
& ${\color{Purple}\bar{\pmb{3}}_\tL \otimes \pmb{3}_\tR}$ 
& ${1\over 3} n_\nu (n_\nu^2-1)\,\langle 8 \rangle$ &
\\\hline 
\end{tabular}
}
\caption{Same as \cref{tab:BNVdim9opeLp1A,tab:BNVdim9opeLp1B} but for $\Delta L=\pm3$ sectors.
Here $A_{[y}B_{z]} \equiv \frac{1}{2}(A_{y}B_{z}-A_{z}B_{y})$ and 
$A_{[x}B_y C_{z]} \equiv \frac{1}{6} [A_{z}B_{y}C_{z} + \text{5 totally antisymmetric perm. of } (x,y,z)]$ for operators involving three neutrinos.
}
\label{tab:BNVdim9opeL3}
\end{table}

Within the LEFT framework, the dim-9 operators contributing to nucleon triple-lepton decays consist of three quarks---composed of up ($u$) and down ($d$) quarks---and three leptons. By applying Fierz transformations, it is always possible to arrange two leptons into a single bilinear and two quarks into another bilinear.
As shown in Ref.~\cite{Liao:2025vlj},
a generic BNV triple-quark operator made of three light quarks,  without being acted upon by covariant derivatives,
can always be expressed in terms of the following four structures 
\begin{align}
{\cal O}_{a}^{uvw} = \overline{\Psi_{a}}{\cal N}_{uvw}^{\tL\tL},\,\,
{\cal O}_{b}^{uvw} =\overline{\Psi_{b}}{\cal N}_{uvw}^{\tR\tL}, \,\,
{\cal O}_{c}^{uvw} = \overline{\Psi_{c,\mu}}
{\cal N}_{uvw}^{\tL\tR,\mu},\,\,
{\cal O}_{d}^{uvw} = \overline{\Psi_{d,\mu\nu}}
{\cal N}_{uvw}^{\tL\tL,\mu\nu}, 
\label{eq:LEFT3qO}
\end{align} 
and their chiral partners with $\tL \leftrightarrow \tR$. 
Here $u,v,w=u,d,s$ denote quark flavor indices, 
and the $\Psi$s are nonquark fields that correspond to combinations of triple-lepton fields relevant to our current focus.  
The ${\cal N}$s are formed with three chiral quark fields in a definite irrep of the chiral group $\rm SU(3)_\tL\otimes SU(3)_\tR$, 
\begin{subequations}
\label{eq:Nyzw}
\begin{align}
{\cal N}_{uvw}^{\tL\tL} & \equiv  q_{\tL, u}^\alpha \big(\overline{ q_{\tL, v}^{\beta \C} }\, q_{\tL, w}^\gamma \big)\epsilon_{\alpha \beta \gamma} \;\in\; \pmb{8}_\tL \otimes  \pmb{1}_\tR , 
\\
{\cal N}_{uvw}^{\tR\tL} & \equiv q_{\tR, u}^\alpha \big(\overline{ q_{\tL, v}^{\beta \C} }\, q_{\tL,w}^\gamma\big)\epsilon_{\alpha \beta \gamma} \;\in\; 
\bar{\pmb{3}}_\tL \otimes \pmb{3}_\tR , 
\\
{\cal N}_{uvw}^{\tL\tR,\mu} & \equiv q_{\tL,\{u}^\alpha \big(\overline{ q_{\tL, v\}}^{\beta \C} }\, \gamma^\mu q_{\tR,w}^\gamma\big)\epsilon_{\alpha \beta \gamma}
\;\in\; \pmb{6}_\tL \otimes \pmb{3}_\tR ,
\\
{\cal N}_{uvw}^{\tL\tL,\mu\nu} & \equiv  q_{\tL, \{u}^\alpha \big(\overline{ q_{\tL, v}^{\beta \C} }\, \sigma^{\mu\nu} q_{\tL, w\} }^\gamma \big)\epsilon_{\alpha \beta \gamma} \;\in\; \pmb{10}_\tL \otimes  \pmb{1}_\tR, 
\end{align}
\end{subequations}  
where curly brackets indicate symmetrization over all flavor indices of fields with the same chirality, viz.,
$A_{\{u}B_{v\}} \equiv \frac{1}{2}(A_{u}B_{v}+A_{v}B_{u})$ and 
$A_{\{u}B_v C_{w\}} \equiv \frac{1}{6} [A_{u}B_{v}C_{w} + \text{5 perm. of } (u,v,w)]$.

In our case, the triple-quark building blocks consist of two field configurations: $uud$ and $ddu$, corresponding to the proton and neutron, respectively. According to the chiral matching convention in \cite{Liao:2025vlj}, the independent triple-quark components can be chosen as follows:
\begin{subequations}
\label{eq:Nqqq}
\begin{align}
\text{proton mode:\quad}& {\cal N}_{uud}^{\tL\tL(\tR\tR)}; \quad 
 {\cal N}_{uud}^{\tR\tL(\tL\tR)};\quad 
 {\cal N}_{uud}^{\tL\tR(\tR\tL),\mu},\,\, 
 {\cal N}_{udu}^{\tL\tR(\tR\tL),\mu};\quad 
 {\cal N}_{uud}^{\tL\tL(\tR\tR),\mu\nu},
 \\
\text{neutron mode:\quad}& {\cal N}_{dud}^{\tL\tL(\tR\tR)};\quad
{\cal N}_{dud}^{\tR\tL(\tL\tR)};\quad 
{\cal N}_{ddu}^{\tL\tR(\tR\tL),\mu},\,\, 
{\cal N}_{udd}^{\tL\tR(\tR\tL),\mu};\quad 
{\cal N}_{udd}^{\tL\tL(\tR\tR),\mu\nu}.
\end{align} 
\end{subequations}
With these objects, a generic dim-9 BNV operator relevant to nucleon triple-lepton decays can be arranged to take one of the following two factorized forms
\begin{align}
\calO_p \sim (\overline{l_x}\Gamma_1 l_y)(\overline{l_z}\Gamma_2 {\cal N}_{uud}^{\Gamma_3}), \quad 
\calO_n \sim (\overline{l_x} \Gamma_1 l_y)(\overline{l_z} \Gamma_2 {\cal N}_{ddu}^{\Gamma_3}),
\end{align}
where $l_{i} \in\{\ell_{\tL(\tR)},\,\ell_{\tL(\tR)}^\C,\,\nu_{\tL},\,\nu_{\tL}^\C\}$ with $i=e,\mu,\tau$ denotes a generic chiral lepton field, and $\ell_{\tL(\tR)}^\C$ and $\nu_{\tL}^\C$ represent the charge-conjugated fields of the charged leptons $\ell_{\tL(\tR)}$ and the left-handed neutrinos $\nu_{\tL}$, respectively. 
For the three lepton fields, Fierz transformations can again be applied to group the same type of leptons into a single current, and we adopt this bilinear field convention. 
The Dirac gamma structures $\Gamma_{1,2,3}$ are determined once all six chiral fields are specified.  

In practice, for each field configuration relevant to each process in \cref{tab:triple_lepton_modes}, 
we use {\tt Basisgen} package~\cite{Criado:2019ugp} to enumerate the number of independent operators for $n_\ell=n_\nu=3$ in that configuration. 
Then, we follow our prescriptions described above to parametrize independent operators.
Our final results are summarized in \cref{tab:BNVdim9opeLp1A,tab:BNVdim9opeLp1B,tab:BNVdim9opeLm1,tab:BNVdim9opeL3}, organized according to the change in lepton number $\Delta L$. 
For each constructed operator, we also show its chiral irrep. and the number of independent operators for general values of $n_\ell$ charged leptons and $n_\nu$ neutrinos, as well as the special case with $n_\ell=n_\nu=3$. The completeness and independence of our results is ensured by both the well-established independent triple-quark structures and consistent counting of the total number of independent operators for each configuration using different methods.  

For demonstration, we construct independent operators corresponding to the field configuration $\ell_{\tR x} \ell_{\tL y} \overline{\ell_{\tR z}} u_\tR u_\tR d_\tR$.
Using the {\tt Basisgen} package for three lepton flavors,
we find that there are 27 independent operators associated with this configuration.
First, for the triple-quark configuration $u_\tR u_\tR d_\tR$, 
two independent structures can be formed: ${\cal N}_{uud}^{\tR\tR}$ and ${\cal N}_{uud}^{\tR\tR,\mu\nu}$. 
Second, the two lepton fields $\ell_{\tR x}$ and $\ell_{\tL y}$ combine into a unique bilinear of the form $(\overline{\ell_{\tR x}^\C}\gamma^\mu\ell_{\tL y})=-(\overline{\ell_{\tL y}^\C}\gamma^\mu\ell_{\tR x})$ due to chirality properties, while the conjugate lepton field $\overline{\ell_{\tR z}}$ forms a bilinear with the remaining unpaired quark field in either ${\cal N}_{uud}^{\tR\tR}$ or ${\cal N}_{uud}^{\tR\tR,\mu\nu}$. As a result, a unique operator $\mathcal{O}_{\ell\ell\bar{\ell},xyz}^{\tV\tL\tV\tR\tS\tR}=(\overline{\ell_{\tR x}^\C}\gamma_\mu\ell_{\tL y})(\overline{\ell_{\tR x}}\gamma^\mu{\cal N}_{uud}^{\tR\tR})$ is obtained after contracting all open Lorentz indices. This construction yields the correct total number of operators when accounting for the three flavors $x,y,z=e,\mu,\tau$. 
Operators involving the other triple-quark structure ${\cal N}_{uud}^{\tR\tR,\mu\nu}$, such as $(\overline{\ell_{\tR x}^\C}\gamma_\mu\ell_{\tL y})(\overline{\ell_{\tR x}}\gamma_\nu{\cal N}_{uud}^{\tR\tR,\mu\nu})$, vanish due to the identity $\gamma_\nu{\cal N}_{yzw}^{\tR\tR,\mu\nu}\equiv0$ \cite{Liao:2025vlj}. 
The remaining operators listed in the tables can be constructed in a similar manner.

Now, we make a few comments on our results. First of all, it should be noted that the operators in $\pmb{10}_\tL \otimes \pmb{1}_\tR$ and $\pmb{1}_\tL \otimes \pmb{10}_\tR$ irreps cannot contribute to the purely triple-lepton modes due to a mismatch of flavor symmetry \cite{Liao:2025vlj}. We include them solely to ensure the completeness of the operator basis corresponding to the relevant field configurations. 
Secondly, unlike the results in \cite{Li:2020tsi}, our operator basis has a lepton-quark factorized form with well-defined chiral irrep properties, which facilitates chiral matching. As will be seen in the next section, these operators can be straightforwardly transformed into four-fermion interactions involving three leptons and a nucleon, enabling easy computation of decay rates. 
Thirdly, given any other dim-9 operators, they can be converted into our standard basis by employing various nonstandard Fierz identities shown in \cite{Liao:2016hru,Liao:2019gex,Liao:2020jmn}.
In \cref{sec:UVmodel}, we will provide several examples to demonstrate how quark-lepton mixed operators can be transformed into our above basis operators. 
Fourthly, for a pair of processes related by $\nu\leftrightarrow\bar\nu$,
i.e., the two processes highlighted by the same symbol in \cref{tab:triple_lepton_modes}, their operator bases are chosen to be related to each other by  $\tL\leftrightarrow\tR$ and $\nu_\tL\leftrightarrow\nu_\tL^\C$.
Lastly, although operators involving the $\tau$ lepton do not contribute to nucleon triple-lepton decay processes due to kinematic constraints, they can still be constrained through BNV $\tau$ decays or through four- and higher-body nucleon decays mediated by a $\tau$ lepton~\cite{Heeck:2024jei}. We leave these possibilities for future work.

Moving onto the SMEFT framework, these dim-9 LEFT operators descend from a variety of SMEFT operators spanning multiple dimensions~\cite{Weinberg:1980bf,Hambye:2017qix,Liao:2020jmn,Li:2020xlh,Harlander:2023psl}. In the $(\Delta L,\Delta F_L)=(1,1)$ and $(-1,1)$ sectors, the corresponding operators can be generated at lowest order by SMEFT dim-6 and dim-7 interactions, respectively, after integrating out heavy SM fields $W,Z,h$, and the top quark. 
However, the dim-9 LEFT operators in \cref{tab:BNVdim9opeLm1}   [\cref{tab:BNVdim9opeL3}] associated with $(\Delta L,\Delta F_L)=(-1,3) [(3,3)]$ can only arise at leading order from SMEFT dim-9 [dim-11] or higher-odd-dimensional operators due to their nontrivial flavor structures. 
Those associated with $(\Delta L,\Delta F_L)=(1,3) [(-3,3)]$ in \cref{tab:BNVdim9opeLp1A,tab:BNVdim9opeLp1B} [\cref{tab:BNVdim9opeL3}] originate from SMEFT dim-10 [dim-12] or higher-even-dimensional operators.
Examples of SMEFT operators in each sector can be found in Weinberg's classical work~\cite{Weinberg:1980bf}. 
Turning to UV models, those involving leptoquarks and  discrete lepton flavor symmetries can generate such higher-dimensional operators at leading order~\cite{Hambye:2017qix,Dorsner:2022twk,Liao:2025vlj,Bhoonah:2025qzd}, while simultaneously forbidding lower-dimensional dim-6 or dim-7 SMEFT operators. 
In \cref{sec:UVmodel}, we will employ the UV-complete model in~\cite{Liao:2025vlj} to demonstrate how the WCs of our LEFT operators relate to the model parameters. 

%%%%%%%%%%%%%%%%%%%%%%%%%%%%%%%%%%%%%%%%%%%%%%%%
\section{Chiral perturbation theory for nucleon decays}
\label{sec:chpt}
%%%%%%%%%%%%%%%%%%%%%%%%%%%%%%%%%%%%%%%%%%%%%%%%

\begin{table}[t]
\centering
\renewcommand{\arraystretch}{1.43}
\resizebox{\linewidth}{!}{
{\Large
\begin{tabular}{|c|l|l|l|l|}
\hhline{|-----|}
%%%%%Delta L = - 1
\multirow{18}*{\rotatebox[origin=c]{90}{
\pmb{$\Delta L = +1 $} } }
& \multicolumn{4}{|c|}{\cellcolor{gray!10} 
$p \to \ell_x^+\ell_y^+\ell_z^-$}
\\\hhline{|~|----|}
&
\multicolumn{4}{|l|}{$ \mathcal{P}_{uud}^{\tL\tL}= C^{\tV\tR\tV\tL\tS\tL}_{\ell\ell\bar\ell,xyz}(\overline{\ell_{\tL x}^\C}\gamma_\mu\ell_{\tR y})\overline{\ell_{\tL z}}\gamma^\mu
+ C^{ \tS\tL\tS\tL\tS\tL}_{\ell\ell\bar\ell,xyz}(\overline{\ell_{\tL x}^\C}\ell_{\tL y})\overline{\ell_{\tR z}}
+ C^{ \tT\tL\tT\tL\tS\tL}_{\ell\ell\bar\ell,xyz}(\overline{\ell_{\tL x}^\C}\sigma_{\mu\nu}\ell_{\tL y})\overline{\ell_{\tR z}}\sigma^{\mu\nu}
+ C^{\tS\tR\tS\tL\tS\tL}_{\ell\ell\bar\ell,xyz}(\overline{\ell_{\tR x}^\C}\ell_{\tR y})\overline{\ell_{\tR z}}$ }
\\
& \multicolumn{4}{|l|}{$ \mathcal{P}_{uud}^{\tR\tR}= C^{\tV\tL\tV\tR\tS\tR}_{\ell\ell\bar\ell,xyz}(\overline{\ell_{\tR x}^\C}\gamma_\mu\ell_{\tL y})\overline{\ell_{\tR z}}\gamma^\mu
+ C^{ \tS\tR\tS\tR\tS\tR}_{\ell\ell\bar\ell,xyz}(\overline{\ell_{\tR x}^\C}\ell_{\tR y})\overline{\ell_{\tL z}}
+ C^{ \tT\tR\tT\tR\tS\tR}_{\ell\ell\bar\ell,xyz}(\overline{\ell_{\tR x}^\C}\sigma_{\mu\nu}\ell_{\tR y})\overline{\ell_{\tL z}}\sigma^{\mu\nu}
+ C^{\tS\tL\tS\tR\tS\tR}_{\ell\ell\bar\ell,xyz}(\overline{\ell_{\tL x}^\C}\ell_{\tL y})\overline{\ell_{\tL z}}$}
\\\hhline{|~|----|}
& \multicolumn{4}{|l|}{$ \mathcal{P}_{uud}^{\tL\tR}=
C^{\tV\tR\tV\tL\tS\tR}_{\ell\ell\bar\ell,xyz}(\overline{\ell_{\tL x}^\C}\gamma_\mu\ell_{\tR y})\overline{\ell_{\tL z}}\gamma^\mu
+ C^{ \tS\tL\tS\tL\tS\tR}_{\ell\ell\bar\ell,xyz}(\overline{\ell_{\tL x}^\C}\ell_{\tL y})\overline{\ell_{\tR z}}
+ C^{ \tT\tL\tT\tL\tS\tR}_{\ell\ell\bar\ell,xyz}(\overline{\ell_{\tL x}^\C}\sigma_{\mu\nu}\ell_{\tL y})\overline{\ell_{\tR z}}\sigma^{\mu\nu}
+ C^{\tS\tR\tS\tL\tS\tR}_{\ell\ell\bar\ell,xyz}(\overline{\ell_{\tR x}^\C}\ell_{\tR y})\overline{\ell_{\tR z}}$ }
\\
& \multicolumn{4}{|l|}{$ \mathcal{P}_{uud}^{\tR\tL}=
C^{\tV\tL\tV\tR\tS\tL}_{\ell\ell\bar\ell,xyz}(\overline{\ell_{\tR x}^\C}\gamma_\mu\ell_{\tL y})\overline{\ell_{\tR z}}\gamma^\mu
+ C^{ \tS\tR\tS\tR\tS\tL}_{\ell\ell\bar\ell,xyz}(\overline{\ell_{\tR x}^\C}\ell_{\tR y})\overline{\ell_{\tL z}}
+ C^{ \tT\tR\tT\tR\tS\tL}_{\ell\ell\bar\ell,xyz}(\overline{\ell_{\tR x}^\C}\sigma_{\mu\nu}\ell_{\tR y})\overline{\ell_{\tL z}}\sigma^{\mu\nu}
+ C^{\tS\tL\tS\tR\tS\tL}_{\ell\ell\bar\ell,xyz}(\overline{\ell_{\tL x}^\C}\ell_{\tL y})\overline{\ell_{\tL z}}$}
\\\hhline{|~|----|}
&
\multicolumn{2}{|l|}{$ \mathcal{P}_{udu}^{\tL\tR,\mu}=\frac{1}{2}
C^{\tT\tL\tV\tL\tV\tR1}_{\ell\ell\bar\ell  ,xyz}(\overline{\ell_{\tL x}^\C}\sigma^{\nu\mu} \ell_{\tL y})\overline{\ell_{\tL z}}\gamma_\nu
+\frac{1}{2}C^{\tV\tL\tS\tL\tV\tR1}_{\ell\ell\bar\ell  ,xyz}(\overline{\ell_{\tR x}^\C}\gamma^{\mu} \ell_{\tL y})\overline{\ell_{\tR z}}$}
&\multicolumn{2}{|l|}{$\mathcal{P}_{uud}^{\tL\tR,\mu}=
C^{\tT\tL\tV\tL\tV\tR2}_{\ell\ell\bar\ell  ,xyz}(\overline{\ell_{\tL x}^\C}\sigma^{\nu\mu} \ell_{\tL y})\overline{\ell_{\tL z}}\gamma_\nu
+C^{\tV\tL\tS\tL\tV\tR2}_{\ell\ell\bar\ell  ,xyz}(\overline{\ell_{\tR x}^\C}\gamma^{\mu} \ell_{\tL y})\overline{\ell_{\tR z}}$}
\\
& \multicolumn{2}{|l|}{$ \mathcal{P}_{udu}^{\tR\tL,\mu}=\frac{1}{2}
C^{\tT\tR\tV\tR\tV\tL1}_{\ell\ell\bar\ell  ,xyz}(\overline{\ell_{\tR x}^\C}\sigma^{\nu\mu} \ell_{\tR y})\overline{\ell_{\tR z}}\gamma_\nu
+\frac{1}{2}C^{\tV\tR\tS\tR\tV\tL1}_{\ell\ell\bar\ell  ,xyz}(\overline{\ell_{\tL x}^\C}\gamma^{\mu} \ell_{\tR y})\overline{\ell_{\tL z}}$}
&\multicolumn{2}{|l|}{$ \mathcal{P}_{uud}^{\tR\tL,\mu}=
C^{\tT\tR\tV\tR\tV\tL2}_{\ell\ell\bar\ell  ,xyz}(\overline{\ell_{\tR x}^\C}\sigma^{\nu\mu} \ell_{\tR y})\overline{\ell_{\tR z}}\gamma_\nu
+C^{\tV\tR\tS\tR\tV\tL2}_{\ell\ell\bar\ell  ,xyz}(\overline{\ell_{\tL x}^\C}\gamma^{\mu} \ell_{\tR y})\overline{\ell_{\tL z}}$}
\\\hhline{|~|----|}%%n 2 l l vb
& \multicolumn{4}{|c|}{\cellcolor{gray!10}
$n \to {\ell}_x^-{\ell}_y^+\bar{\nu}_z$}
\\\hhline{|~|----|}
&\multicolumn{2}{|l|}{
$\makecell[l]{\mathcal{P}_{dud}^{\tL\tL} = 
C^{\tS\tL\tS\tL\tS\tL}_{\bar\ell\ell\nu,xyz}(\overline{\ell_{\tR x}}\ell_{\tL y})\overline{\nu_{\tL z}^\C}
+C^{\tT\tL\tT\tL\tS\tL}_{\bar\ell\ell\nu,xyz}(\overline{\ell_{\tR x}}\sigma_{\mu\nu}\ell_{\tL y})\overline{\nu_{\tL z}^\C}\sigma^{\mu\nu}
\\%
\qquad\quad\,+ C^{\tS\tR\tS\tL\tS\tL}_{\bar\ell\ell\nu,xyz}(\overline{\ell_{\tL x}}\ell_{\tR y})\overline{\nu_{\tL z}^\C}} $ }
&\multicolumn{2}{l|}{$ \makecell[l]{ \mathcal{P}_{dud}^{\tL\tR} = C^{\tS\tL\tS\tL\tS\tR}_{\bar\ell\ell\nu,xyz}(\overline{\ell_{\tR x}}\ell_{\tL y})\overline{\nu_{\tL z}^\C}
+C^{\tT\tL\tT\tL\tS\tR}_{\bar\ell\ell\nu,xyz}(\overline{\ell_{\tR x}}\sigma_{\mu\nu}\ell_{\tL y})\overline{\nu_{\tL z}^\C}\sigma^{\mu\nu}
\\%
\qquad\quad\,+C^{\tS\tR\tS\tL\tS\tR}_{\bar\ell\ell\nu,xyz}(\overline{\ell_{\tL x}}\ell_{\tR y})\overline{\nu_{\tL z}^\C}}$ }
\\
& \multicolumn{2}{|l|}{$ \mathcal{P}_{dud}^{\tR\tR}= C^{\tV\tR\tV\tR\tS\tR}_{\bar\ell\ell\nu,xyz}(\overline{\ell_{\tR x}}\gamma_\mu\ell_{\tR y})\overline{\nu_{\tL z}^\C}\gamma^\mu
+C^{\tV\tL\tV\tR\tS\tR}_{\bar\ell\ell\nu,xyz}(\overline{\ell_{\tL x}}\gamma_\mu\ell_{\tL y})\overline{\nu_{\tL z}^\C}\gamma^\mu $ }
&\multicolumn{2}{l|}{$ \mathcal{P}_{dud}^{\tR\tL} = C^{\tV\tR\tV\tR\tS\tL}_{\bar\ell\ell\nu,xyz}(\overline{\ell_{\tR x}}\gamma_\mu\ell_{\tR y})\overline{\nu_{\tL z}^\C}\gamma^\mu
+C^{\tV\tL\tV\tR\tS\tL}_{\bar\ell\ell\nu,xyz}(\overline{\ell_{\tL x}}\gamma_\mu\ell_{\tL y})\overline{\nu_{\tL z}^\C}\gamma^\mu $ }
\\\hhline{|~|----|}
& \multicolumn{2}{|l|}{$ \mathcal{P}_{udd}^{\tL\tR,\mu}=
\frac{1}{2}C^{\tV\tR\tS\tL\tV\tR1}_{\bar\ell\ell\nu,xyz}(\overline{\ell_{\tR x}}\gamma^\mu\ell_{\tR y})\overline{\nu_{\tL z}^\C}
+\frac{1}{2}C^{\tV\tL\tS\tL\tV\tR1}_{\bar\ell\ell\nu,xyz}(\overline{\ell_{\tL x}}\gamma^\mu\ell_{\tL y})\overline{\nu_{\tL z}^\C} $ }
&\multicolumn{2}{l|}{$ \mathcal{P}_{ddu}^{\tL\tR,\mu}=
C^{\tV\tR\tS\tL\tV\tR2}_{\bar\ell\ell\nu,xyz}(\overline{\ell_{\tR x}}\gamma^\mu\ell_{\tR y})\overline{\nu_{\tL z}^\C}
+C^{\tV\tL\tS\tL\tV\tR2}_{\bar\ell\ell\nu,xyz}(\overline{\ell_{\tL x}}\gamma^\mu\ell_{\tL y})\overline{\nu_{\tL z}^\C} $ }
\\
& \multicolumn{2}{|l|}{$\mathcal{P}_{udd}^{\tR\tL,\mu}=
\frac{1}{2}C^{\tT\tR\tV\tR\tV\tL1}_{\bar\ell\ell\nu,xyz}(\overline{\ell_{\tL x}}\sigma^{\nu\mu}\ell_{\tR y})\overline{\nu_{\tL z}^\C}\gamma_\nu $ }
&\multicolumn{2}{l|}{$\mathcal{P}_{ddu}^{\tR\tL,\mu}=
C^{\tT\tR\tV\tR\tV\tL2}_{\bar\ell\ell\nu,xyz}(\overline{\ell_{\tL x}}\sigma^{\nu\mu}\ell_{\tR y})\overline{\nu_{\tL z}^\C}\gamma_\nu$}
\\\hhline{|~|----|}%%p 2 v vb l
&\multicolumn{4}{|c|}{\cellcolor{gray!10}
$p \to {\nu}_x\bar{\nu}_y{\ell}_z^+$}
\\\hhline{|~|----|}
& \multicolumn{1}{|l|}{ $\mathcal{P}_{uud}^{\tL\tL}= C^{\tV\tL\tV\tL\tS\tL}_{\bar\nu\nu\ell,xyz}(\overline{\nu_{\tL x}}\gamma_\mu\nu_{\tL y})\overline{\ell_{\tR z}^\C}\gamma^\mu $ }
& \multicolumn{1}{|l|}{$\mathcal{P}_{uud}^{\tL\tR}= C^{\tV\tL\tV\tL\tS\tR}_{\bar\nu\nu\ell,xyz}(\overline{\nu_{\tL x}}\gamma_\mu\nu_{\tL y})\overline{\ell_{\tR z}^\C}\gamma^\mu $ }
& \multicolumn{1}{|l|}{$\mathcal{P}_{udu}^{\tL\tR,\mu}=
\frac{1}{2}C^{\tV\tL\tS\tL\tV\tR1}_{\bar\nu\nu\ell,xyz}(\overline{\nu_{\tL x}}\gamma^\mu\nu_{\tL y})\overline{\ell_{\tL z}^\C}$ }
& \multicolumn{1}{|l|}{$\mathcal{P}_{uud}^{\tL\tR,\mu}=
C^{\tV\tL\tS\tL\tV\tR2}_{\bar\nu\nu\ell,xyz}(\overline{\nu_{\tL x}}\gamma^\mu\nu_{\tL y})\overline{\ell_{\tL z}^\C}$ }
\\
& \multicolumn{1}{|l|}{ $\mathcal{P}_{uud}^{\tR\tR}= C^{\tV\tL\tV\tR\tS\tR}_{\bar\nu\nu\ell,xyz}(\overline{\nu_{\tL x}}\gamma_\mu\nu_{\tL y})\overline{\ell_{\tL z}^\C}\gamma^\mu $ }
& \multicolumn{1}{|l|}{$\mathcal{P}_{uud}^{\tR\tL}= C^{\tV\tL\tV\tR\tS\tL}_{\bar\nu\nu\ell,xyz}(\overline{\nu_{\tL x}}\gamma_\mu\nu_{\tL y})\overline{\ell_{\tL z}^\C}\gamma^\mu $ }
& \multicolumn{1}{|l|}{$\mathcal{P}_{udu}^{\tR\tL,\mu}= \frac{1}{2}C^{\tV\tL\tS\tR\tV\tL1}_{\bar\nu\nu\ell,xyz}(\overline{\nu_{\tL x}}\gamma^\mu\nu_{\tL y})\overline{\ell_{\tR z}^\C} $ }
& \multicolumn{1}{|l|}{ $\mathcal{P}_{uud}^{\tR\tL,\mu}= C^{\tV\tL\tS\tR\tV\tL2}_{\bar\nu\nu\ell,xyz}(\overline{\nu_{\tL x}}\gamma^\mu\nu_{\tL y})\overline{\ell_{\tR z}^\C} $ }
\\\hhline{|~|----|}%% n 2 vb vb v
& \multicolumn{4}{|c|}{\cellcolor{gray!10}
$n \to \bar{\nu}_x\bar{\nu}_y{\nu}_z$}
\\\hhline{|~|----|}
&\multicolumn{1}{|l|}{ $\mathcal{P}_{dud}^{\tR\tR}= C^{\tS\tL\tS\tR\tS\tR}_{\nu\nu\bar\nu,xyz}(\overline{\nu_{\tL x}^\C}\nu_{\tL y})\overline{\nu_{\tL z}} $ }
&\multicolumn{1}{|l|}{$\mathcal{P}_{dud}^{\tR\tL}=C^{\tS\tL\tS\tR\tS\tL}_{\nu\nu\bar\nu,xyz}(\overline{\nu_{\tL x}^\C}\nu_{\tL y})\overline{\nu_{\tL z}} $ }
&\multicolumn{1}{|l|}{ $\mathcal{P}_{udd}^{\tL\tR,\mu}= \frac{1}{2}C^{\tT\tL\tV\tL\tV\tR1}_{\nu\nu\bar\nu,xyz}(\overline{\nu_{\tL x}^\C}\sigma^{\nu\mu}\nu_{\tL y})\overline{\nu_{\tL z}}\gamma_\nu $ }
&\multicolumn{1}{|l|}{ $ \mathcal{P}_{ddu}^{\tL\tR,\mu}=C^{\tT\tL\tV\tL\tV\tR2}_{\nu\nu\bar\nu,xyz}(\overline{\nu_{\tL x}^\C}\sigma^{\nu\mu}\nu_{\tL y})\overline{\nu_{\tL z}}\gamma_\nu $ }
\\\hhline{|-----|}
%%%%%Delta L = - 1
\multirow{11}*{\rotatebox[origin=c]{90}{ 
\pmb{$\Delta L = -1 $} } }
&%%n 2 l l v
\multicolumn{4}{|c|}{\cellcolor{gray!10}
$n \to {\ell}_x^-{\ell}_y^+{\nu}_z$}
\\\hhline{|~|----|}
&\multicolumn{2}{|l|}{$\makecell[l]{\mathcal{P}_{dud}^{\tR\tR}=C^{\tS\tR\tS\tR\tS\tR}_{\bar\ell\ell\bar\nu,xyz}(\overline{\ell_{\tL x}}\ell_{\tR y})\overline{\nu_{\tL z}}
+C^{\tT\tR\tT\tR\tS\tR}_{\bar\ell\ell\bar\nu,xyz}(\overline{\ell_{\tL x}}\sigma_{\mu\nu}\ell_{\tR y})\overline{\nu_{\tL z}}\sigma^{\mu\nu}
\\%
\qquad\quad\,+C^{\tS\tL\tS\tR\tS\tR}_{\bar\ell\ell\bar\nu,xyz}(\overline{\ell_{\tR x}}\ell_{\tL y})\overline{\nu_{\tL z}} } $ }
&\multicolumn{2}{l|}{$ \makecell[l]{\mathcal{P}_{dud}^{\tR\tL}=C^{\tS\tR\tS\tR\tS\tL}_{\bar\ell\ell\bar\nu,xyz}(\overline{\ell_{\tL x}}\ell_{\tR y})\overline{\nu_{\tL z}}
+C^{\tT\tR\tT\tR\tS\tL}_{\bar\ell\ell\bar\nu,xyz}(\overline{\ell_{\tL x}}\sigma_{\mu\nu}\ell_{\tR y})\overline{\nu_{\tL z}}\sigma^{\mu\nu}
\\%
\qquad\quad\,+C^{\tS\tL\tS\tR\tS\tL}_{\bar\ell\ell\bar\nu,xyz}(\overline{\ell_{\tR x}}\ell_{\tL y})\overline{\nu_{\tL z}}}
$ }
\\
& \multicolumn{2}{|l|}{$ \mathcal{P}_{dud}^{\tL\tL}=C^{\tV\tL\tV\tL\tS\tL}_{\bar\ell\ell\bar\nu,xyz}(\overline{\ell_{\tL x}}\gamma_\mu\ell_{\tL y})\overline{\nu_{\tL z}}\gamma^\mu
+C^{\tV\tR\tV\tL\tS\tL}_{\bar\ell\ell\bar\nu,xyz}(\overline{\ell_{\tR x}}\gamma_\mu\ell_{\tR y})\overline{\nu_{\tL z}}\gamma^\mu
$ }
&\multicolumn{2}{l|}{$ \mathcal{P}_{dud}^{\tL\tR}=C^{\tV\tL\tV\tL\tS\tR}_{\bar\ell\ell\bar\nu,xyz}(\overline{\ell_{\tL x}}\gamma_\mu\ell_{\tL y})\overline{\nu_{\tL z}}\gamma^\mu
+C^{\tV\tR\tV\tL\tS\tR}_{\bar\ell\ell\bar\nu,xyz}(\overline{\ell_{\tR x}}\gamma_\mu\ell_{\tR y})\overline{\nu_{\tL z}}\gamma^\mu
$ }
\\\hhline{|~|----|}
& \multicolumn{2}{|l|}{$ \mathcal{P}_{udd}^{\tR\tL,\mu}=\frac{1}{2}C^{\tV\tL\tS\tR\tV\tL1}_{\bar\ell\ell\bar\nu,xyz}(\overline{\ell_{\tL x}}\gamma^\mu\ell_{\tL y})\overline{\nu_{\tL z}}
+\frac{1}{2}C^{\tV\tR\tS\tR\tV\tL1}_{\bar\ell\ell\bar\nu,xyz}(\overline{\ell_{\tR x}}\gamma^\mu\ell_{\tR y})\overline{\nu_{\tL z}} $ }
&\multicolumn{2}{l|}{$ \mathcal{P}_{ddu}^{\tR\tL,\mu}=C^{\tV\tL\tS\tR\tV\tL2}_{\bar\ell\ell\bar\nu,xyz}(\overline{\ell_{\tL x}}\gamma^\mu\ell_{\tL y})\overline{\nu_{\tL z}}
+C^{\tV\tR\tS\tR\tV\tL2}_{\bar\ell\ell\bar\nu,xyz}(\overline{\ell_{\tR x}}\gamma^\mu\ell_{\tR y})\overline{\nu_{\tL z}} $ }
\\
& \multicolumn{2}{|l|}{$\mathcal{P}_{udd}^{\tL\tR,\mu}= \frac{1}{2}C^{\tT\tL\tV\tL\tV\tR1}_{\bar\ell\ell\bar\nu,xyz}(\overline{\ell_{\tR x}}\sigma^{\nu\mu}\ell_{\tL y})\overline{\nu_{\tL z}}\gamma_\nu
$ }
&\multicolumn{2}{l|}{$\mathcal{P}_{ddu}^{\tL\tR,\mu}= C^{\tT\tL\tV\tL\tV\tR2}_{\bar\ell\ell\bar\nu,xyz}(\overline{\ell_{\tR x}}\sigma^{\nu\mu}\ell_{\tL y})\overline{\nu_{\tL z}}\gamma_\nu
$}
\\\hhline{|~|----|}%% p 2 v v l
& \multicolumn{4}{|c|}{\cellcolor{gray!10}$p \to {\nu}_x{\nu}_y\ell_z^+$}
\\\hhline{|~|----|}
& \multicolumn{2}{|l|}{$\mathcal{P}_{uud}^{\tR\tR}= C^{\tS\tR\tS\tR\tS\tR}_{\bar\nu\bar\nu\ell,xyz}(\overline{\nu_{\tL x}}\nu_{\tL y}^\C)\overline{\ell_{\tR z}^\C}
+ C^{\tT\tR\tT\tR\tS\tR}_{\bar\nu\bar\nu\ell,xyz}(\overline{\nu_{\tL x}}\sigma_{\mu\nu}\nu_{\tL y}^\C)\overline{\ell_{\tR z}^\C}\sigma^{\mu\nu}
$ }
& \multicolumn{2}{|l|}{$ \mathcal{P}_{uud}^{\tR\tL}=C^{\tS\tR\tS\tR\tS\tL}_{\bar\nu\bar\nu\ell,xyz}(\overline{\nu_{\tL x}}\nu_{\tL y}^\C)\overline{\ell_{\tR z}^\C}
+ C^{\tT\tR\tT\tR\tS\tL}_{\bar\nu\bar\nu\ell,xyz}(\overline{\nu_{\tL x}}\sigma_{\mu\nu}\nu_{\tL y}^\C)\overline{\ell_{\tR z}^\C}\sigma^{\mu\nu}
$ }
\\\hhline{|~|----|}
&\multicolumn{1}{|l|}{$\mathcal{P}_{uud}^{\tL\tL}=C^{\tS\tR\tS\tL\tS\tL}_{\bar\nu\bar\nu\ell,xyz}(\overline{\nu_{\tL x}}\nu_{\tL y}^\C)\overline{\ell_{\tL z}^\C}$ }
&\multicolumn{1}{|l|}{$\mathcal{P}_{uud}^{\tL\tR}= C^{\tS\tR\tS\tL\tS\tR}_{\bar\nu\bar\nu\ell,xyz}(\overline{\nu_{\tL x}}\nu_{\tL y}^\C)\overline{\ell_{\tL z}^\C}$ }
&\multicolumn{1}{|l|}{$\mathcal{P}_{udu}^{\tR\tL,\mu}= \frac{1}{2}C^{\tT\tR\tV\tR\tV\tL1}_{\bar\nu\bar\nu\ell,xyz}(\overline{\nu_{\tL x}}\sigma^{\nu\mu}\nu_{\tL y}^\C)\overline{\ell_{\tL z}^\C}\gamma_\nu$ }
&\multicolumn{1}{|l|}{$\mathcal{P}_{uud}^{\tR\tL,\mu}= C^{\tT\tR\tV\tR\tV\tL2}_{\bar\nu\bar\nu\ell,xyz}(\overline{\nu_{\tL x}}\sigma^{\nu\mu}\nu_{\tL y}^\C)\overline{\ell_{\tL z}^\C}\gamma_\nu
$ }
\\\hhline{|~|----|}%%n 2 v v vb
& \multicolumn{4}{|c|}{\cellcolor{gray!10}$n \to {\nu}_x{\nu}_y\bar{\nu}_z$}
\\\hhline{|~|----|}
&\multicolumn{1}{|l|}{$\mathcal{P}_{dud}^{\tL\tL}= C^{\tS\tR\tS\tL\tS\tL}_{\bar\nu\bar\nu\nu,xyz}(\overline{\nu_{\tL x}}\nu_{\tL y}^\C)\overline{\nu_{\tL z}^\C}
$ }
&\multicolumn{1}{|l|}{$\mathcal{P}_{dud}^{\tL\tR}= C^{\tS\tR\tS\tL\tS\tR}_{\bar\nu\bar\nu\nu,xyz}(\overline{\nu_{\tL x}}\nu_{\tL y}^\C)\overline{\nu_{\tL z}^\C}
$ }
&\multicolumn{1}{|l|}{$ \mathcal{P}_{udd}^{\tR\tL,\mu}=\frac{1}{2}C^{\tT\tR\tV\tR\tV\tL1}_{\bar\nu\bar\nu\nu,xyz}(\overline{\nu_{\tL x}}\sigma^{\nu\mu}\nu_{\tL y}^\C)\overline{\nu_{\tL z}^\C}\gamma_\nu
$ }
&\multicolumn{1}{|l|}{$ \mathcal{P}_{ddu}^{\tR\tL,\mu}=C^{\tT\tR\tV\tR\tV\tL2}_{\bar\nu\bar\nu\nu,xyz}(\overline{\nu_{\tL x}}\sigma^{\nu\mu}\nu_{\tL y}^\C)\overline{\nu_{\tL z}^\C}\gamma_\nu
$ }
\\\hhline{|-----|}
%%%%%Delta L = +3
\multirow{5}*{\rotatebox[origin=c]{90}{
\pmb{$\Delta L = + 3 $} } }
&%%p 2 vb vb l
\multicolumn{4}{|c|}{\cellcolor{gray!10}
$p \to \bar{\nu}_x\bar{\nu}_y\ell_z^+$}
\\\hhline{|~|----|}
& \multicolumn{2}{|l|}{$ \mathcal{P}_{uud}^{\tL\tL}=C^{\tS\tL\tS\tL\tS\tL}_{\nu\nu\ell,xyz}(\overline{\nu_{\tL x}^\C}\nu_{\tL y})\overline{\ell_{\tL z}^\C}
+ C^{\tT\tL\tT\tL\tS\tL}_{\nu\nu\ell,xyz}(\overline{\nu_{\tL x}^\C}\sigma_{\mu\nu}\nu_{\tL y})\overline{\ell_{\tL z}^\C}\sigma^{\mu\nu}
$ }
& \multicolumn{2}{|l|}{$ \mathcal{P}_{uud}^{\tL\tR}=C^{\tS\tL\tS\tL\tS\tR}_{\nu\nu\ell,xyz}(\overline{\nu_{\tL x}^\C}\nu_{\tL y})\overline{\ell_{\tL z}^\C}
+ C^{\tT\tL\tT\tL\tS\tR}_{\nu\nu\ell,xyz}(\overline{\nu_{\tL x}^\C}\sigma_{\mu\nu}\nu_{\tL y})\overline{\ell_{\tL z}^\C}\sigma^{\mu\nu}
$ }
\\\hhline{|~|----|}
&\multicolumn{1}{|l|}{$\mathcal{P}_{uud}^{\tR\tR}=C^{\tS\tL\tS\tR\tS\tR}_{\nu\nu\ell,xyz}(\overline{\nu_{\tL x}^\C}\nu_{\tL y})\overline{\ell_{\tR z}^\C}
$}
&\multicolumn{1}{|l|}{$\mathcal{P}_{uud}^{\tR\tL}=C^{\tS\tL\tS\tR\tS\tL}_{\nu\nu\ell,xyz}(\overline{\nu_{\tL x}^\C}\nu_{\tL y})\overline{\ell_{\tR z}^\C}
$}
&\multicolumn{1}{|l|}{$ \mathcal{P}_{udu}^{\tL\tR,\mu}=\frac{1}{2}C^{\tT\tL\tV\tL\tV\tR1}_{\nu\nu\ell,xyz}(\overline{\nu_{\tL x}^\C}\sigma^{\nu\mu}\nu_{\tL y})\overline{\ell_{\tR z}^\C}\gamma_\nu
$ }
&\multicolumn{1}{|l|}{$ \mathcal{P}_{uud}^{\tL\tR,\mu}=C^{\tT\tL\tV\tL\tV\tR2}_{\nu\nu\ell,xyz}(\overline{\nu_{\tL x}^\C}\sigma^{\nu\mu}\nu_{\tL y})\overline{\ell_{\tR z}^\C}\gamma_\nu
$ }
\\\hhline{|~|----|}%%n2 vb vb vb
& \multicolumn{4}{|c|}{\cellcolor{gray!10}$n \to \bar{\nu}_x\bar{\nu}_y\bar{\nu}_z$}
\\\hhline{|~|----|}
& \multicolumn{2}{|l|}{$ \mathcal{P}_{dud}^{\tL\tL}=C^{\tS\tL\tS\tL\tS\tL}_{\nu\nu\nu,xyz}(\overline{\nu_{\tL x}^\C}\nu_{\tL [y})\overline{\nu_{\tL z]}^\C} $ }
&\multicolumn{2}{|l|}{$\mathcal{P}_{dud}^{\tL\tR}= C^{\tS\tL\tS\tL\tS\tR}_{\nu\nu\nu,xyz}(\overline{\nu_{\tL x}^\C}\nu_{\tL [y})\overline{\nu_{\tL z]}^\C} $ }
%%%%%Delta L = -3
\\\hhline{|-----|}%%n2v v v
\multirow{2}*{\rotatebox[origin=c]{90}{
\pmb{$\Delta L = - 3 $}} }
& \multicolumn{4}{|c|}{\cellcolor{gray!10}$n \to {\nu}_x{\nu}_y{\nu}_z$}
\\\hhline{|~|----|}
& \multicolumn{2}{|l|}{$\mathcal{P}_{dud}^{\tR\tR}= C^{\tS\tR\tS\tR\tS\tR}_{\bar\nu\bar\nu\bar\nu,xyz}(\overline{\nu_{\tL x}}\nu_{\tL [y}^\C)\overline{\nu_{\tL z]}} $ }
& \multicolumn{2}{|l|}{$ \mathcal{P}_{dud}^{\tR\tL}=C^{\tS\tR\tS\tR\tS\tL}_{\bar\nu\bar\nu\bar\nu,xyz}(\overline{\nu_{\tL x}}\nu_{\tL [y}^\C)\overline{\nu_{\tL z]}} $ }
\\\hhline{|-----|}
\end{tabular} } }
\caption{Explicit expressions of the spurion fields for each nucleon triple-lepton decay.
Note that the summation over the flavor indices $x,y,z$ is implied; and for lepton bilinears involving identical chiral fields, we impose the condition $x\leq y$ 
to restrict ourselves to independent flavor combinations.
}
\label{tab:spurions}
\end{table}

To calculate the nucleon decay amplitudes induced by the LEFT dim-9 operators constructed in the previous section, a systematic and convenient way is to work within the baryon ChPT framework and perform a chiral matching to translate the quark degrees of freedom into hadronic building blocks such as nucleons and pseudoscalar mesons.
Matching to ChPT has the advantage that all interactions at some chiral order in the hadron level are collected in the same term (or terms) as long as they share the same symmetry as their quark-level operator. Those interactions can induce different decays which are actually related. The lattice calculation of the hadronic matrix elements of quark operators fixes the low-energy constant(LEC) as an overall factor in the matching.
Such a chiral matching for the general triple-quark operators has recently been developed in \cite{Liao:2025vlj}. 
In terms of ${\cal N}_i$ in \cref{eq:Nqqq}, all dim-9 operators relevant to nucleon triple-lepton decays can be written as
\begin{align}
{\cal L}_{q^3l^3}^{\Delta B=1} =   
& {\cal P}_{uud}^{\tL\tL} {\cal N}_{uud}^{\tL\tL}
+ {\cal P}_{uud}^{\tL\tR} {\cal N}_{uud}^{\tL\tR}
+ {\cal P}_{uud,\mu}^{\tL\tR} {\cal N}_{uud}^{\tL\tR,\mu}
+ {\cal P}_{udu,\mu}^{\tL\tR} {\cal N}_{udu}^{\tL\tR,\mu}
+ {\cal P}_{duu,\mu}^{\tL\tR} {\cal N}_{udu}^{\tL\tR,\mu}
\label{eq:Lagq3l3}
\\
+ &{\cal P}_{dud}^{\tL\tL} {\cal N}_{dud}^{\tL\tL}
+ {\cal P}_{dud}^{\tL\tR} {\cal N}_{dud}^{\tL\tR}
+ {\cal P}_{ddu,\mu}^{\tL\tR} {\cal N}_{ddu}^{\tL\tR,\mu}
+ {\cal P}_{udd,\mu}^{\tL\tR} {\cal N}_{udd}^{\tL\tR,\mu}
+ {\cal P}_{dud,\mu}^{\tL\tR} {\cal N}_{udd}^{\tL\tR,\mu}
+\tL\leftrightarrow \tR.
\nonumber
\end{align}
Here, ${\cal P}_i$ are the so-called spurion fields, which are linear combinations of the products of the non-quark component of each operator in \cref{tab:BNVdim9opeLp1A,tab:BNVdim9opeLp1B,tab:BNVdim9opeLm1,tab:BNVdim9opeL3} and its corresponding WC.
${\cal P}_{duu,\mu}^{\tL\tR}={\cal P}_{udu,\mu}^{\tL\tR}$ and ${\cal P}_{udd,\mu}^{\tL\tR}={\cal P}_{udd,\mu}^{\tL\tR}$.
In \cref{tab:spurions}, we summarize the relevant spurion fields arising from these dim-9 operators, where $C_i^j$ denotes the WC corresponding to the operator $\calO_i^j$.

The leading-order chiral realizations of all triple-quark interactions relevant to our analysis are given in Eq.\,(12) of Ref.~\cite{Liao:2025vlj}. By expanding the pseudoscalar matrix to the zeroth order in the meson fields, the relevant four-fermion interactions contributing to nucleon triple-lepton decays are as follows~\cite{Fan:2025xhi}: 
\begin{align}
\label{eq:chiral_Lagrangian}
{\cal L}_{{\tt N}lll}^{\Delta B=1}
& = 
\Big\{ 
\big[ c_1 {\cal P}_{uud}^{\tL\tR} + c_2 {\cal P}_{uud}^{\tL\tL} 
+ c_3 \Lambda_\chi^{-1} ( {\cal P}_{uud}^{\tL\tR,\mu} - {\cal P}_{udu}^{\tL\tR,\mu} ) 
i\tilde{\partial}_\mu \big]p_\tL
 \notag\\%p
&
+ \big[ c_1 {\cal P}_{dud}^{\tL\tR} + c_2 {\cal P}_{dud}^{\tL\tL} 
- c_3 \Lambda_\chi^{-1} ({\cal P}_{ddu}^{\tL\tR,\mu} -{\cal P}_{udd}^{\tL\tR,\mu} ) 
i\tilde{\partial}_\mu \big] n_\tL
\Big\} 
- \tL \leftrightarrow \tR,
\end{align}
where the derivative operator is defined as 
$\tilde{\partial}_\mu = \partial_\mu - \tfrac{1}{4}\gamma_\mu 
\slashed{\partial}$, 
and $\Lambda_\chi\approx1.2\,\rm GeV$ denotes the chiral symmetry breaking scale. The LECs $c_1=-0.01257(111)\,{\rm GeV}^3$ and $c_2=0.01269(107)\,{\rm GeV}^3$ are taken from recent lattice QCD calculations~\cite{Yoo:2021gql}, and are also referred to as $\alpha$ and $\beta$, respectively. For the LEC $c_{3}$, only a naive dimensional analysis estimation is available, with a value of $c_3=0.011\,{\rm GeV}^3$~\cite{Liao:2025vlj}.  For later convenience, we define the ratios $\kappa_2\equiv c_2/c_1$ and $\kappa_3 \equiv c_3/c_1$ to organize our results.

%%%%%%%%%%%%%%%%%%%%%%%%%%%%%%%%%%%%%%%%%%%%%%
\section{Calculation of nucleon decay widths and experimental constraints}
\label{sec:calc_nucleon_decay}
%%%%%%%%%%%%%%%%%%%%%%%%%%%%%%%%%%%%%%%%%%%%%%

Denoting a generic lepton by $l\equiv\ell^\pm, ~\nu, ~\bar\nu$, 
the decay amplitude for $\texttt{N}(p)\to l_1(p_1) l_2(p_2) l_3(p_3)$ can be calculated directly from the four-fermion interactions given in \cref{eq:chiral_Lagrangian}.
When identical particles appear in the final state, one has to account for the contributions from crossed diagrams. 
For reference, the full amplitudes are summarized in \cref{app:decay_width_WCsmplitude}. 
After finishing the spin-summed and -averaged squared amplitude, $\overline{|{\cal M}_{\texttt{N}\to l_1l_2l_3}|^2}$, the decay width takes the form
\begin{align}
\Gamma_{\texttt{N}\to l_1l_2l_3}
&= \frac{1}{S!} \frac{1}{256\pi^3 m_{\texttt N}^3}  
\int d s \int_{t_-}^{t_+} d t \;
\overline{|{\cal M}_{\texttt{N}\to l_1l_2l_3}|^2},
\end{align}
where the prefactor $1/S!$ removes double counting of phase space integration for $S~(=1,2,3)$ identical particles in the final state. 
The invariant masses squared
are defined as $s\equiv(p_1+p_2)^2$ and $t\equiv(p_2+p_3)^2$, and their integration domains are specified as follows
\begin{align}
& (m_1 + m_2)^2 \leq s \leq (m_{\texttt N} - m_3)^2, \notag\\
& t_\pm = (E_2^* + E_3^*)^2 - \Big(\sqrt{E_2^{*2} - m_2^2} 
          \mp \sqrt{E_3^{*2} - m_3^2}\Big)^2, \notag\\
& E_2^* = \frac{s - m_1^2 + m_2^2}{2\sqrt{s}},\qquad
  E_3^* = \frac{m_{\texttt N}^2 - s - m_3^2}{2\sqrt{s}},
\end{align}
where $m_{\tt N},~m_{1,2,3}$ are the masses of the nucleon and three leptons respectively. 
After carrying out the full phase-space integration, the decay widths are expressed in terms of the LEFT WCs, and the final results are provided in \cref{app:decay_width_WCs}. 

%%%%%%%%%%%%%%%%%%%%%%%%%%%%%%%%%%%%%%
\subsection{Momentum distribution}
%%%%%%%%%%%%%%%%%%%%%%%%%%%%%%%%%%%%%%

\begin{figure}[t]
\centering
\includegraphics[width=0.328\linewidth]{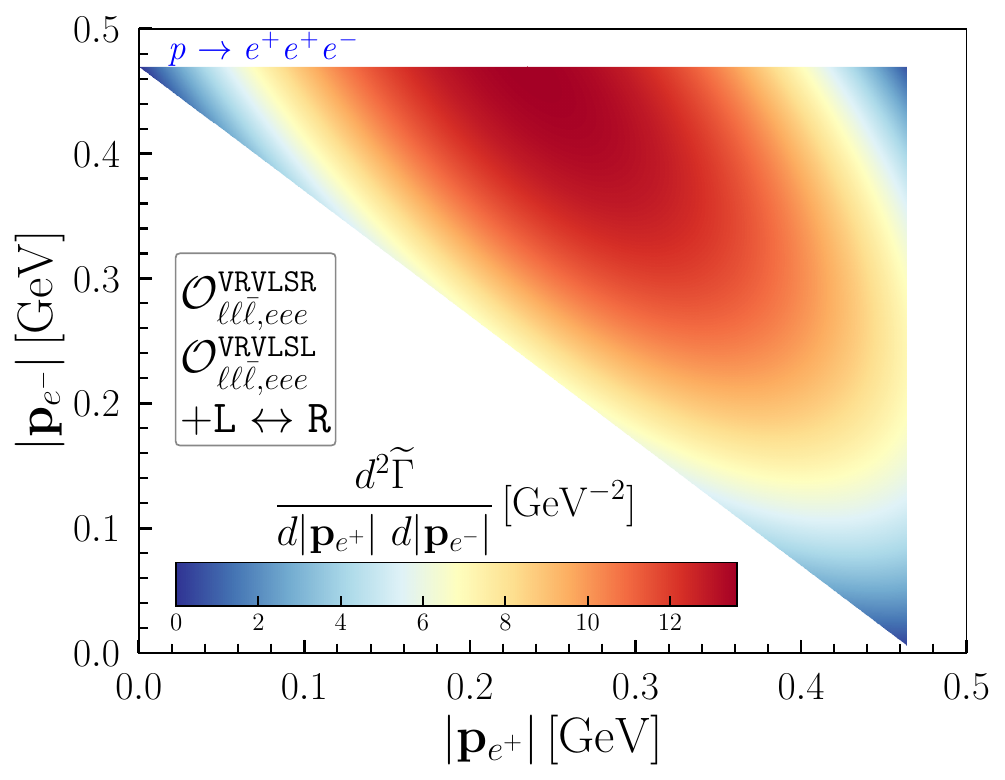}
\includegraphics[width=0.328\linewidth]{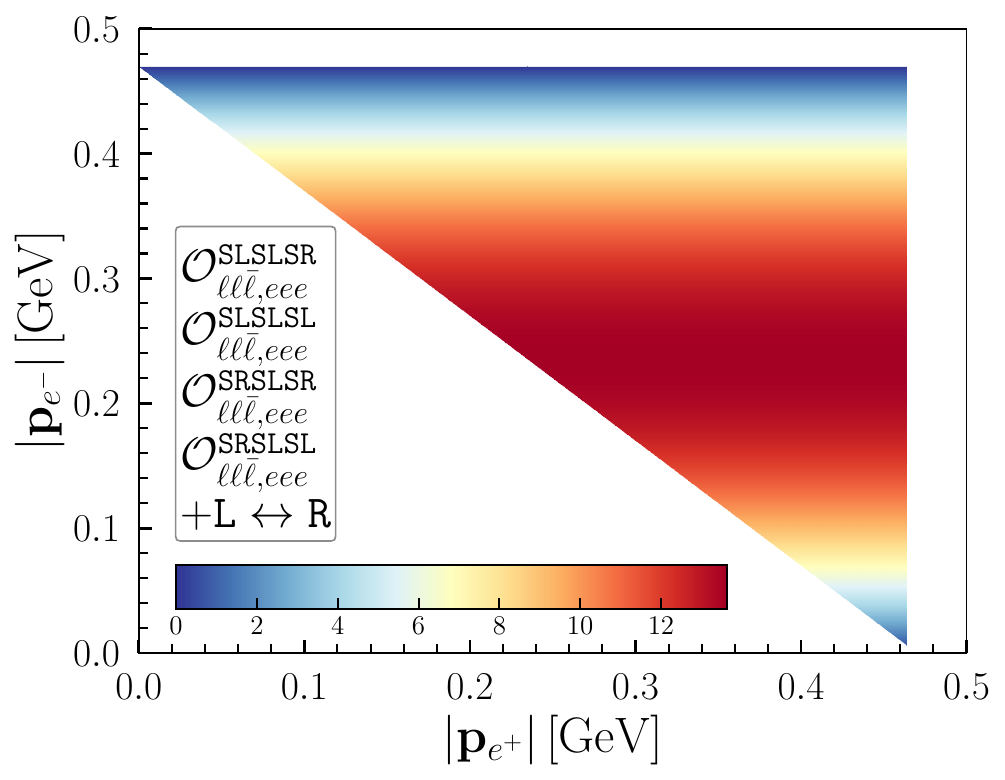}
\includegraphics[width=0.328\linewidth]{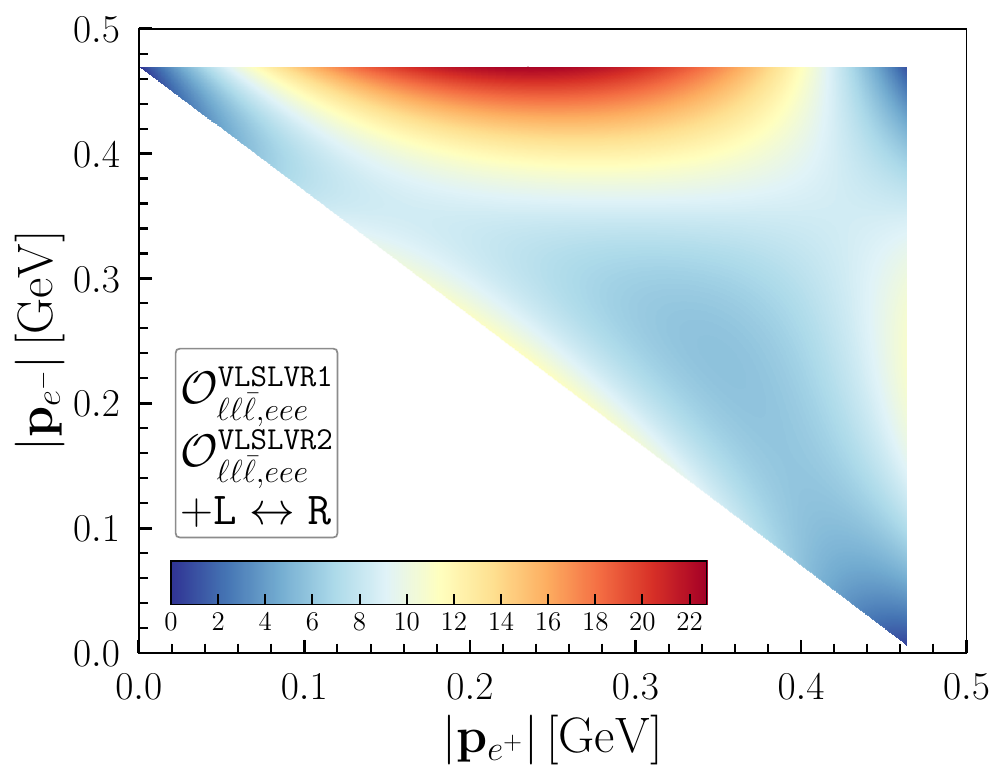}
\\[1em]
\includegraphics[width=0.328\linewidth]{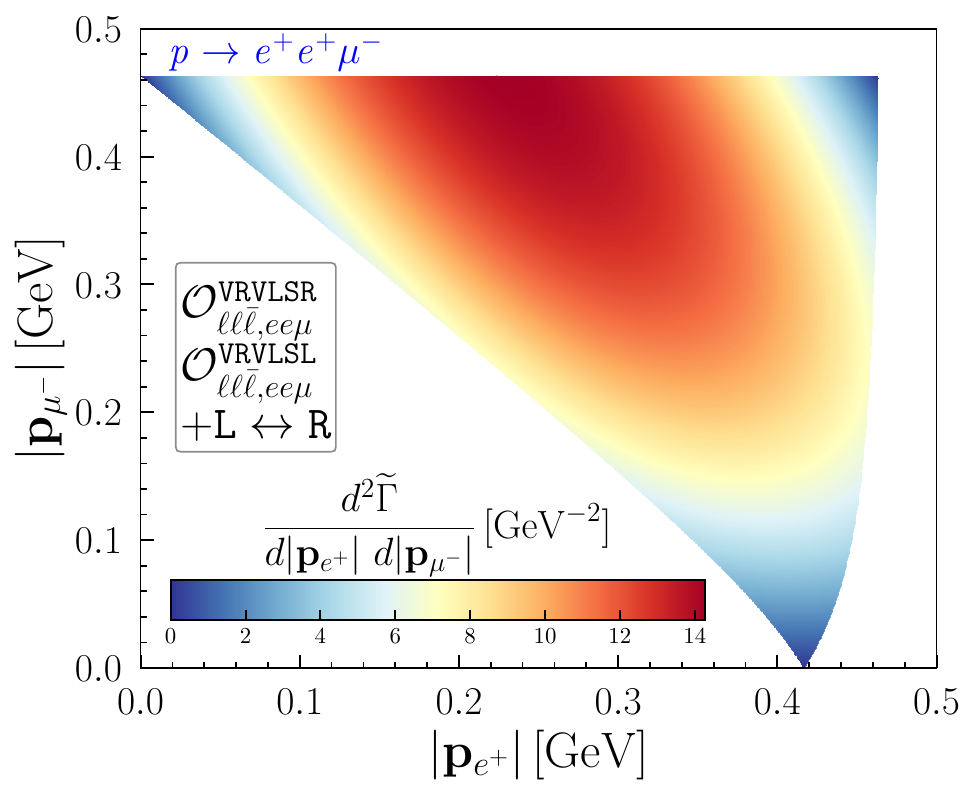}
\includegraphics[width=0.328\linewidth]{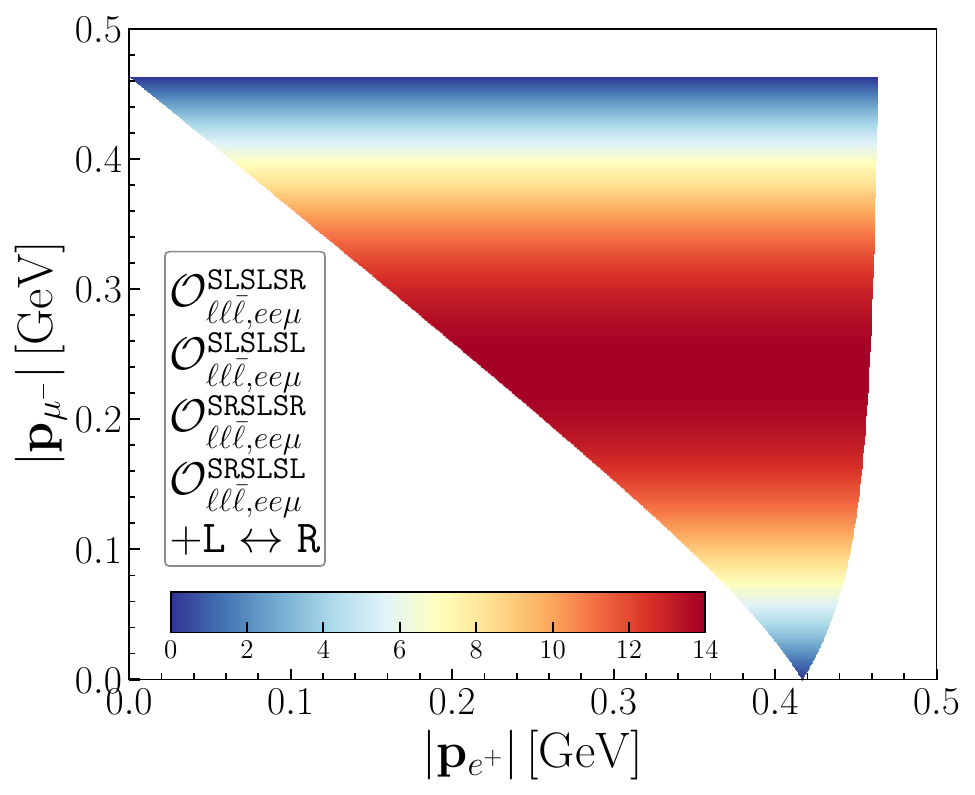}
\includegraphics[width=0.328\linewidth]{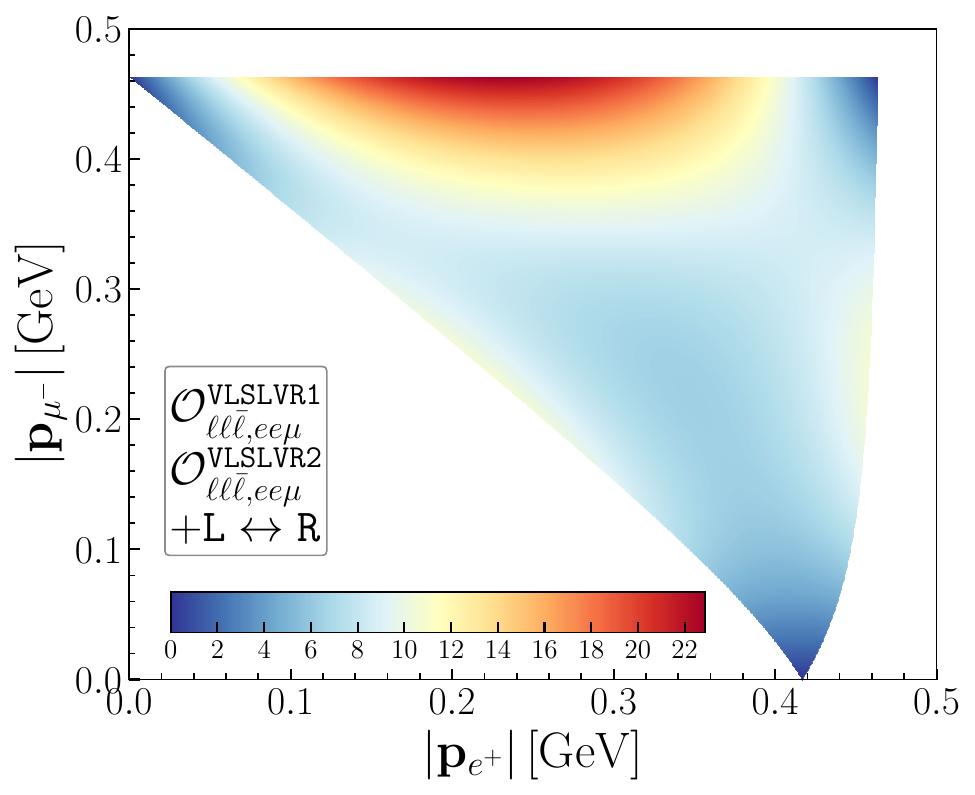}
\\[1em]
\includegraphics[width=0.328\linewidth]{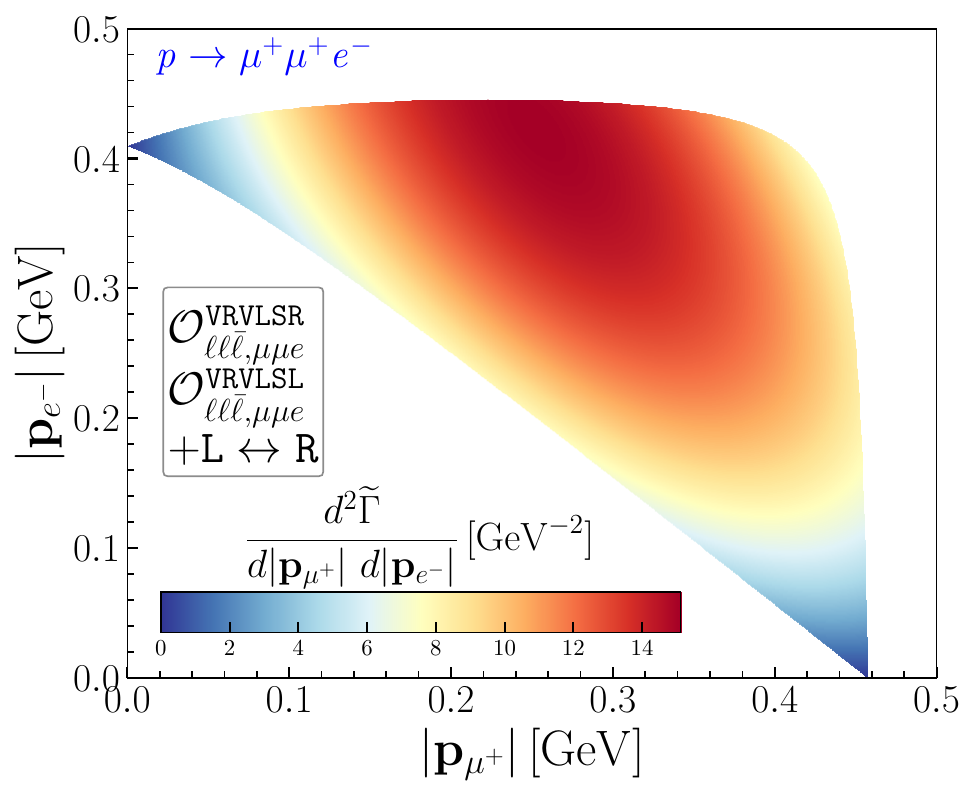}
\includegraphics[width=0.328\linewidth]{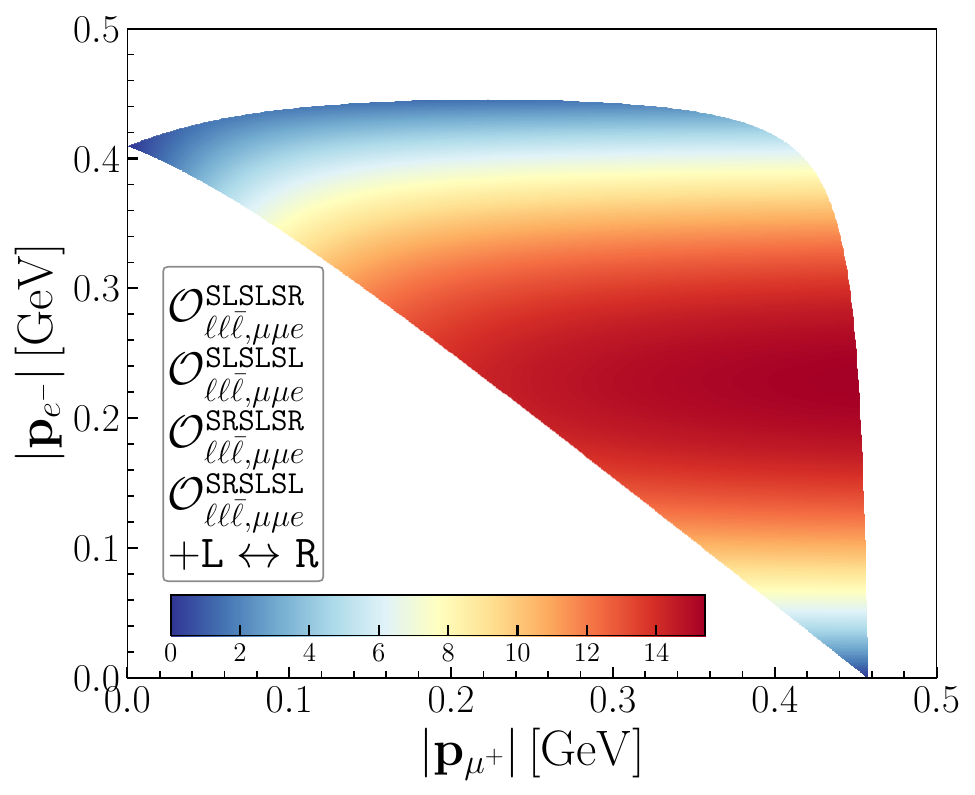}
\includegraphics[width=0.328\linewidth]{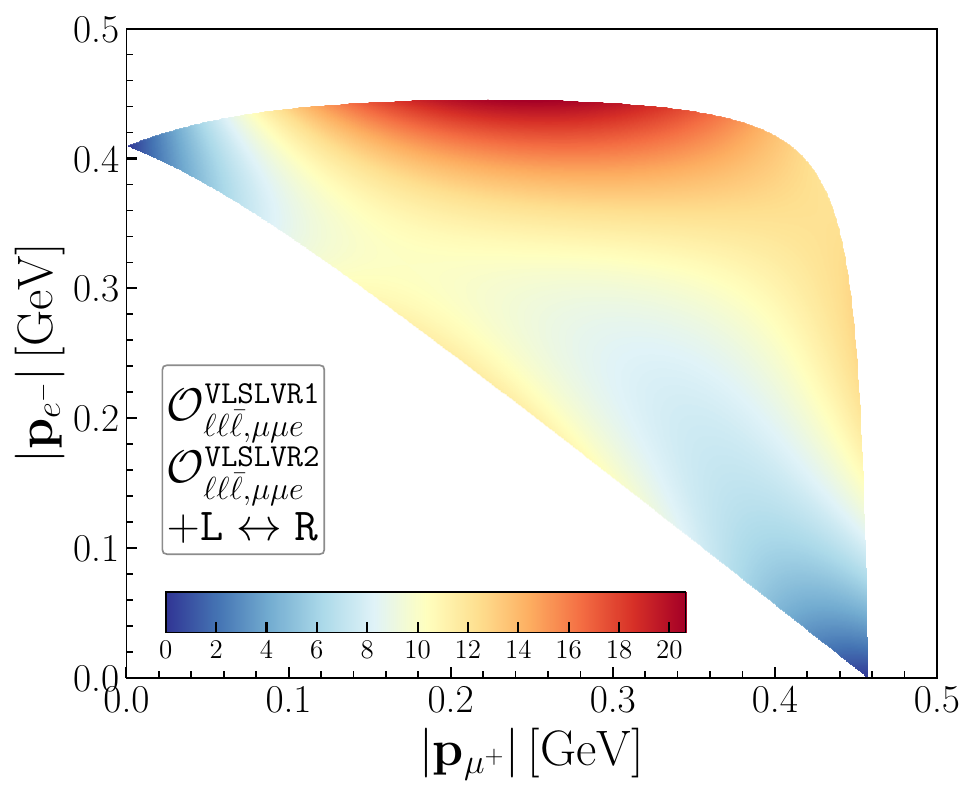}
\\[1em]
\includegraphics[width=0.328\linewidth]{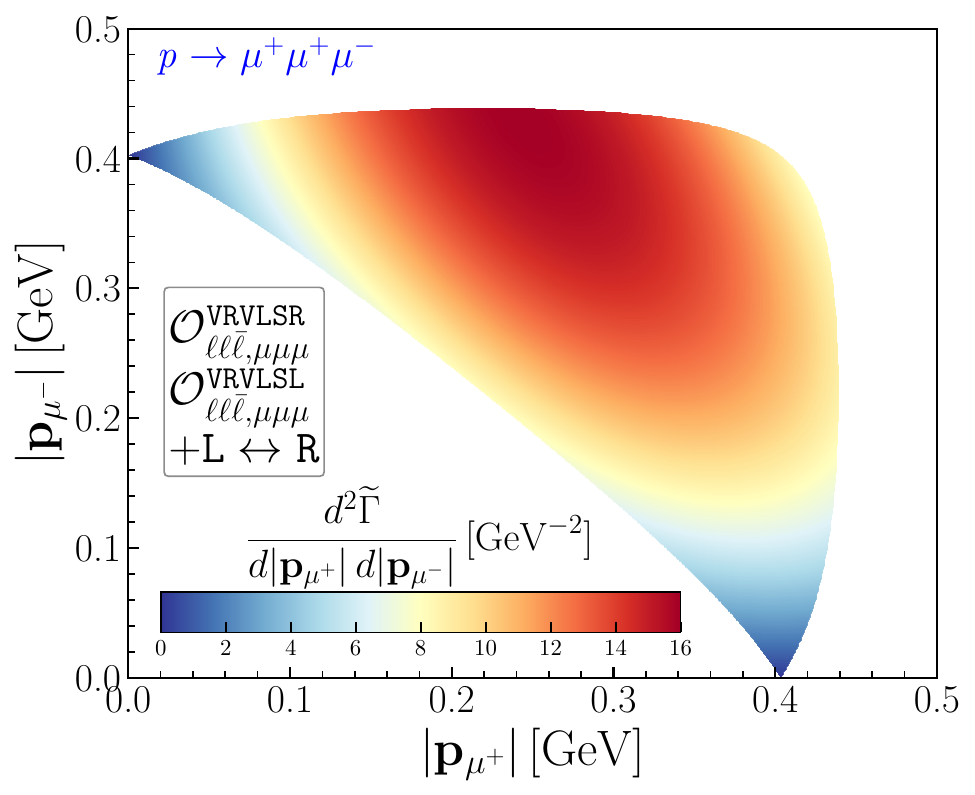}
\includegraphics[width=0.328\linewidth]{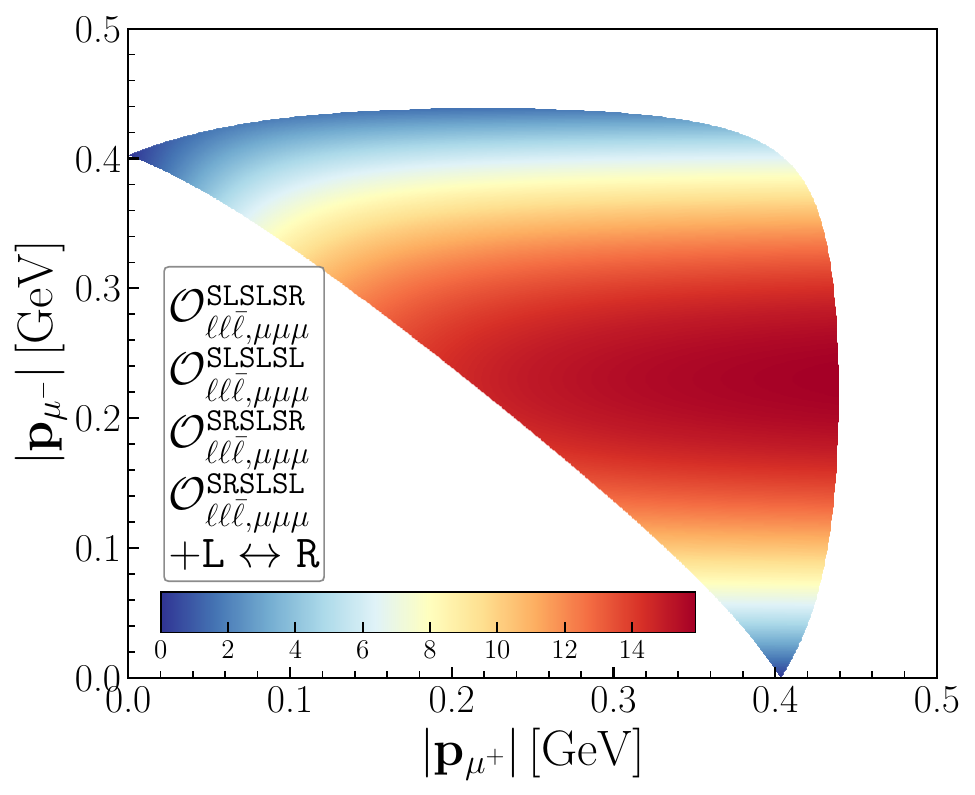}
\includegraphics[width=0.328\linewidth]{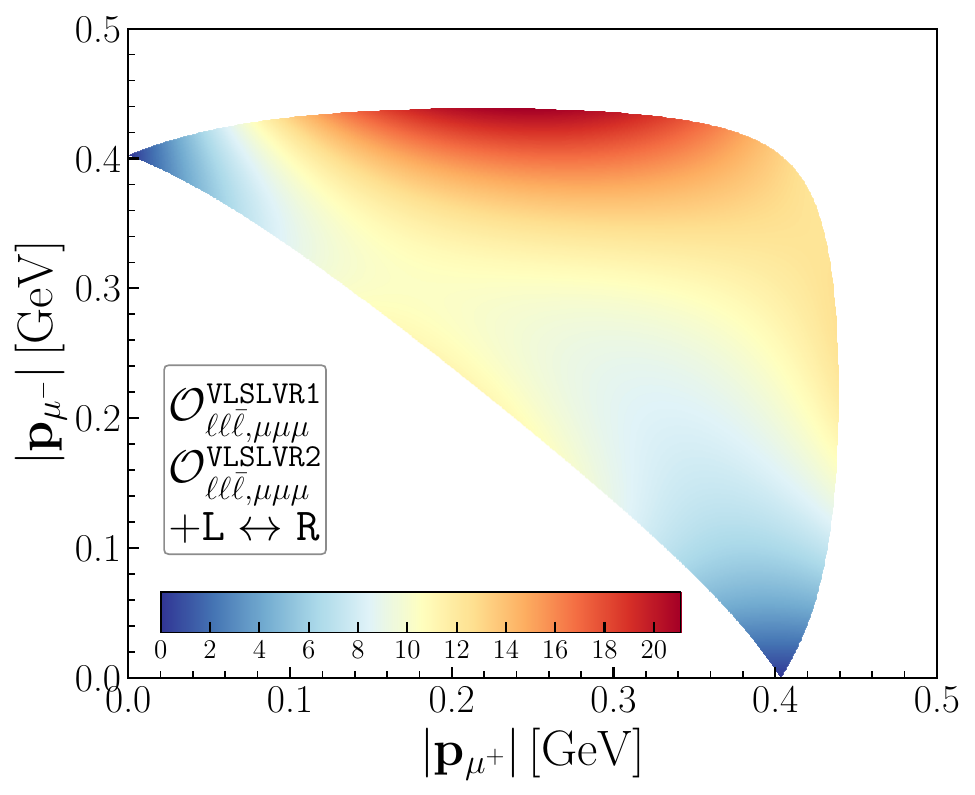}
\caption{Normalized double-momenta distributions of the processes $p\to e^+e^+e^-$ (top row), $p\to e^+e^+\mu^-$ (second row), $p\to \mu^+\mu^+e^-$ (third row), and $p\to \mu^+\mu^+\mu^-$ (bottom row).}
\label{fig:double_distribution}
\end{figure}

For a three-body decay process, the momentum distributions are key observables that encode rich information about the underlying dynamics driving the decay. Although no positive signals have been observed for nucleon decays so far, theoretical studies of these observables remain valuable for guiding future experimental searches and for identifying the operator structures once a positive signal is detected. 
For illustration, here, we focus on the four nucleon decay modes $p\to \ell_x^+\ell_x^+ \ell_y^-$ (with $xy=ee, e\mu, \mu e, \mu\mu$)
and study a normalized double differential distribution with respect to the momenta of the final-state $\ell_x^+$ and $\ell_y^-$, 
\begin{align}
\frac{d^2\tilde{\Gamma}}{d|\mathbf{p}_{\ell_x^+}|d|\mathbf{p}_{\ell_y^-}|}\equiv
\frac{1}{\Gamma}\frac{d^2\Gamma}{d|\mathbf{p}_{\ell_x^+}|d|\mathbf{p}_{\ell_y^-}|} ,
\end{align} 
where $|\mathbf{p}_{\ell_x^+}|$ and $|\mathbf{p}_{\ell_y^-}|$ denote the momentum magnitudes of $\ell_x^+$ and $\ell_y^-$, respectively.  

There are 16 dim-9 operators contributing to each decay $p\to \ell_x^+\ell_x^+ \ell_y^-$, half of which are chiral partners of the other half under the interchange $\tL\leftrightarrow\tR$.  
In our analysis with the insertion of one operator at a time, we find that the 16 operators fall into three distinct classes, each exhibiting a characteristic behavior, as shown in each row of \cref{fig:double_distribution}. 
Specifically, the operators $\calO_{\ell\ell\bar\ell,xxy}^{\tV\tR\tV\tL\tS\tR}, \calO_{\ell\ell\bar\ell,xxy}^{\tV\tR\tV\tL\tS\tL}$, and their chiral partners share the same distribution (left panels), as they share similar vector-like Lorentz structures, in which only the axial-vector component in the first lepton current contributes. As shown in the left panel of each row, the distribution is concentrated mainly in the region with $0.1\lesssim|\mathbf{p}_{\ell_x^+} |\lesssim 0.4\,{\rm GeV}$ and $|\mathbf{p}_{\ell_y^-}|\gtrsim 0.2\,{\rm GeV}$.
For the scalar bilinear structure operators $
\calO_{\ell\ell\bar\ell,xxy}^{\tS\tL\tS\tL\tS\tR},
\calO_{\ell\ell\bar\ell,xxy}^{\tS\tL\tS\tL\tS\tL},
\calO_{\ell\ell\bar\ell,xxy}^{\tS\tR\tS\tL\tS\tR},
\calO_{\ell\ell\bar\ell,xxy}^{\tS\tR\tS\tL\tS\tL}$, and their chiral partners, 
they have the same distribution (the middle panel of each row) because the amplitude squared is independent of chirality. 
The distribution is concentrated in the region $0.1\lesssim|\mathbf{p}_{\ell_y^-}|\lesssim 0.4\,{\rm GeV}$ and shows only a mild dependence on $|\mathbf{p}_{\ell_x^+}|$. This behavior arises because the  squared matrix element is independent of $|\mathbf{p}_{\ell_x^+}|$, with the dependence on $|\mathbf{p}_{\ell_x^+}|$ entering solely through the Jacobian factor $|\mathbf{p}_{\ell_x^+}|/E_{\ell_x^+}$. 
Finally, for the two operators $\calO_{\ell\ell\bar\ell,xxy}^{\tV\tL\tS\tL\tV\tL1,2}$ and their chiral partners, which belong to the chiral irreps $\pmb{3}_{\tL(\tR)}\otimes\pmb{6}_{\tR(\tL)}$,
the distributions shown in the right panels peak along the largest value of  
$|\mathbf{p}_{\ell^-_y}|$.
The main difference among the four processes lies in the muon mass, which results in distinct distribution boundaries, as illustrated in the plots.

%%%%%%%%%%%%%%%%%%%%%%%%%%%%%%%%%%%%%
\subsection{Constraints}
%%%%%%%%%%%%%%%%%%%%%%%%%%%%%%%%%%%%%

\begin{table}[h]
\centering
\renewcommand{\arraystretch}{1.}
\resizebox{\linewidth}{!}{
\begin{tabular}{|c|c|c|c|c|c|c|c|}
\hline
Process & Exp. bound $\Gamma^{-1}[\rm yr]$ & Process  & Exp. bound $\Gamma^{-1}[\rm yr]$ & Process  & Exp. bound $\Gamma^{-1}[\rm yr]$
\\\hline
$p \to e^+ e^+ e^-$ 
& $3.4\times 10^{34}$~\cite{Super-Kamiokande:2020tor}
&$n \to e^- e^+\hat\nu $ 
&$2.57\times 10^{32}$~\cite{McGrew:1999nd} 
& \multirow{2}*{$p \to e^+\hat\nu \hat\nu$}
&$1.7\times 10^{32}$~\cite{Super-Kamiokande:2014pqx}
\\
$p \to e^+ e^+ \mu^- $ 
& $1.9\times 10^{34}$~\cite{Super-Kamiokande:2020tor}
&$n \to e^- \mu^+\hat\nu $ 
& $8.3\times 10^{31}$~\cite{McGrew:1999nd}
& 
& \cellcolor{gray!15}($10.2\times 10^{32}$~\cite{Hyper-Kamiokande:2018ofw})
\\
$p \to e^+ \mu^+  e^-$ 
& $2.3\times 10^{34}$~\cite{Super-Kamiokande:2020tor}
&$n \to \mu^- e^+\hat\nu $ 
& $0.6\times 10^{30}$~\cite{Learned:1979gp}
& \multirow{2}*{$p \to \mu^+\hat\nu \hat\nu$}
&  $2.2\times 10^{32}$~\cite{Super-Kamiokande:2014pqx}
\\
$p \to  e^+ \mu^+ \mu^- $ 
& $9.2\times 10^{33}$~\cite{Super-Kamiokande:2020tor}
&$n \to \mu^- \mu^+\hat\nu $ 
& $7.9\times 10^{31}$~\cite{McGrew:1999nd}
& 
&   \cellcolor{gray!15}($10.7\times 10^{32}$~\cite{Hyper-Kamiokande:2018ofw})
\\
$p \to \mu^+ \mu^+ e^- $ 
& $1.1\times 10^{34}$~\cite{Super-Kamiokande:2020tor}
& 
&
& \multirow{2}*{$n \to \hat\nu\hat\nu \hat\nu$}
&  $0.9\times 10^{30}$ \cite{SNO:2022trz}
\\
$p \to \mu^+ \mu^+ \mu^- $ 
& $1.0\times 10^{34}$~\cite{Super-Kamiokande:2020tor}
& 
&
& 
&  \cellcolor{gray!15}($5\times 10^{31}$~\cite{JUNO:2024pur})
\\\hline
\end{tabular}}
\caption{Current 90\% C.L. experimental lower bounds on various inverse partial decay widths of nucleons into three leptons.  
$\hat\nu$ denotes either a neutrino or an antineutrino. The numbers shown in parentheses indicate the expected sensitivity for single-charged-lepton decays of the proton at Hyper-Kamiokande~\cite{Hyper-Kamiokande:2018ofw} and the neutron invisible decay at JUNO~\cite{JUNO:2024pur}. }
\label{tab:exp_bounds}
\end{table}

Using the decay width expressions provided in \cref{app:decay_width_WCs}, we can apply the existing experimental bounds on partial lifetimes to constrain the EFT operators. In \cref{tab:exp_bounds}, we summarize the current 90\% C.L. experimental bounds for all possible nucleon triple-lepton modes. 
It is evident that the constraints are particularly stringent for proton decays into three charged leptons, with partial lifetimes on the order of $10^{34}\,\rm yr$, whereas modes involving neutrinos are approximately two orders of magnitude weaker. 
Additionally, we include the Hyper-K and JUNO projections for proton single-charged lepton decays~\cite{Hyper-Kamiokande:2018ofw} and neutron invisible decay~\cite{JUNO:2024pur}, respectively.  
Since neutrinos are treated as missing energy in experimental searches, the bounds on neutrino-involved modes are flavor-independent. Consequently, these constraints apply universally to all neutrino or antineutrino flavors, as well as to decay modes involving other visible particles.
 
For each dim-9 operator $\calO_i$, we define an effective scale $\Lambda_{i,\tt eff}$ through its WC $C_i$ by the relation $\Lambda_{i,\rm eff}\equiv |C_i|^{-1/5}$.
By requiring that the theoretically predicted partial lifetime exceed the corresponding experimental limit, we derive a lower bound on $\Lambda_{i,\tt eff}$. To reduce the number of free parameters, we present results in terms of linear combinations of WCs defined in  \cref{eq:WCcomp2lll,eq:WCcomn2llvb,eq:WCcomp2vvbl,eq:WCcomn2vbvbv,eq:WCcomp2vvl,eq:WCcomn2vbvbvb}, since the decay widths depend only on these combinations. The results for specific dim-9 operators can be obtained by appropriately rescaling the bounds according to the relevant powers of the $\kappa_{2,3}$ factors.
For three-body decay modes involving charged leptons, the experimental bounds on $\Gamma^{-1}$s are obtained by assuming a uniform phase space distribution and imposing specific kinematic cuts~\cite{Super-Kamiokande:2020tor}, implying that our derived bounds on the effective scales from the total rates are approximate. A more dedicated analysis that incorporates distributions as in \cref{fig:double_distribution} and experimental cuts, is beyond the scope of this work. We therefore encourage experiments such as Super-K to take these distributions into consideration in their future analysis to obtain more realistic bounds.

\begin{figure}[t]
\centering
\includegraphics[width=\linewidth]{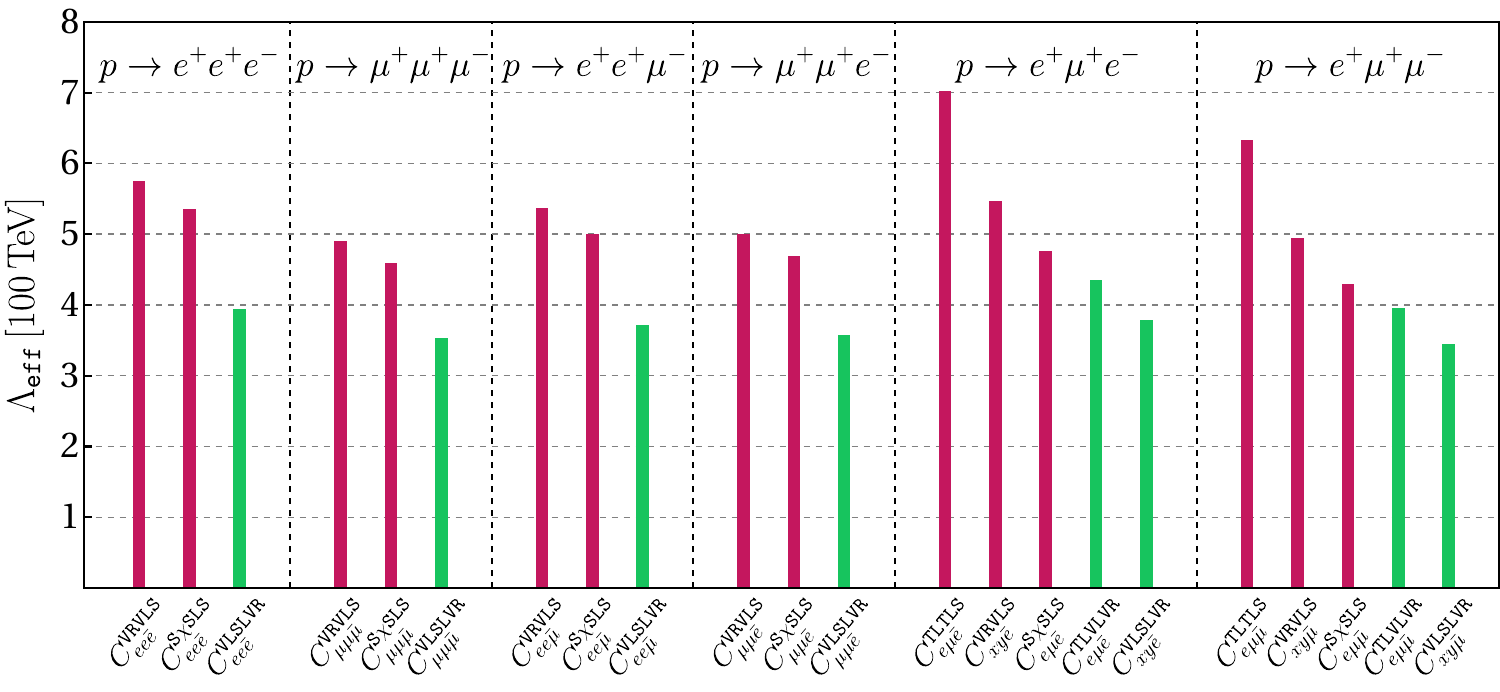}
\caption{Constraints on the effective scales of the LEFT dim-9 operators from proton decays into three charged leptons. 
The red and green bars indicate operators belonging to the chiral irreps ${\bar{\pmb{3}}_{\tL(\tR)}\otimes\pmb{3}_{\tR(\tL)}}/{\pmb{8}_{\tL(\tR)}\otimes\pmb{1}_{\tR(\tL)}}$ and ${\pmb{6}_{\tL(\tR)}\otimes\pmb{3}_{\tR(\tL)}}$, respectively. 
Here, the superscript $\chi$ denotes either $\tL$ or $\tR$, and the flavor indices $xy$ correspond to $e\mu$ or $\mu e$.}
\label{fig:cons_WCs1}
\end{figure}

\begin{figure}[t]
\centering
\includegraphics[width=\linewidth]{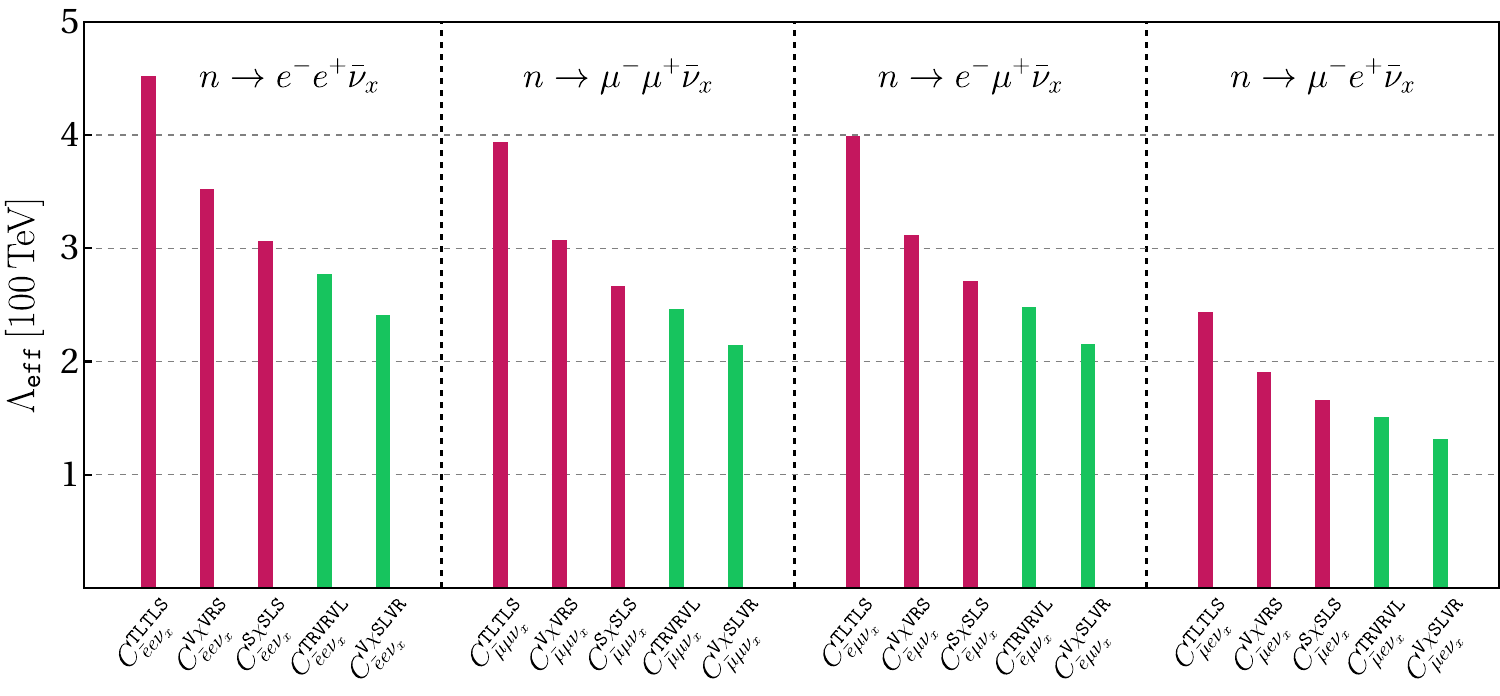}
\\[1em]
\includegraphics[width=\linewidth]{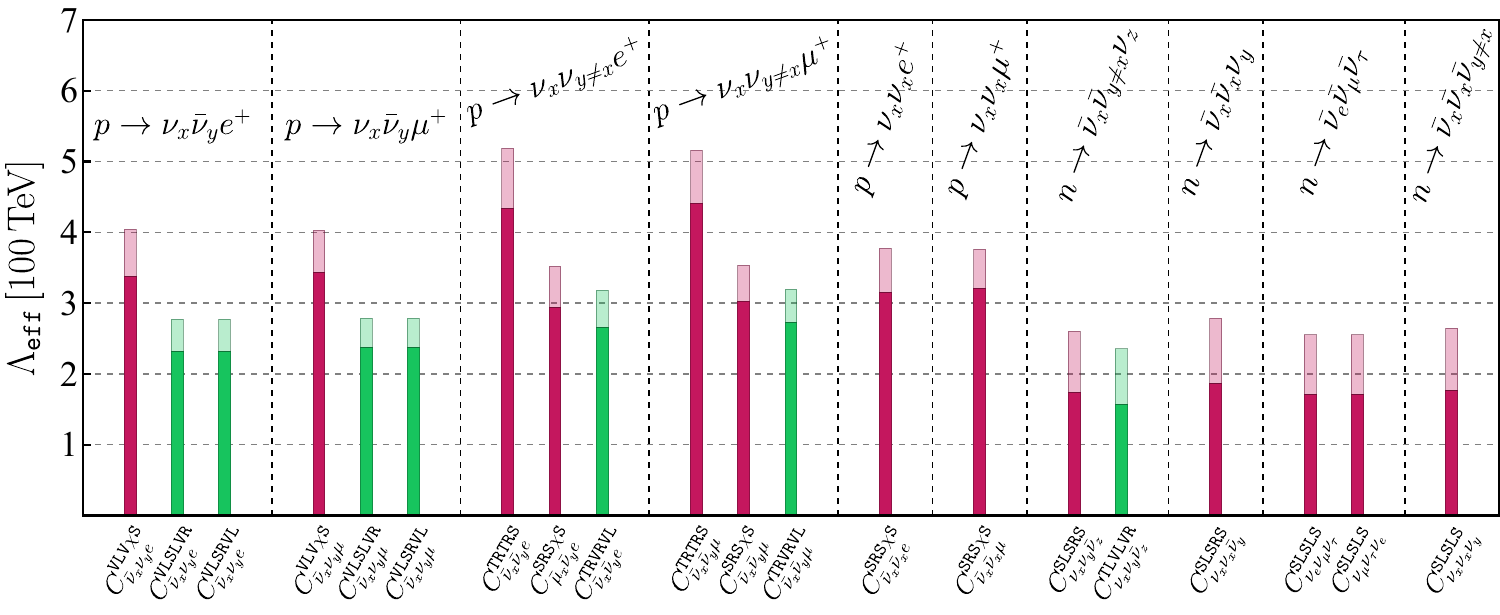}
\caption{Same as \cref{fig:cons_WCs1} but for the operators involving neutrino fields. 
The expected Hyper-K sensitivities for $p\to\hat{\nu}\hat{\nu}e^+(\mu^+)$ and JUNO sensitivities for $n\to\hat{\nu}\hat{\nu}\hat{\nu}$ are denoted by light color bars. Here, $\chi=\tL,\tR$, and the neutrino flavor indices $x,y=e,\mu,\tau$.}
\label{fig:cons_WCs2}
\end{figure}

Figure~\ref{fig:cons_WCs1} presents the independent lower bounds on the effective scales $\Lambda_{\tt eff}$ associated with operators involving three-charged leptons, derived from experimental searches for the six proton decay modes illustrated in the figure.
For brevity, the subscripts of the WCs are abbreviated directly by the leptons involved; for instance, $C^{\tV\tR\tV\tL\tS}_{e e\bar e}\equiv C^{\tV\tR\tV\tL\tS}_{\ell\ell\bar\ell,e ee}$.
The constraints are shown for half of the operators; the other half, which are their parity-conjugate counterparts obtained by exchanging $\tL\leftrightarrow\tR$, are subject to identical constraints. 
The red and green bars indicate operators in chiral irreps ${\bar{\pmb{3}}_{\tL(\tR)}\otimes\pmb{3}_{\tR(\tL)}}$ or ${\pmb{8}_{\tL(\tR)}\otimes\pmb{1}_{\tR(\tL)}}$, and ${\pmb{6}_{\tL(\tR)}\otimes\pmb{3}_{\tR(\tL)}}$, respectively.
As shown, the effective scales are constrained to lie approximately above 300 to 700 TeV, varying with decays and operators involved. 
Generally, for each process, the bounds for operators in the chiral irreps ${\pmb{6}_{\tL(\tR)}\otimes\pmb{3}_{\tR(\tL)}}$ are slightly weaker than those for the chiral irreps ${\bar{\pmb{3}}_{\tL(\tR)}\otimes\pmb{3}_{\tR(\tL)}}$ or ${\pmb{8}_{\tL(\tR)}\otimes\pmb{1}_{\tR(\tL)}}$. This difference arises partly from the somewhat arbitrary choice of the normalization factor in the related chiral Lagrangian~\cite{Liao:2025vlj}, as well as from our limited knowledge of the associated LEC $c_3$.   
According to the relations in \cref{eq:WCcomp2lll}, the effective scales can be rescaled by a factor of $\kappa_2^{1/5}$ and $\kappa_3^{1/5}$ 
for the specific dim-9 operators belonging to the chiral irreps ${\pmb{8}_{\tL(\tR)}\otimes\pmb{1}_{\tR(\tL)}}$ and ${\pmb{6}_{\tL(\tR)}\otimes\pmb{3}_{\tR(\tL)}}$, respectively.

Figure~\ref{fig:cons_WCs2} presents analogous results for operators involving neutrino fields related to the following five classes of processes: 
\begin{align}
n \to \ell_x^- \ell_y^+ \bar{\nu}_z,\,\,
p \to \nu_x \bar\nu_y\ell_z^+, \,\,
n \to \bar\nu_x\bar\nu_y \nu_z;\quad 
p \to \nu_x \nu_y\ell_z^+;\quad 
n \to \bar\nu_x\bar\nu_y \bar\nu_z;
\end{align}
where the first three decay modes have $\Delta L=1$, the fourth mode has $\Delta L=-1$, and the final mode has $\Delta L=3$.
Constraints on operators contributing to the other four classes of decay modes--- 
$n \to \ell_x^- \ell_y^+{\nu}_z$ and $n \to \nu_x\nu_y\bar{\nu}_z$ ($\Delta L=-1$),
$p \to \bar{\nu}_x \bar{\nu}_y\ell_z^+$ ($\Delta L=3$),
and $n\to{\nu}_x {\nu}_y{\nu}_z$ ($\Delta L=-3$)---are not explicitly shown. 
This is because their WC combinations receive identical constraints to those related by $\nu\leftrightarrow\bar\nu$ and $\tL\leftrightarrow\tR$ that are already shown in \cref{fig:cons_WCs2}. 
Note that similar abbreviations for WCs are employed, for example, 
$C^{\tT\tL\tT\tL\tS}_{\bar e e\nu_x}\equiv C^{\tT\tL\tT\tL\tS}_{\bar\ell\ell\nu,eex}$ and $C^{\tS\tL\tS\tL\tS}_{\nu_x\nu_y\nu_z}\equiv C^{\tS\tL\tS\tL\tS}_{\nu\nu\nu,xyz}$.
The constraints on these operators are less stringent than those on the operators shown in
\cref{fig:cons_WCs1}. This difference arises from the comparatively weaker experimental limits on neutrino-related modes, as summarized in \cref{tab:exp_bounds}. 
The bound on the process $n\to \mu^- e^+\bar\nu_x$ is derived from the inclusive limit in~\cite{Learned:1979gp}. Alternatively, one may employ the bound on $n\to e^-\mu^+\hat \nu$ from the IMB-3 experiment~\cite{McGrew:1999nd}, which is a water-Cherenkov detector with very limited charge identification, so that the constraints on the two processes with oppositely-charged final particles should be approximately the same.
For neutron decay modes involving three neutrinos, the upcoming JUNO data are expected to probe effective scales up to twice the current limits set by SNO+~\cite{SNO:2022trz}.
It should be noted that the constraints on the effective scales of decay modes involving a single charged lepton $e^+$ or $\mu^-$ are expected to be further improved by a factor of 1.4 in the upcoming Hyper-K experiment~\cite{Hyper-Kamiokande:2018ofw}, whose fiducial mass is approximately eight times that of the Super-K.

\begin{figure}[t]
\centering
\includegraphics[width=0.49\linewidth]{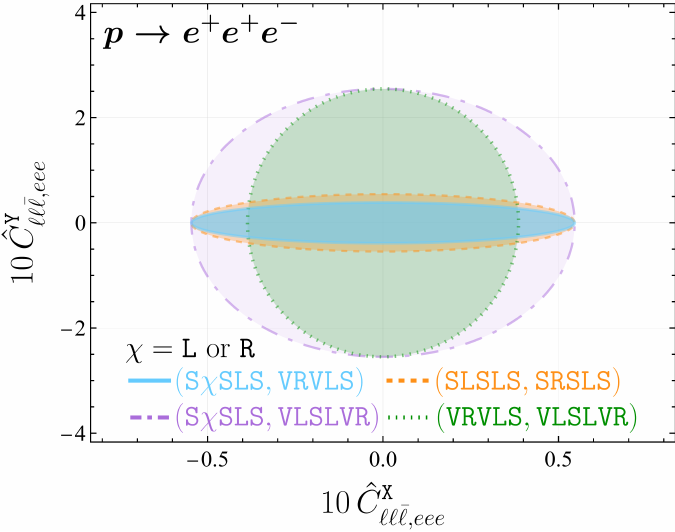}~
\includegraphics[width=0.49\linewidth]{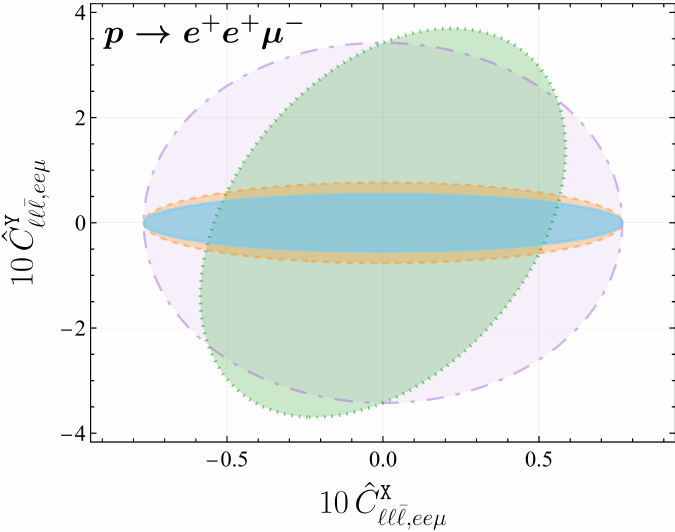}
\\[1em]
\includegraphics[width=0.49\linewidth]{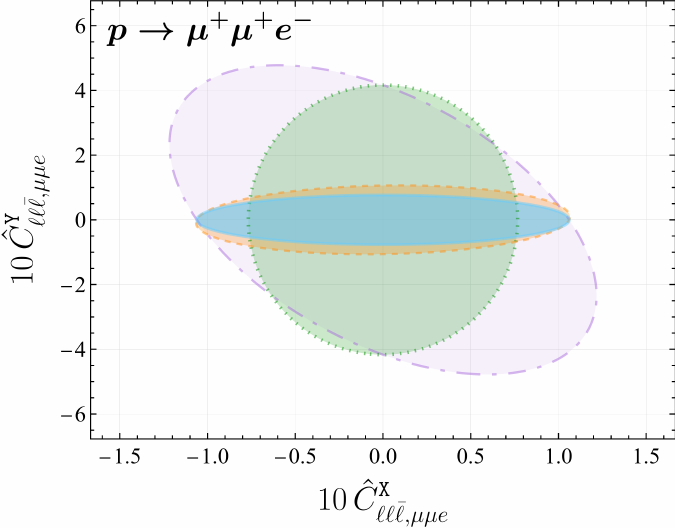}~
\includegraphics[width=0.49\linewidth]{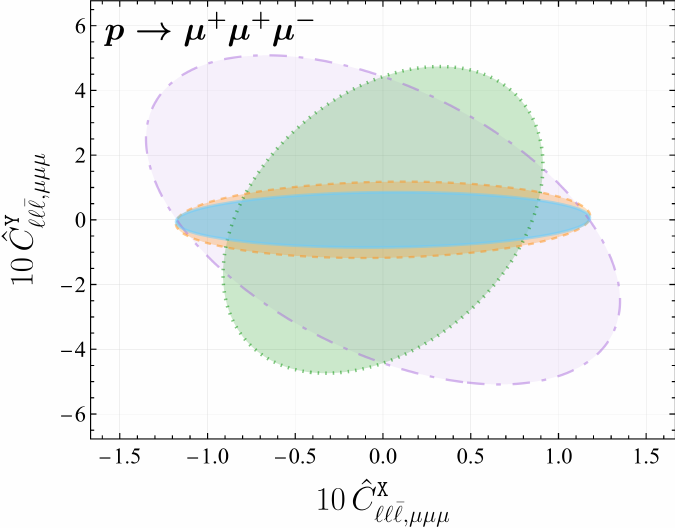}
\caption{
Constraints in the planes of pairs of WCs from proton decay modes 
$p \to e^+ e^+ e^-$ (top left), 
$p \to e^+ e^+ \mu^-$ (top right), 
$p \to \mu^+ \mu^+ e^-$ (bottom left),
and $p \to \mu^+ \mu^+ \mu^-$ (bottom right).}
\label{fig:2Dconstraint}
\end{figure}

Finally, we study the constraints in two-dimensional planes by analyzing pairs of WCs that contribute simultaneously. Using the four proton decay modes analyzed in \cref{fig:double_distribution} as examples, we illustrate the key features. 
For better visualization, we define a dimensionless counterpart for each WC $C_{\ell\ell\bar\ell,xxy}^i$ 
through the relation $\hat C_{xx\bar y}^i\equiv (300\,{\rm TeV})^5 C_{\ell\ell\bar\ell,xxy}^i$, where the effective scale is fixed at $300\,\rm TeV$.
Furthermore, for each process $p\to \ell_x^+\ell_x^+\ell_y^-\,(xy=ee,e\mu,\mu e, \mu\mu)$, we focus on the following four pairs of WC combinations: 
$(\hat C_{xx\bar y}^{\tS\chi\tS\tL\tS},\hat C_{xx\bar y}^{\tV\tR\tV\tL\tS})$,
$(\hat C_{xx\bar y}^{\tS\chi\tS\tL\tS},\hat C_{xx\bar y}^{\tV\tL\tS\tL\tV\tR})$,
$(\hat C_{xx\bar y}^{\tS\tL\tS\tL\tS},\hat C_{xx\bar y}^{\tS\tR\tS\tL\tS})$,
and $(\hat C_{xx\bar y}^{\tV\tR\tV\tL\tS},\hat C_{xx\bar y}^{\tV\tL\tS\tL\tV\tR})$, where $\chi=\tL,\tR$. 
Since a pair of identical lepton fields is involved in each relevant operator, the decay widths from the first two pairs of WC combinations are independent of chirality in the lepton scalar bilinear, leading to identical constraints for the cases $\chi=\tL$ and $\tR$. 
Our final results are presented in \cref{fig:2Dconstraint}, where the WCs are assumed to be real. As shown in the plots, these processes constrain the WC pairs in closed contours. The bounds on the two pairs $(\hat C_{xx\bar y}^{\tS\chi\tS\tL\tS},\hat C_{xx\bar y}^{\tV\tR\tV\tL\tS})$ and $(\hat C_{xx\bar y}^{\tS\tL\tS\tL\tS},\hat C_{xx\bar y}^{\tS\tR\tS\tL\tS})$ are stronger than those on the other two pairs. This is because the operators associated with these WCs 
belong exclusively to the chiral irreps ${\bar{\pmb{3}}_{\tL(\tR)}\otimes\pmb{3}_{\tR(\tL)}}$ and ${\pmb{8}_{\tL(\tR)}\otimes\pmb{1}_{\tR(\tL)}}$. Similarly, constraints on pairs of WCs from other processes can be studied using the full numerical results provided in \cref{app:decay_width_WCs}, where comparable elliptical or band-like constraints are anticipated.

%%%%%%%%%%%%%%%%%%%%%%%%%%%%%%%%%%%%%%
\section{A UV-complete model}
\label{sec:UVmodel}
%%%%%%%%%%%%%%%%%%%%%%%%%%%%%%%%%%%%%%

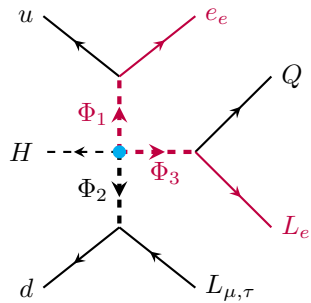
\begin{figure}[t]
\centering
\begin{tikzpicture}[decoration={markings, 
mark= at position 0.6 with {\arrow{stealth}},
mark= at position 3cm with {\arrow{stealth}}},scale=1.2]
\begin{scope}[shift={(1,1)}]
\draw[postaction={decorate},thick] (0, 0.7) --(-1.0,1.5) node[left]{$u$};
\draw[postaction={decorate},thick, purple] (0, 0.7) -- (1.0,1.5) node[right]{$e_e$};
\draw[dashed,postaction={decorate},ultra thick, purple] (0,-0.3) -- (0,0.7) node[midway,xshift = -10 pt]{$\Phi_{1}$};
\draw[dashed,postaction={decorate}, ultra thick,purple] (0,-0.3) -- (1.0,-0.3) node[midway, xshift=3pt, yshift = - 8 pt, purple]{$\Phi_3 $};
\draw[postaction={decorate},thick] (1,-0.3)--(2,0.7) node[right]{$Q$};
\draw[postaction={decorate},thick, purple] (1,-0.3)--(2,-1.3) node[right]{$L_e$};
\draw[dashed,postaction={decorate}, ultra thick] (0,-0.3) -- (0,-1.3) node[midway,xshift = -10 pt]{$\Phi_{2}$};
\draw[dashed,postaction={decorate},thick] (0,-0.3) -- (-1,-0.3)node[left]{$H$};
\draw[postaction={decorate},thick] (0,-1.3)--(-1.0,-2.1) node[left]{$d$};
\draw[postaction={decorate},thick](1.0,-2.1)node[right]{$L_{\mu,\tau}$}--(0,-1.3);
\draw[draw=cyan,fill=cyan] (0,-0.3) circle (0.08cm);
\end{scope}
\end{tikzpicture}
\caption{The diagram that induces dim-10 SMEFT BNV interactions.
There are two additional topologies involving new vertices $\overline{L_e}\Phi_{1} i\sigma_2Q_p^\C$ and $\epsilon_{\alpha\beta\gamma}  \Phi_2^{\alpha,\tt T} i \sigma_2 [\Phi_3^\beta,\Phi_3^\gamma] H$, respectively, which we have neglected for the sake of a minimal analysis.
}
\label{fig:BNVdim10}
\end{figure}

In this section, we take the UV model from~\cite{Liao:2025vlj} as an example to demonstrate the matching procedure and transformation of operators into our chosen LEFT operator basis.
We denote the SM Higgs by $H$, the left-handed quark and lepton doublets by $Q$ and $L$, and the right-handed up- and down-type quarks and leptons by $u,~d,~e$, respectively. 
In addition to the SM fields, the model introduces three leptoquarks $\Phi_{1,2,3}$, whose masses are denoted by $M_{1,2,3}$, respectively. Under the SM gauge group $\rm SU(3)_c\otimes SU(2)_L\otimes U(1)_Y$, their representations are: $\Phi_1\sim (\pmb{3},\pmb{1},-1/3)$, 
$\Phi_2\sim ({\pmb 3}, \pmb{2},1/6)$, and 
$\Phi_3\sim (\pmb{3}, \pmb{3},-1/3)$. 
Furthermore, a lepton-flavored $\mathbb{Z}_2$ symmetry is imposed, under which $\Phi_{1,3}$ and the electron-type leptons $e_e$ and $L_e$ are odd, while all other fields are even.
Due to the $\mathbb{Z}_2$ symmetry, the model generates BNV SMEFT interactions at dimension 10 at leading order, as illustrated by the diagram shown in \cref{fig:BNVdim10}.
The same symmetry implies that, as in the SM, this model preserves $\Delta(B-L)=0$, which forbids the generation of any odd-dimensional operators such as those at dimension 7 or 9.

The relevant Lagrangian terms involved in the diagram are 
\begin{align}
\mathcal{L}=
y_{1e,p}\,\overline{e_e}u_p^\C \Phi_1
+y_{2x,p}\,\Phi_2^{\tt T}i\sigma_2\overline{d_p} L_x
+y_{3e,p}\,\overline{L_e}\Phi_{3} i\sigma_2Q_p^\C
+\kappa\,\epsilon_{\alpha\beta\gamma} \Phi_1^\alpha \Phi_2^{\beta,\tt T} i \sigma_2 \Phi_3^\gamma H+\hc,
\end{align}
where $\alpha,\beta,\gamma$ are the color indices.
The lepton index $x$ is summed over the $\mu$ and $\tau$ flavors, 
while the quark generation index $p=1,2,3$.
The $\rm SU(2)_L$ triplet field $\Phi_3$ is represented by $\Phi_3=\Phi_{3}^I \sigma^I$ with $\sigma^I$ ($I=1,2,3$) being the Pauli matrices. The parameters $y_{1,2,3}$ denote new Yukawa couplings, 
and $\kappa$ represents the quartic Higgs coupling involving the four different scalar fields. 
After integrating out the leptoquarks $\Phi_{1,2,3}$, \cref{fig:BNVdim10} yields the following dim-10 SMEFT BNV interaction 
\begin{align}
\label{eq:model2SMEFT}
\mathcal{L}_{\tt SMEFT}^{\rm dim10}=C_{x,prs}^{(10)}
\mathcal{O}_{x,prs}^{(10)}+\hc,
\end{align}
where the operator and its corresponding WC are
\begin{align}
\mathcal{O}_{x,prs}^{(10)}=
\epsilon_{\alpha\beta\gamma}(\overline{u_p^{\alpha \C}}e_e)\,
(\overline{L_x}d_r^\beta \sigma_I H)\,
(\overline{Q_s^{\gamma\C}}i\sigma_2\sigma^I L_e),\quad 
C_{x,prs}^{(10)}&= -\frac{y_{1e,p}^* y_{2x,r}^{*} y_{3e,s}^*\kappa}{M_1^2 M_2^2 M_3^2}.
\end{align}
Using the relation,
$ (\sigma^I)_{ij}(\epsilon \sigma^I)_{kl} = 2\delta_{il}\epsilon_{kj} - \delta_{ij} \epsilon_{kl}$, 
where $\epsilon = i\sigma^2$ is the rank-two anti-symmetric tensor with $\epsilon_{12}\equiv1$, we have 
\begin{align}
\mathcal{O}_{x,prs}^{(10)}=
\epsilon_{\alpha\beta\gamma}
(2\delta_{il}\epsilon_{kj} - \delta_{ij} \epsilon_{kl})
(\overline{u_p^{\alpha \C}}e_e)\,
(\overline{L_x^i}d_r^\beta)\,
(\overline{Q_s^{\gamma k \C}} L_e^l) H^j. 
\end{align}

Next, we consider the matching onto LEFT at the electroweak scale. 
We work on the up-quark flavor basis, where the left- and right-handed up-type quark fields, as well as the right-handed down-type quark fields, are already in their mass eigenstates. The left-handed down-type quark weak eigenstates, $d'_{\tL}$, and mass eigenstates, $d_{\tL}$, are related by the CKM matrix $V$ via $d'_{\tL}=V d_{\tL}$. By sending $H\to v/\sqrt{2}$, \cref{eq:model2SMEFT} leads to the following dim-9 LEFT interactions involving up and down quarks that contribute to nucleon triple-lepton decays, 
\begin{align}
\label{eq:LEFT1}
\mathcal{L}_{\tt LEFT}^{(9)}=C_1^x\mathcal{O}_1^x+C_2^x\mathcal{O}_2^x+C_3^x\mathcal{O}_3^x+\hc,
\end{align}
with the LEFT operators and the corresponding WCs being
\begin{subequations}
\begin{align}
\label{eq:LEFT_UV}
\mathcal{O}_1^x&=\epsilon_{\alpha\beta\gamma}(\overline{u^{\alpha\C}_\tR}e_\tR)(\overline{\nu_{\tL x}}d_\tR^\beta)(\overline{u^{\gamma\C}_\tL}\nu_{ \tL e}), &
C_1^x(\Lambda_{\tt EW}) &= \sqrt{2} v C_{x,111}^{(10)},
\\
\mathcal{O}_2^x&=\epsilon_{\alpha\beta\gamma}(\overline{u^{\alpha\C}_\tR}e_\tR)(\overline{\ell_{\tL x}}d_\tR^\beta)(\overline{u^{\gamma\C}_\tL}e_\tL), &
C_2^x(\Lambda_{\tt EW}) &= \frac{v}{\sqrt{2}} C_{x,111}^{(10)}, 
\\
\mathcal{O}_3^x&=\epsilon_{\alpha\beta\gamma}(\overline{u^{\alpha\C}_\tR}e_\tR)(\overline{\ell_{\tL x}}d_\tR^\beta)(\overline{d^{\gamma\C}_\tL}\nu_{\tL e}), &
C_3^x(\Lambda_{\tt EW}) &= \frac{v}{\sqrt{2} }V_{w1} C_{x,11w}^{(10)}. 
\end{align}
\end{subequations}

Now, we transform the operators $\calO_{1,2,3}^x$ into our established basis in \cref{sec:dim9basis} through various Fierz identities.
According to the organization of the lepton fields for each operator in \cref{tab:BNVdim9opeLp1A,tab:BNVdim9opeLp1B}, we combine the two neutrino fields in $\calO_1^x$, the two like-sign charged leptons in $\calO_2^x$, and the two opposite-sign charged leptons in $\calO_3^x$ into a single fermion bilinear. This can be finished by using the following Fierz identities for $(\overline{\nu_{\tL x}}d_\tR^\beta)(\overline{u^{\gamma\C}_\tL}\nu_{ \tL e})$, $(\overline{u^{\alpha\C}_\tR}e_\tR)(\overline{u^{\gamma\C}_\tL}e_\tL)$, and $(\overline{u^{\alpha\C}_\tR}e_\tR)(\overline{\ell_{\tL x}}d_\tR^\beta)$ respectively,
\begin{subequations}
 \begin{align}
(\overline{\psi_{1\tL}} \psi_{2\tR})
(\overline{\psi_{3\tL}^\C} \psi_{4\tL})
& = + \frac{1}{2}(\overline{\psi_{1\tL}}\gamma_\mu \psi_{4\tL})
(\overline{\psi_{2\tR}^\C}\gamma^\mu \psi_{3\tL}),
\\
(\overline{\psi_{1\tR}^\C} \psi_{2\tR})
(\overline{\psi_{3\tL}^\C} \psi_{4\tL})
& =- \frac{1}{2} (\overline{\psi_{4\tL}^\C}\gamma_\mu \psi_{2\tR})(\overline{\psi_{1\tR}^\C}\gamma^\mu \psi_{3\tL}),
\\
(\overline{\psi_{1\tR}^\C} \psi_{2\tR})
(\overline{\psi_{3\tL}} \psi_{4\tR})
&=-\frac{1}{2}
(\overline{\psi_{3\tL}} \psi_{2\tR})
(\overline{\psi_{1\tR}^\C} \psi_{4\tR})
-\frac{1}{8}
(\overline{\psi_{3\tL}}\sigma_{\mu\nu}\psi_{2\tR})
(\overline{\psi_{1\tR}^\C}\sigma^{\mu\nu}\psi_{4\tR}).
\end{align}   
\end{subequations}
Using the identities  
$\overline{\psi_{1\tL}} \psi_{2\tR}=\overline{\psi_{1\tR}^\C} \psi_{2\tL}^\C$ and 
$(\overline{\psi_{1\tL}}\gamma^\mu \psi_{2\tL})=-(\overline{\psi_{2\tL}^\C}\gamma^\mu \psi_{1\tL}^\C)$, as well as their counterparts with $\tL\leftrightarrow\tR$,
the three operators can be factorized into products of lepton and quark sectors:
\begin{subequations}
\begin{align}
 \calO_1^x & = \frac{1}{2}(\overline{\nu_{\tL x}}\gamma_\mu \nu_{\tL e})
 (\overline{e_\tR^\C} u_\tR^\alpha)
 (\overline{d_\tR^{\beta\C}}\gamma^\mu u_\tL^\gamma)
 \epsilon_{\alpha\beta\gamma},
 \\
 \calO_2^x & = \frac{1}{2}(\overline{e_\tL^\C}\gamma_\mu e_\tR)
 (\overline{\ell_{\tL x}}d_\tR^\beta)
 (\overline{u_\tR^{\beta\C}}\gamma^\mu u_\tL^\gamma)
 \epsilon_{\alpha\beta\gamma},
 \\
 \calO_3^x & = -\frac{1}{2}(\overline{\ell_{\tL x}} e_\tR)
 (\overline{\nu_{\tL e}^\C}d^{\alpha}_\tL)
(\overline{u_\tR^{\beta\C}}d_\tR^\gamma)
\epsilon_{\alpha\beta\gamma}
-\frac{1}{8}(\overline{\ell_{\tL x}}\sigma_{\mu\nu} e_\tR)
 (\overline{\nu_{\tL e}^\C}d^{\alpha}_\tL)
(\overline{u_\tR^{\beta\C}}\sigma^{\mu\nu}d_\tR^\gamma)
\epsilon_{\alpha\beta\gamma}.
\end{align}
\end{subequations}

Applying the identities given in Eq.\,(S1) of the Supplemental Material in Ref.~\cite{Liao:2025vlj} and the definitions in \cref{eq:Nyzw},
we finally obtain  
\begin{subequations}
\begin{align}
\calO_1^x &=
\frac{1}{2}\calO_{\bar\nu\nu\ell,xee}^{\tV\tL\tS\tR\tV\tL1}
-\frac{1}{4}\calO_{\bar\nu\nu\ell,xee}^{\tV\tL\tV\tL\tS\tR},
\\
\calO_2^x &=
\frac{1}{2}\calO_{\ell\ell\bar\ell,eex}^{\tV\tR\tS\tR\tV\tL1}
+ \frac{1}{4}\calO_{\ell\ell\bar\ell,eex}^{\tV\tR\tV\tL\tS\tR},
\\
\calO_3^x &=\frac{i}{4}\calO_{\bar\ell\ell\nu,x ee}^{\tT\tR\tV\tR\tV\tL1}
 -\frac{1}{2}\calO_{\bar\ell\ell\nu,x ee}^{\tS\tR\tS\tL\tS\tR},
\end{align}
\end{subequations}
where the operators on the right-hand side of the equalities are listed in  \cref{tab:BNVdim9opeLp1A,tab:BNVdim9opeLp1B}.
After including the renormalization group (RG) effects from the electroweak scale down to the chiral symmetry breaking scale $\Lambda_\chi$, the resulting LEFT Lagrangian responsible for nucleon triple-lepton decays from this model is given by
\begin{align}
\mathcal{L}_{{\tt N}\to l_1 l_2 l_3}^{(9)}=&
C^{\tV\tR\tV\tL\tS\tR}_{\ell\ell\bar{\ell},eex}
\calO^{\tV\tR\tV\tL\tS\tR}_{\ell\ell\bar{\ell},eex}
+C^{\tV\tR\tS\tR\tV\tL1}_{\ell\ell\bar{\ell}, eex}
\calO^{\tV\tR\tS\tR\tV\tL1}_{\ell\ell\bar{\ell},eex}
+C^{\tT\tR\tV\tR\tV\tL1}_{\bar{\ell}\ell\nu,x ee}
\calO^{\tT\tR\tV\tR\tV\tL1}_{\bar{\ell}\ell\nu,x ee}
+C^{\tS\tR\tS\tL\tS\tR}_{\bar{\ell}\ell\nu,x ee}
\calO^{\tS\tR\tS\tL\tS\tR}_{\bar{\ell}\ell\nu,x ee}
\nn\\
&+C^{\tV\tL\tS\tR\tV\tL1}_{\bar{\nu}\nu\ell,x ee}
\calO^{\tV\tL\tS\tR\tV\tL1}_{\bar{\nu}\nu\ell,x ee}
+C^{\tV\tL\tV\tL\tS\tR}_{\bar{\nu}\nu\ell,x ee}
\calO^{\tV\tL\tV\tL\tS\tR}_{\bar{\nu}\nu\ell,x ee},
\end{align}
where $x=\mu,\tau$ and 
\begin{subequations}
\begin{align}
C^{\tV\tR\tV\tL\tS\tR}_{\ell\ell\bar{\ell},eex} &
= 1.32\frac{\sqrt{2}v}{8}C_{x,111}^{(10)}, 
&C^{\tV\tR\tS\tR\tV\tL1}_{\ell\ell\bar{\ell}, eex} &
= 0.91\frac{\sqrt{2}v}{4}C_{x,111}^{(10)},
\\ 
C^{\tS\tR\tS\tL\tS\tR}_{\bar{\ell}\ell\nu,x ee} & 
= -1.32\frac{\sqrt{2}v}{4}V_{w1}C_{x,11w}^{(10)}, 
&C^{\tT\tR\tV\tR\tV\tL1}_{\bar{\ell}\ell\nu,x ee} &
= 0.91\frac{i\sqrt{2}v}{8}V_{w1}C_{x,11w}^{(10)}, 
\\ 
C^{\tV\tL\tV\tL\tS\tR}_{\bar{\nu}\nu\ell,x ee} 
&=- 1.32\frac{\sqrt{2}v}{4}C_{x,111}^{(10)}, 
& C^{\tV\tL\tS\tR\tV\tL1}_{\bar{\nu}\nu\ell,x ee} &=
0.91\frac{\sqrt{2}v}{2}C_{x,111}^{(10)}.
\end{align}
\end{subequations}
The numerical factors $1.32$ and $0.91$ in the WCs are the RG running factors associated with the operators in the chiral irreps $\pmb{3}_{\tL(\tR)}\otimes \bar{\pmb{3}}_{\tR(\tL)}$ and $\pmb{3}_{\tL(\tR)}\otimes \pmb{6}_{\tR(\tL)}$~\cite{Liao:2025vlj}, respectively. 
These interactions will induce the four nucleon triple-lepton decay processes: 
$p\to e^+e^+\mu^-$, $n\to \mu^- e^+\bar{\nu}_e$, and $p\to \nu_{\mu,\tau} \bar{\nu}_e e^+$.
From the complete expressions given in \cref{eq:Gammap2eemu,eq:Gamman2muevb,eq:Gammap2vvbe},
this model leads to 
\begin{subequations}
\begin{align}
\Gamma_{p\to e^+e^+\mu^-}
& \approx \big( 2.2|C_{\ell\ell\bar\ell,ee\mu}^{\tV\tR\tV\tL\tS\tR}|^2
+0.01\kappa_3^2 |C_{\ell\ell\bar\ell,ee\mu}^{\tV\tR\tS\tR\tV\tL1}|^2\big) 
\big(10^{-9}{\rm GeV}^{11}\big)
\nonumber
\\
& \approx 7.2\cdot 10^{-6}\big(1+0.01\kappa_3^2\big) \big|C_{\mu,111}^{(10)}{\rm GeV}^6\big|^2{\rm GeV},
\\%%
\Gamma_{n\to \mu^-e^+ \bar{\nu}_e} & \approx 
\big( 0.6|C_{\bar\ell \ell \nu,\mu e e}^{\tS\tR\tS\tL\tS\tR}|^2
+ 0.05 \kappa_3^2 |C_{\bar\ell \ell \nu,\mu e e}^{\tT\tR\tV\tR\tV\tL1}|^2 \big)
\big(10^{-9}{\rm GeV}^{11}\big)
\nonumber
\\
& \approx 7.9\cdot 10^{-6}\big(1+0.01\kappa_3^2\big) \big|V_{w1}C_{\mu,11w}^{(10)}{\rm GeV}^6\big|^2{\rm GeV},
\\%%
\Gamma_{p\to \nu_x\bar{\nu}_e e^+} & \approx   
\big( 2.4|C_{\bar\nu \nu \ell,xee}^{\tV\tL\tV\tL\tS\tR}|^2
+ 0.01 \kappa_3^2 |C_{\bar\nu \nu \ell,xee}^{\tV\tL\tS\tR\tV\tL1}|^2 \big)
\big(10^{-9}{\rm GeV}^{11}\big)
\nonumber
\\
& \approx 3.2\cdot 10^{-5}\big(1+0.01\kappa_3^2\big) \big|C_{x,111}^{(10)}{\rm GeV}^6\big|^2{\rm GeV}, \quad x=\mu,\tau. 
\end{align}
\end{subequations}

\begin{figure}[t]
\centering
\includegraphics[width=0.49\linewidth]{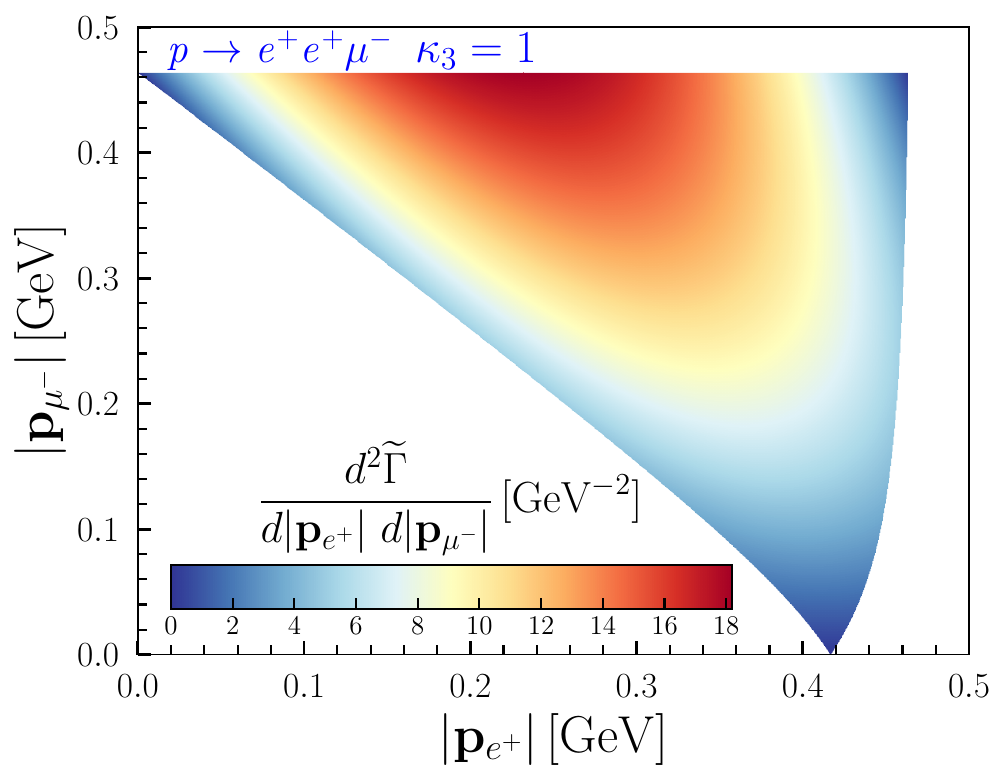}~
\includegraphics[width=0.49\linewidth]{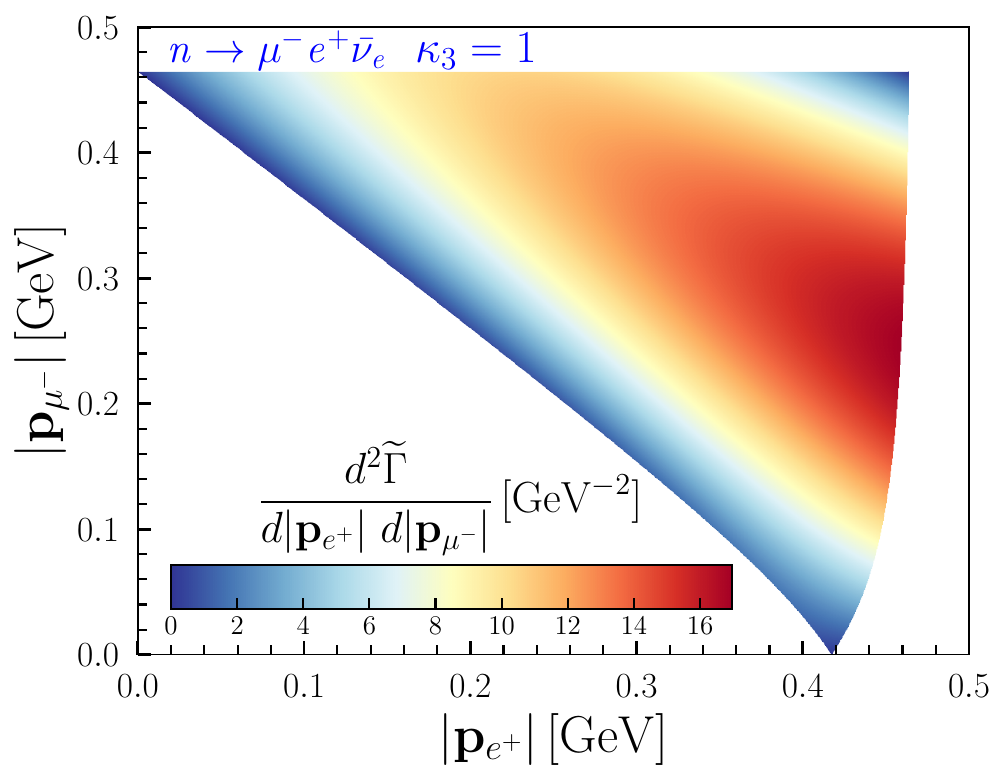}
\caption{
Normalized double-momenta distributions of the processes $p\to e^+ e^+\mu^-$ (left panel) and $n\to \mu^-e^+\bar{\nu}_e$ (right panel) predicted by the leptoquark model.}
\label{fig:distribution_model}
\end{figure}

\begin{figure}[t]
\centering
\includegraphics[width=0.6\linewidth]
{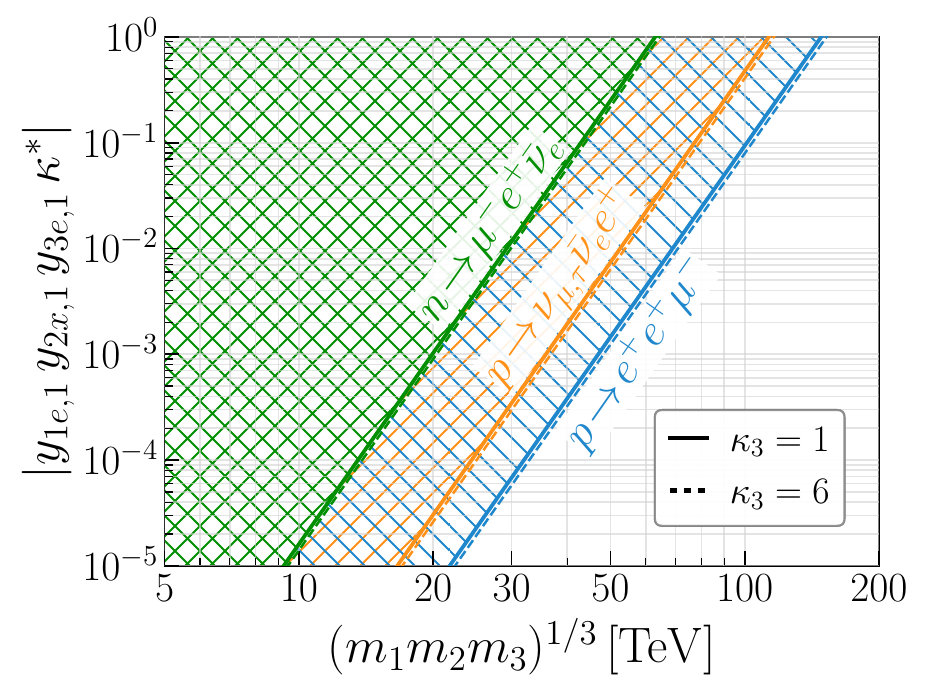}
\caption{
Excluded regions in the $\Lambda_{\tt NP}{\rm-}c_x$ plane. 
}
\label{fig:exclude_region}
\end{figure}

In this specific model, each nucleon decay mode depends only on one independent parameter, either $C_{x,111}^{(10)}$ or $V_{w1}C_{x,11w}^{(10)}$. Therefore, it is useful to analyze the charged-lepton's momentum distributions and compare them with those shown in \cref{fig:double_distribution}.  
We focus on the two modes, $p\to e^+e^+\mu^-$ and 
$n\to \mu^-e^+ \bar{\nu}_e$, both involving at least two charged leptons. The corresponding normalized distributions are shown in \cref{fig:distribution_model}.
In these plots, we set $\kappa_3=1$.
Increasing $\kappa_3$ causes the distributions to become more concentrated in the red regions, while decreasing $\kappa_3$ results in a more diffuse distribution away from those regions.
For the process $p\to e^+e^+\mu^-$, 
we observe that the distribution differs from those of the individual operators shown in \cref{fig:double_distribution},
which is attributed to interference effects between the two dim-9 operators. 
Defining a new physics scale $\Lambda_{\tt NP}\equiv (M_1 M_2 M_3)^{1/3}$ and a dimensionless coupling $c_x \equiv |y_{1e,1}y_{2x,1}y_{3e,1}\kappa^*|$, we obtain $C_{x,111}^{(10)} = - c_x/\Lambda_{\tt NP}^6$.  \cref{fig:exclude_region} displays the excluded regions in the $\Lambda_{\tt NP}{\rm-}c_x$ plane based on the current experimental bounds on the four processes illustrated in the plot. For the decay $n\to\mu^- e^+\bar\nu_e$, the CKM matrix has been approximated by the identity matrix. Taking Yukawa and scalar quartic couplings of $\calO(0.1)$, the new physics scale can be probed up to several tens of TeV, indicating that such models are potentially testable at future collider experiments.  

%%%%%%%%%%%%%%%%%%%%%%%%%%%%%%%%%%%%%%
\section{Summary}
\label{sec:summary}
%%%%%%%%%%%%%%%%%%%%%%%%%%%%%%%%%%%%%%

We have systematically investigated BNV nucleon decays into three leptons induced by contact dim-9 operators within the LEFT framework. These processes exhibit distinctive experimental signatures and can probe BNV new physics scenarios beyond the conventional two-body decays. 
We classified all such BNV decay modes and constructed the complete set of dim-9 LEFT operators that contribute to them.
These operators are organized into four generic triple-quark structures 
identified in~\cite{Liao:2025vlj}, corresponding to the chiral irreducible representations, 
\{$\pmb{8}_{\tt L}\otimes \pmb{1}_{\tt R}$,
$\bar{\pmb{3}}_{\tt L}\otimes \pmb{3}_{\tt R}$,
$\pmb{6}_{\tt L}\otimes \pmb{3}_{\tt R}$,
$\pmb{10}_{\tt L}\otimes \pmb{1}_{\tt R}$\}, along with their chiral partners with ${\tt L}\leftrightarrow {\tt R}$.
In particular, for the $\Delta F_L=3$ nucleon triple-lepton decays, the leading contributions arise from these dim-9 LEFT operators. 

We applied the spurion field method to match these dim-9 operators onto chiral perturbation theory, therefore obtaining their hadronic four-fermion counterparts. 
Using the resulting hadronic Lagrangian, we calculated all relevant nucleon triple-lepton decay widths, which are expressed as functions of the WCs associated with these dim-9 operators.  
Since these decays are inherently three-body processes, we analyzed their momentum distributions to demonstrate how such information can help to reveal the underlying interaction structures, and thus guide experimental searches.
Using existing experimental data, we established stringent limits on the effective scales associated with these LEFT operators. 
These scales are constrained to around $\mathcal{O}(100)\,\mathrm{TeV}$, rendering related new physics scenarios testable at future collider facilities.

Finally, we presented a UV-complete model that generates a subset of these dim-9 operators while simultaneously forbidding the lower-dimensional operators contributing to nucleon two-body decays. 
By employing various Fierz and group-theoretical identities, we demonstrated how our formalism can be used to constrain the relevant new physics models. 
In view of current and upcoming neutrino experiments,  
our results will facilitate further experimental searches and theoretical studies of these exotic nucleon decay modes.

%%%%%%%%%%%%%%%%%%%%%%%%%%
\acknowledgments
%%%%%%%%%%%%%%%%%%%%%%%%%%

We thank Ilja Doršner and two anonymous referees for several comments and suggestions that have helped us
improve the presentation of our work. 
This work was supported 
by Grants 
No.\,NSFC-12305110 
and No.\,NSFC-12035008.  

\appendix

%%%%%%%%%%%%%%%%%%%%%%%%%%%%%%%%%%%%%%%%%
\section{Nucleon decay amplitudes} 
\label{app:decay_width_WCsmplitude}
%%%%%%%%%%%%%%%%%%%%%%%%%%%%%%%%%%%%%%%%%

In this appendix, we provide the amplitude of each nucleon triple-lepton decay mode induced by the LEFT operators listed in \cref{tab:BNVdim9opeLp1A,tab:BNVdim9opeLp1B,tab:BNVdim9opeLm1,tab:BNVdim9opeL3}. 
Since multiple LEFT operators can contribute to the same four-fermion operators in chiral matching, it is convenient to introduce the following combinations of WCs for each decay to simplify the calculations.
\begin{itemize}
\item $p\to\ell_x^+\ell_y^+\ell_z^-$:
\begin{align}
C^{\tS\tL\tS\tL\tS}_{\ell\ell\bar\ell,xyz}&
\equiv C^{\tS\tL\tS\tL\tS\tR}_{\ell\ell\bar\ell,xyz}
+\kappa_2 C^{\tS\tL\tS\tL\tS\tL}_{\ell\ell\bar\ell,xyz},
&
C^{\tS\tR\tS\tL\tS}_{\ell\ell\bar\ell,xyz}&
\equiv C^{\tS\tR\tS\tL\tS\tR}_{\ell\ell\bar\ell,xyz}
+\kappa_2 C^{\tS\tR\tS\tL\tS\tL}_{\ell\ell\bar\ell,xyz},
\nn\\
C^{\tV\tR\tV\tL\tS}_{\ell\ell\bar\ell,xyz}&
\equiv C^{\tV\tR\tV\tL\tS\tR}_{\ell\ell\bar\ell,xyz}
+\kappa_2 C^{\tV\tR\tV\tL\tS\tL}_{\ell\ell\bar\ell,xyz},
&
C^{\tV\tL\tS\tL\tV\tR}_{\ell\ell\bar\ell,xyz}&
\equiv \kappa_3\big(C^{\tV\tL\tS\tL\tV\tR2}_{\ell\ell\bar\ell,xyz}
-\frac{1}{2} C^{\tV\tL\tS\tL\tV\tR1}_{\ell\ell\bar\ell,xyz}\big),
\nn\\
C^{\tT\tL\tT\tL\tS}_{\ell\ell\bar\ell,xyz}&
\equiv C^{\tT\tL\tT\tL\tS\tR}_{\ell\ell\bar\ell,xyz}
+\kappa_2 C^{\tT\tL\tT\tL\tS\tL}_{\ell\ell\bar\ell,xyz},
& 
C^{\tT\tL\tV\tL\tV\tR}_{\ell\ell\bar\ell,xyz}&
\equiv \kappa_3\big(C^{\tT\tL\tV\tL\tV\tR2}_{\ell\ell\bar\ell,xyz}
-\frac{1}{2} C^{\tT\tL\tV\tL\tV\tR1}_{\ell\ell\bar\ell,xyz}\big),
\label{eq:WCcomp2lll}
\end{align}  
and their chiral partners with $\tL\leftrightarrow\tR$.
\item $n\to\ell_x^-\ell_y^+\bar\nu_z$:
\begin{align}
C^{\tS\tL\tS\tL\tS}_{\bar\ell\ell\nu,xyz}&
\equiv C^{\tS\tL\tS\tL\tS\tR}_{\bar\ell\ell\nu,xyz}
 +\kappa_2 C^{\tS\tL\tS\tL\tS\tL}_{\bar\ell\ell\nu,xyz},
&
C^{\tS\tR\tS\tL\tS}_{\bar\ell\ell\nu,xyz}&
\equiv C^{\tS\tR\tS\tL\tS\tR}_{\bar\ell\ell\nu,xyz}
+\kappa_2 C^{\tS\tR\tS\tL\tS\tL}_{\bar\ell\ell\nu,xyz},
\nn\\
C^{\tV\tL\tV\tR\tS}_{\bar\ell\ell\nu,xyz}&
\equiv C^{\tV\tL\tV\tR\tS\tL}_{\bar\ell\ell\nu,xyz}
+\kappa_2 C^{\tV\tL\tV\tR\tS\tR}_{\bar\ell\ell\nu,xyz},
&
C^{\tV\tR\tV\tR\tS}_{\bar\ell\ell\nu,xyz}&
\equiv C^{\tV\tR\tV\tR\tS\tL}_{\bar\ell\ell\nu,xyz}
+\kappa_2 C^{\tV\tR\tV\tR\tS\tR}_{\bar\ell\ell\nu,xyz},
\nn\\
C^{\tT\tL\tT\tL\tS}_{\bar\ell\ell\nu,xyz}&
\equiv C^{\tT\tL\tT\tL\tS\tR}_{\bar\ell\ell\nu,xyz}
 +\kappa_2 C^{\tT\tL\tT\tL\tS\tL}_{\bar\ell\ell\nu,xyz},
&
C^{\tV\tR\tS\tL\tV\tR}_{\bar\ell\ell\nu,xyz}&
\equiv \kappa_3\big(C^{\tV\tR\tS\tL\tV\tR2}_{\bar\ell\ell\nu,xyz}
-\frac{1}{2} C^{\tV\tR\tS\tL\tV\tR1}_{\bar\ell\ell\nu,xyz}\big),
\nn\\
C^{\tV\tL\tS\tL\tV\tR}_{\bar\ell\ell\nu,xyz}&
\equiv \kappa_3\big(C^{\tV\tL\tS\tL\tV\tR2}_{\bar\ell\ell\nu,xyz}
-\frac{1}{2} C^{\tV\tL\tS\tL\tV\tR1}_{\bar\ell\ell\nu,xyz}\big),
&
C^{\tT\tR\tV\tR\tV\tL}_{\bar\ell\ell\nu,xyz}&
\equiv \kappa_3\big(C^{\tT\tR\tV\tR\tV\tL2}_{\bar\ell\ell\nu,xyz}
-\frac{1}{2} C^{\tT\tR\tV\tR\tV\tL1}_{\bar\ell\ell\nu,xyz}\big).
\label{eq:WCcomn2llvb}
\end{align}  
\item $p\to\nu_x\bar\nu_y\ell_z^+$:
\begin{align}
C^{\tV\tL\tV\tL\tS}_{\bar\nu \nu \ell,xyz}&
=C_{\bar\nu \nu \ell,xyz}^{\tV\tL\tV\tL\tS\tR}
+\kappa_2C_{\bar\nu \nu \ell,xyz}^{\tV\tL\tV\tL\tS\tL},
&
C^{\tV\tL\tV\tR\tS}_{\bar\nu \nu \ell,xyz}&
=C_{\bar\nu \nu \ell,xyz}^{\tV\tL\tV\tR\tS\tL}
+\kappa_2C_{\bar\nu \nu \ell,xyz}^{\tV\tL\tV\tR\tS\tR},
\nn\\
C_{\bar\nu \nu \ell,xyz}^{\tV\tL\tS\tL\tV\tR}&
=\kappa_3\big(C_{\bar\nu \nu \ell,xyz}^{\tV\tL\tS\tL\tV\tR2}
-\frac{1}{2}C_{\bar\nu \nu \ell,xyz}^{\tV\tL\tS\tL\tV\tR1}\big),
&
C_{\bar\nu \nu \ell,xyz}^{\tV\tL\tS\tR\tV\tL}&
=\kappa_3\big(C_{\bar\nu \nu \ell,xyz}^{\tV\tL\tS\tR\tV\tL2}
-\frac{1}{2}C_{\bar\nu \nu \ell,xyz}^{\tV\tL\tS\tR\tV\tL1}\big).
\label{eq:WCcomp2vvbl}
\end{align}
\item $n\to\bar\nu_x\bar\nu_y\nu_z$:
\begin{align}
C^{\tS\tL\tS\tR\tS}_{\nu\nu\bar\nu,xyz}&
=C^{\tS\tL\tS\tR\tS\tL}_{\nu\nu\bar\nu,xyz}
+\kappa_2 C^{\tS\tL\tS\tR\tS\tR}_{\nu\nu\bar\nu,xyz},
&
C^{\tT\tL\tV\tL\tV\tR}_{\nu\nu\bar\nu,xyz}&
=\kappa_3\big(C^{\tT\tL\tV\tL\tV\tR2}_{\nu\nu\bar\nu,xyz}
- \frac{1}{2}C^{\tT\tL\tV\tL\tV\tR1}_{\nu\nu\bar\nu,xyz}\big).
\label{eq:WCcomn2vbvbv}
\end{align}
\item $p\to\nu_x\nu_y\ell_z^+$:
\begin{align}
C^{\tS\tR\tS\tR\tS}_{\bar\nu\bar\nu\ell,xyz}&
=C^{\tS\tR\tS\tR\tS\tL}_{\bar\nu\bar\nu\ell,xyz}
+\kappa_2 C^{\tS\tR\tS\tR\tS\tR}_{\bar\nu\bar\nu\ell,xyz},
&
C^{\tS\tR\tS\tL\tS}_{\bar\nu\bar\nu\ell,xyz}&
=C^{\tS\tR\tS\tL\tS\tR}_{\bar\nu\bar\nu\ell,xyz}
+\kappa_2 C^{\tS\tR\tS\tL\tS\tL}_{\bar\nu\bar\nu\ell,xyz},
\nn\\
C^{\tT\tR\tT\tR\tS}_{\bar\nu\bar\nu\ell,xyz}&
=C^{\tT\tR\tT\tR\tS\tL}_{\bar\nu\bar\nu\ell,xyz}
+\kappa_2 C^{\tT\tR\tT\tR\tS\tR}_{\bar\nu\bar\nu\ell,xyz},
&
C^{\tT\tR\tV\tR\tV\tL}_{\bar\nu\bar\nu\ell,xyz}&
=\kappa_3\big(C^{\tT\tR\tV\tR\tV\tL2}_{\bar\nu\bar\nu\ell,xyz}
- \frac{1}{2}C^{\tT\tR\tV\tR\tV\tL1}_{\bar\nu\bar\nu\ell,xyz}\big).
\label{eq:WCcomp2vvl}
\end{align}
\item 
$n\to\bar\nu_x\bar\nu_y\bar\nu_z$:
\begin{align}
C^{\tS\tL\tS\tL\tS}_{\nu\nu\nu,xyz}=
C_{\nu\nu\nu,xyz}^{\tS\tL\tS\tL\tS\tR}
+\kappa_2C_{\nu\nu\nu,xyz}^{\tS\tL\tS\tL\tS\tL}.
\label{eq:WCcomn2vbvbvb}
\end{align}
\item For the other four classes of processes: 
$n\to\ell_x^-\ell_y^+\nu_z$, 
$n\to\nu_x\nu_y\bar\nu_z$,
$p\to\bar\nu_x\bar\nu_y\ell_z^+$,
and $n\to\nu_x\nu_y\nu_z$, their abbreviated WCs can be obtained respectively from those of 
the corresponding processes in 
\cref{eq:WCcomn2llvb}, \cref{eq:WCcomn2vbvbv}, \cref{eq:WCcomp2vvl}, and \cref{eq:WCcomn2vbvbvb} by $\tL\leftrightarrow\tR$ and $\nu\leftrightarrow\bar\nu$ simultaneously. 
\end{itemize}
With the above WC abbreviations, the amplitude for each process can be expressed as follows:\\
$\bullet\,\,p(p)\to\ell_x^+(p_1)\ell_y^+(p_2)\ell_z^-(p_3)$:
\begin{subequations}
\label{eq:Amp_p2lll}
\begin{align}
\mathcal{M}_{p\to e^+\mu^+\ell_z^-}
= & c_1\Big\{
\big[\overline{u(p_1)} 
(C_{\ell\ell\bar\ell,e\mu z}^{\tS\tL\tS\tL\tS} P_\tL
+C_{\ell\ell\bar\ell,e\mu z}^{\tS\tR\tS\tL\tS} P_\tR) v(p_2)\big]
\big[\overline{u(p_3)} P_\tL u(p) \big]
\nn\\
&+C_{\ell\ell\bar\ell,e\mu z}^{\tT\tL\tT\tL\tS}
\big[\overline{u(p_1)} \sigma_{\mu\nu}P_\tL v(p_2)\big]
\big[\overline{u(p_3)} \sigma^{\mu\nu}P_\tL u(p)\big]
\nn\\
&+\Lambda_\chi^{-1}
C_{\ell\ell\bar\ell,e\mu z}^{\tT\tL\tV\tL\tV\tR}
\big[\overline{u(p_1)} \sigma_{\nu\mu}P_\tL v(p_2)\big]
\big[\overline{u(p_3)} \gamma^\nu  \tilde p^\mu P_\tL u(p) \big]
\nn\\
&+\Big[\Lambda_\chi^{-1}C_{\ell\ell\bar\ell,e\mu z}^{\tV\tL\tS\tL\tV\tR}
\big[\overline{u(p_1)} \gamma_{\mu}P_\tL v(p_2)\big]
\big[\overline{u(p_3)} \tilde p^\mu P_\tL u(p)\big]
\nn\\
&+C_{\ell\ell\bar\ell,e\mu z}^{\tV\tR\tV\tL\tS}
\big[\overline{u(p_1)}\gamma_\mu P_\tR v(p_2)\big] 
\big[\overline{u(p_3)}\gamma^\mu P_\tL u(p) \big]
-(e,p_1)\leftrightarrow (\mu,p_2)\Big]
\Big\}
\nn\\
&-\tL\leftrightarrow\tR,
\\%%
\mathcal{M}_{p\to \ell_x^+\ell_x^+\ell_y^-}
=& c_1\Big\{
\big[\overline{u(p_1)} 
(C_{\ell\ell\bar\ell,xxy}^{\tS\tL\tS\tL\tS} P_\tL
+C_{\ell\ell\bar\ell,xxy}^{\tS\tR\tS\tL\tS} P_\tR) v(p_2)\big]
\big[\overline{u(p_3)} P_\tL u(p) \big]
\nn\\
&+\Lambda_\chi^{-1} C_{\ell\ell\bar\ell,xxy}^{\tV\tL\tS\tL\tV\tR}
\big[\overline{u(p_1)} \gamma_{\mu}P_\tL v(p_2)\big]
\big[\overline{u(p_3)} \tilde p^\mu P_\tL u(p)\big]
\nn\\
&+C_{\ell\ell\bar\ell,xxy}^{\tV\tR\tV\tL\tS}
\big[\overline{u(p_1)}\gamma_\mu P_\tR v(p_2)\big] 
\big[\overline{u(p_3)}\gamma^\mu P_\tL u(p)\big] -(p_1\leftrightarrow p_2)
\Big\}
-\tL\leftrightarrow\tR.
\end{align}
\end{subequations}
where $\tilde p^\mu = p^\mu - \frac{1}{4}\gamma^\mu\slashed{p}$.
\\%%
$\bullet\,\,n(p)\to\ell_x^-(p_1)\ell_y^+(p_2)\bar\nu_z(p_3)$:
\begin{align}
\mathcal{M}_{n\to\ell_x^-\ell_y^+\bar\nu_z}
=& c_1\Big\{
\big[\overline{u(p_1)}( C_{\bar\ell \ell \nu,xyz}^{\tS\tL\tS\tL\tS} P_\tL 
+ C_{\bar\ell \ell \nu,xyz}^{\tS\tR\tS\tL\tS} P_\tR )v(p_2)\big]
\big[\overline{u(p_3)}P_\tL u(p_\N) \big]
\nn\\
&- \big[\overline{u(p_1)}\gamma_\mu 
(C_{\bar\ell \ell \nu,xyz}^{\tV\tL\tV\tR\tS} P_\tL
+C_{\bar\ell \ell \nu,xyz}^{\tV\tR\tV\tR\tS} P_\tR) 
v(p_2)\big]
\big[\overline{u(p_3)}\gamma^\mu P_\tR u(p) \big]
\nn\\
&+C_{\bar\ell \ell \nu,xyz}^{\tT\tL\tT\tL\tS}
\big[\overline{u(p_1)} \sigma_{\mu\nu}P_\tL v(p_2)\big]
\big[\overline{u(p_3)}\sigma^{\mu\nu}P_\tL u(p) \big]
\nn\\
&-\Lambda_\chi^{-1}
\big[\overline{u(p_1)} \gamma_{\mu}
( C_{\bar\ell \ell \nu,xyz}^{\tV\tL\tS\tL\tV\tR}P_\tL
+ C_{\bar\ell \ell \nu,xyz}^{\tV\tR\tS\tL\tV\tR} P_\tR)
v(p_2)\big]
\big[\overline{u(p_3)} \tilde p^\mu P_\tL u(p) \big]
\nn\\
&+\Lambda_\chi^{-1}C_{\bar\ell \ell \nu,xyz}^{\tT\tR\tV\tR\tV\tL}
\big[\overline{u(p_1)} \sigma_{\nu\mu}P_\tR v(p_2)\big]
\big[\overline{u(p_3)}\gamma^\nu \tilde p^\mu P_\tR u(p) \big] \Big\}.
\end{align}
$\bullet\,\,p(p)\to\nu_x(p_1)\bar\nu_y(p_2)\ell_z^+(p_3)$:
\begin{align}
\mathcal{M}_{p\to\nu_x\bar\nu_y\ell_z^+}
= & c_1 \big[\overline{u(p_1)} \gamma_\mu P_\tL v(p_2)\big]\,
\Big\{\overline{u(p_3)} \Big[
\gamma^\mu ( C_{\bar\nu\nu\ell,xyz}^{\tV\tL\tV\tL\tS} P_\tL 
-C_{\bar\nu\nu\ell,xyz}^{\tV\tL\tV\tR\tS} P_\tR )
\nn\\
&+\Lambda_\chi^{-1} \tilde p^\mu
( C_{\bar\nu\nu\ell,xyz}^{\tV\tL\tS\tL\tV\tR} P_\tL
- C_{\bar\nu\nu\ell,xyz}^{\tV\tL\tS\tR\tV\tL} P_\tR )\Big]u(p)\Big\}.
\end{align}
$\bullet\,\,n(p)\to\bar\nu_x(p_1)\bar\nu_{y}(p_2)\nu_z(p_3)$:
\begin{align}
\mathcal{M}_{n\to\bar\nu_x\bar\nu_y\nu_z}
= & -c_1\Big\{
\Lambda_\chi^{-1}C_{\nu\nu\bar\nu,xyz}^{\tT\tL\tV\tL\tV\tR}
\big[\overline{u(p_1)} \sigma_{\nu\mu}P_\tL v(p_2)\big]
\big[\overline{u(p_3)}\gamma^\nu \tilde p^\mu P_\tL u(p_\N) \big] 
\nn\\
& +C_{\nu\nu\bar\nu,xyz}^{\tS\tL\tS\tR\tS}
\big[\overline{u(p_1)} P_\tL v(p_2)\big]\,\big[\overline{u(p_3)} P_\tR u(p_\N)\big] \Big\}
-\delta_{x,y}(p_1\leftrightarrow p_2).
\end{align}
$\bullet\,\,p(p)\to \nu_x(p_1) \nu_y(p_2) \ell_z^+(p_3)$:
\begin{align}
\mathcal{M}_{p\to\nu_x\nu_y\ell_z^+}
=& c_1\Big\{
\big[\overline{u(p_1)} P_\tR v(p_2)\big]\,\big[\overline{u(p_3)}
( C_{\bar\nu\bar\nu\ell,xyz}^{\tS\tR\tS\tL\tS}P_\tL
-C_{\bar\nu\bar\nu\ell,xyz}^{\tS\tR\tS\tR\tS} P_\tR) u(p)\big]
\nn\\
&- \Lambda_\chi^{-1}C_{\bar\nu\bar\nu\ell,xyz}^{\tT\tR\tV\tR\tV\tL}
\big[\overline{u(p_1)} \sigma_{\nu\mu}P_\tR v(p_2)\big]
\big[\overline{u(p_3)} \gamma^\nu \tilde p^\mu P_\tR u(p) \big]
\nn\\
& - C_{\bar\nu\bar\nu\ell,xyz}^{\tT\tR\tT\tR\tS}
\big[ \overline{u(p_1)} \sigma_{\mu\nu}P_\tR v(p_2)\big]
\big[ \overline{u(p_3)}\sigma^{\mu\nu} P_\tR u(p) \big] \Big\}
-\delta_{x,y}(p_1\leftrightarrow p_2).
\end{align}
$\bullet\,\,n(p)\to \bar\nu_x(p_1) \bar\nu_y(p_2) \bar\nu_z(p_3)$:
\begin{subequations}
\label{eq:Amp_n2vbvbvb}
\begin{align}
\mathcal{M}_{n\to\bar\nu_e\bar\nu_\mu\bar\nu_\tau}
=&\frac{c_1}{2}\Big\{C_{\nu\nu\nu,e\mu\tau}^{\tS\tL\tS\tL\tS}
\Big[\big[\overline{u(p_1)} P_\tL v(p_2)\big] 
\big[\overline{u(p_3)} P_\tL u(p) \big]
+p_2\leftrightarrow p_3\Big]
\nn\\
& + C_{\nu\nu\nu,\mu\tau e}^{\tS\tL\tS\tL\tS}
\Big[\big[\overline{u(p_2)} P_\tL v(p_3)\big]
\big[\overline{u(p_1)} P_\tL u(p) \big]+p_1\leftrightarrow p_3\Big]\Big\},
\\
\mathcal{M}_{n\to\bar\nu_x\bar\nu_x\bar\nu_{y\neq x}}
=&\frac{c_1}{2} C_{\nu\nu\nu,xxy}^{\tS\tL\tS\tL\tS} \Big\{
\big[\overline{u(p_1)} P_\tL v(p_2)\big] 
\big[\overline{u(p_3)} P_\tL u(p) \big]
\nn\\
& + \big[\overline{u(p_1)} P_\tL v(p_3)\big]
\big[\overline{u(p_2)} P_\tL u(p) \big]\Big\}-p_1\leftrightarrow p_2.
\end{align}
\end{subequations}
For the other four classes of processes 
$n\to\ell_x^-\ell_y^+\nu_z$
$n\to\nu_x\nu_y\bar\nu_z$,
$p\to\bar\nu_x\bar\nu_y\ell_z^+$,
and $n\to\nu_x\nu_y\nu_z$, their amplitudes, up to a minus sign,
are related to the corresponding modes above with $\nu\leftrightarrow\bar\nu$ by a simultaneous interchange of $\tL\leftrightarrow\tR$ and $\nu\leftrightarrow\bar\nu$. For brevity, we therefore do not show their amplitudes explicitly. 

For the two types of dim-9 operators $\calO_{\nu\nu\nu,xyz}^{\tS\tL\tS\tL\tS\tL}$   
and $\calO_{\nu\nu\nu,xyz}^{\tS\tL\tS\tL\tS\tR}$ contributing to the $\Delta L=3$ processes $n\to\bar\nu_x\bar\nu_y\bar\nu_z$, each type contains eight independent flavor-specific operators, as indicated in \cref{tab:BNVdim9opeL3}. 
For the process $n\to\bar\nu_e\bar\nu_\mu\bar\nu_\tau$, two flavor-specific operators in each type contribute, while for $n\to\bar\nu_x\bar\nu_x\bar\nu_{y\neq x}$, only one flavor-specific operator contributes. 
Therefore, for the amplitudes in \cref{eq:Amp_n2vbvbvb}, we choose the following eight independent flavor combinations:   
$xyz=e\mu\tau,\mu\tau e, ee\mu,ee\tau,\mu\mu e,\mu\mu\tau,\tau\tau e,\tau\tau\mu$. 
A similar convention applies to the $\Delta L=-3$ processes  $n\to \nu_x \nu_y \nu_z$. 
Note that there is no contribution from dim-9 LEFT operators to the neutron decay into three identical neutrinos or antineutrinos, i.e., no $n\to \nu_x\nu_x\nu_x$ or $n\to \bar\nu_x\bar\nu_x\bar\nu_x$. 
At the hadronic level, the only possible operator is $(\overline{\nu_{\tL,x}}\nu_{\tL,x}^\C)(\overline{\nu_{\tL,x}}n_\tR)$, which vanishes due to the Fierz identity:
\begin{align}
(\overline{\psi_{1\tL}} \psi_{2\tL}^\C)
(\overline{\psi_{3\tL}} \psi_{4\tR})=
-(\overline{\psi_{1\tL}} \psi_{3\tL}^\C)
(\overline{\psi_{2\tL}} \psi_{4\tR})
-(\overline{\psi_{3\tL}} \psi_{2\tL}^\C)
(\overline{\psi_{1\tL}} \psi_{4\tR}). 
\end{align}

%%%%%%%%%%%%%%%%%%%%%%%%%%%%%%%%%%%%%%
\section{Complete expressions for decay widths in terms of LEFT WCs}
\label{app:decay_width_WCs}
%%%%%%%%%%%%%%%%%%%%%%%%%%%%%%%%%%%%%%

With the amplitudes given in \cref{app:decay_width_WCsmplitude}, 
the calculation of the squared matrix elements can be completed using the {\tt FeynCalc} package~\cite{Shtabovenko:2016sxi}, followed by numerical integration over the phase space to obtain the final decay widths expressed in terms of the combinations of WCs in \cref{eq:WCcomp2lll,eq:WCcomn2llvb,eq:WCcomp2vvbl,eq:WCcomn2vbvbv,eq:WCcomp2vvl,eq:WCcomn2vbvbvb}.
For numerical integration, we use the central mass values of SM particles taken from the PDG~\cite{ParticleDataGroup:2024cfk}.
The chiral symmetry breaking scale is taken as $\Lambda_\chi\approx 1.2\,\rm{GeV}$,
and the central value of LEC $c_1$ is adopted.
Note that the LECs $c_{2,3}$ are incorporated through the parameters $\kappa_{2,3}$ in the WC combinations. 
In the following, we present the complete expressions for each decay mode, including the interference terms.\\
%%%%%
%%%%%p2lll
$\bullet\,\,p\to \ell^+\ell^+\ell^-$:
\begin{subequations}
\begin{align}
{\Gamma_{p\to e^+e^+e^-} \over 10^{-9}{\rm GeV}^{11}}  
& = 2.4|C_{\ell\ell\bar\ell, eee}^{\tV\tR\tV\tL\tS}|^2
+1.2\big(|C_{\ell\ell\bar\ell,eee}^{\tS\tL\tS\tL\tS}|^2
+|C_{\ell\ell\bar\ell,eee}^{\tS\tR\tS\tL\tS}|^2\big) 
+0.06|C_{\ell\ell\bar\ell,eee}^{\tV\tL\tS\tL\tV\tR}|^2
\nn\\
&+0.005\Re\big[(C_{\ell\ell\bar\ell, eee}^{\tS\tL\tS\tR\tS}
-C_{\ell\ell\bar\ell, eee}^{\tS\tR\tS\tR\tS})
C_{\ell\ell\bar\ell,eee}^{\tV\tR\tV\tL\tS*}
-C_{\ell\ell\bar\ell,eee}^{\tS\tL\tS\tL\tS}C_{\ell\ell\bar\ell, eee}^{\tS\tL\tS\tR\tS*}\big]
-0.003\Re\big(C_{\ell\ell\bar\ell,eee}^{\tV\tR\tV\tL\tS}C_{\ell\ell\bar\ell, eee}^{\tV\tL\tV\tR\tS*}\big)
\nn\\
&+0.002\Re\big[(C_{\ell\ell\bar\ell, eee}^{\tS\tL\tS\tL\tS}-C_{\ell\ell\bar\ell, ee e}^{\tS\tR\tS\tL\tS})C_{\ell\ell\bar\ell, eee}^{\tV\tL\tS\tL\tV\tR*}-C_{\ell\ell\bar\ell,eee}^{\tV\tL\tS\tL\tV\tR}
C_{\ell\ell\bar\ell,eee}^{\tV\tR\tV\tL\tS*}\big]
\nn\\
& +6\cdot10^{-6}\Re\big[ (C_{\ell\ell\bar\ell, eee}^{\tS\tR\tS\tL\tS}-C_{\ell\ell\bar\ell, eee}^{\tS\tL\tS\tL\tS})C_{\ell\ell\bar\ell, eee}^{\tV\tR\tV\tL\tS*}
-C_{\ell\ell\bar\ell, eee}^{\tS\tL\tS\tL\tS}C_{\ell\ell\bar\ell, eee}^{\tS\tR\tS\tL\tS*}\big]
\nn\\ 
&
+3\cdot10^{-6}\Re\big[(C_{\ell\ell\bar\ell, eee}^{\tS\tL\tS\tL\tS}-C_{\ell\ell\bar\ell, eee}^{\tS\tR\tS\tL\tS})C_{\ell\ell\bar\ell, eee}^{\tV\tR\tS\tR\tV\tL*}
-C_{\ell\ell\bar\ell, eee}^{\tV\tR\tS\tR\tV\tL}
C_{\ell\ell\bar\ell, eee}^{\tV\tR\tV\tL\tS*}\big]
\nn\\
&+5\cdot10^{-9}\Re\big(C_{\ell\ell\bar\ell, eee}^{\tS\tR\tS\tL\tS}C_{\ell\ell\bar\ell, eee}^{\tS\tL\tS\tR\tS*}
+C_{\ell\ell\bar\ell, eee}^{\tS\tL\tS\tL\tS}C_{\ell\ell\bar\ell, eee}^{\tS\tR\tS\tR\tS*}\big)
+4\cdot10^{-9}\Re\big(C_{\ell\ell\bar\ell, eee}^{\tV\tL\tS\tL\tV\tR}C_{\ell\ell\bar\ell, eee}^{\tV\tR\tS\tR\tV\tL*}\big)
\nn\\
&+\tL\leftrightarrow\tR,
\\
{\Gamma_{p\to \mu^+\mu^+\mu^-}\over 10^{-9}{\rm GeV}^{11}} 
& = 1.7|C_{\ell\ell\bar\ell,\mu\mu \mu}^{\tV\tR\tV\tL\tS}|^2
+0.9\big(|C_{\ell\ell\bar\ell,\mu\mu \mu}^{\tS\tL\tS\tL\tS}|^2
+|C_{\ell\ell\bar\ell,\mu\mu \mu}^{\tS\tR\tS\tL\tS}|^2\big) 
+0.06|C_{\ell\ell\bar\ell,\mu\mu \mu}^{\tV\tL\tS\tL\tV\tR}|^2
\nn\\
&+0.7\Re\big[(C_{\ell\ell\bar\ell, \mu\mu\mu}^{\tS\tL\tS\tR\tS}-C_{\ell\ell\bar\ell, \mu\mu\mu}^{\tS\tR\tS\tR\tS})C_{\ell\ell\bar\ell, \mu\mu \mu}^{\tV\tR\tV\tL\tS*}-C_{\ell\ell\bar\ell, \mu\mu \mu}^{\tS\tL\tS\tL\tS}C_{\ell\ell\bar\ell, \mu\mu\mu}^{\tS\tL\tS\tR\tS*}\big]
-0.3\Re\big(C_{\ell\ell\bar\ell, \mu\mu\mu}^{\tV\tR\tV\tL\tS}C_{\ell\ell\bar\ell, \mu\mu \mu}^{\tV\tL\tV\tR\tS*}\big)
\nn\\
&+0.2\Re\big[(C_{\ell\ell\bar\ell, \mu\mu\mu}^{\tS\tL\tS\tL\tS}-C_{\ell\ell\bar\ell, \mu\mu \mu}^{\tS\tR\tS\tL\tS})C_{\ell\ell\bar\ell, \mu\mu\mu}^{\tV\tL\tS\tL\tV\tR*}
-C_{\ell\ell\bar\ell,\mu\mu\mu}^{\tV\tL\tS\tL\tV\tR}
C_{\ell\ell\bar\ell,\mu\mu\mu}^{\tV\tR\tV\tL\tS*}\big]
\nn\\
&+0.2\Re\big[(C_{\ell\ell\bar\ell, \mu\mu\mu}^{\tS\tR\tS\tL\tS}-C_{\ell\ell\bar\ell,\mu\mu\mu}^{\tS\tL\tS\tL\tS})C_{\ell\ell\bar\ell, \mu\mu\mu}^{\tV\tR\tV\tL\tS*}-C_{\ell\ell\bar\ell, \mu\mu\mu}^{\tS\tL\tS\tL\tS}C_{\ell\ell\bar\ell,\mu\mu \mu}^{\tS\tR\tS\tL\tS*}\big]
\nn\\
&
+0.1\Re\big[(C_{\ell\ell\bar\ell, \mu\mu\mu}^{\tS\tL\tS\tL\tS}-C_{\ell\ell\bar\ell, \mu\mu \mu}^{\tS\tR\tS\tL\tS})C_{\ell\ell\bar\ell, \mu\mu\mu}^{\tV\tR\tS\tR\tV\tL*}
-C_{\ell\ell\bar\ell, \mu\mu\mu}^{\tV\tR\tS\tR\tV\tL}
C_{\ell\ell\bar\ell, \mu\mu\mu}^{\tV\tR\tV\tL\tS*}\big]
\nn\\
&+0.03\Re\big(C_{\ell\ell\bar\ell, \mu\mu \mu}^{\tS\tR\tS\tL\tS}C_{\ell\ell\bar\ell, \mu\mu \mu}^{\tS\tL\tS\tR\tS*}
+C_{\ell\ell\bar\ell, \mu\mu\mu}^{\tS\tL\tS\tL\tS}C_{\ell\ell\bar\ell, \mu\mu\mu}^{\tS\tR\tS\tR\tS*}\big)
+0.01\Re\big(C_{\ell\ell\bar\ell, \mu\mu \mu}^{\tV\tL\tS\tL\tV\tR}C_{\ell\ell\bar\ell, \mu\mu\mu}^{\tV\tR\tS\tR\tV\tL*}\big)
\nn\\
&+\tL\leftrightarrow\tR,
\\
{\Gamma_{p\to e^+e^+\mu^-} \over 10^{-9}{\rm GeV}^{11}}  
& = 2.2|C_{\ell\ell\bar\ell,ee\mu}^{\tV\tR\tV\tL\tS}|^2
+1.1\big(|C_{\ell\ell\bar\ell,ee\mu}^{\tS\tL\tS\tL\tS}|^2
+|C_{\ell\ell\bar\ell,ee\mu}^{\tS\tR\tS\tL\tS}|^2\big) 
+0.06|C_{\ell\ell\bar\ell,ee\mu}^{\tV\tL\tS\tL\tV\tR}|^2
\nn\\
&-0.8\Re\big(C_{\ell\ell\bar\ell, ee\mu}^{\tS\tL\tS\tL\tS}C_{\ell\ell\bar\ell, ee\mu}^{\tS\tL\tS\tR\tS*}\big)
-0.4\Re\big(C_{\ell\ell\bar\ell, ee\mu}^{\tV\tR\tV\tL\tS}C_{\ell\ell\bar\ell, ee\mu}^{\tV\tL\tV\tR\tS*}\big)
-0.3\Re\big(C_{\ell\ell\bar\ell,ee\mu}^{\tV\tR\tV\tL\tS}C_{\ell\ell\bar\ell, ee\mu}^{\tV\tL\tS\tL\tV\tR*}\big)
\nn\\
&+0.005\Re\big[(C_{\ell\ell\bar\ell, ee  \mu}^{\tS\tL\tS\tR\tS}-C_{\ell\ell\bar\ell, ee\mu}^{\tS\tR\tS\tR\tS})C_{\ell\ell\bar\ell,ee\mu}^{\tV\tR\tV\tL\tS*}\big]
+0.003\Re\big(C_{\ell\ell\bar\ell, ee\mu}^{\tV\tL\tS\tL\tV\tR}C_{\ell\ell\bar\ell, ee \mu}^{\tV\tR\tS\tR\tV\tL*}\big)
\nn\\
&+0.001\Re\big[(C_{\ell\ell\bar\ell, ee\mu}^{\tS\tL\tS\tL\tS}-C_{\ell\ell\bar\ell, ee \mu}^{\tS\tR\tS\tL\tS})C_{\ell\ell\bar\ell, ee\mu}^{\tV\tL\tS\tL\tV\tR*}\big]
+0.001\Re\big[(C_{\ell\ell\bar\ell, ee  \mu}^{\tS\tR\tS\tL\tS}-C_{\ell\ell\bar\ell, ee\mu}^{\tS\tL\tS\tL\tS})C_{\ell\ell\bar\ell, ee\mu}^{\tV\tR\tV\tL\tS*}\big]
\nn\\
&+6\cdot10^{-4}\Re\big[(C_{\ell\ell\bar\ell, ee\mu}^{\tS\tL\tS\tL\tS}-C_{\ell\ell\bar\ell, ee \mu}^{\tS\tR\tS\tL\tS})C_{\ell\ell\bar\ell, ee\mu}^{\tV\tR\tS\tR\tV\tL*}\big]
-6\cdot10^{-6}\Re\big(C_{\ell\ell\bar\ell, ee \mu}^{\tS\tL\tS\tL\tS}C_{\ell\ell\bar\ell, ee \mu}^{\tS\tR\tS\tL\tS*}\big)
\nn\\
&-3\cdot10^{-6}\Re\big(C_{\ell\ell\bar\ell, ee \mu}^{\tV\tR\tV\tL\tS}C_{\ell\ell\bar\ell, e e\mu}^{\tV\tR\tS\tR\tV\tL*}\big)
+8\cdot10^{-7}\Re\big(C_{\ell\ell\bar\ell, ee \mu}^{\tS\tR\tS\tL\tS}C_{\ell\ell\bar\ell, ee \mu}^{\tS\tL\tS\tR\tS*}+C_{\ell\ell\bar\ell, ee \mu}^{\tS\tL\tS\tL\tS}C_{\ell\ell\bar\ell, ee \mu}^{\tS\tR\tS\tR\tS*}\big)
\nn\\
&+\tL\leftrightarrow\tR,
\label{eq:Gammap2eemu}
\\
{\Gamma_{p\to \mu^+\mu^+e^-} \over 10^{-9}{\rm GeV}^{11}}
& = 1.9|C_{\ell\ell\bar\ell,\mu\mu e}^{\tV\tR\tV\tL\tS}|^2
+1.0\big(|C_{\ell\ell\bar\ell,\mu\mu e}^{\tS\tL\tS\tL\tS}|^2
+|C_{\ell\ell\bar\ell,\mu\mu e}^{\tS\tR\tS\tL\tS}|^2\big) 
+0.06|C_{\ell\ell\bar\ell,\mu\mu e}^{\tV\tL\tS\tL\tV\tR}|^2
\nn\\
&+0.8\Re\big[(C_{\ell\ell\bar\ell, \mu\mu  e}^{\tS\tL\tS\tR\tS}-C_{\ell\ell\bar\ell, \mu\mu  e}^{\tS\tR\tS\tR\tS})C_{\ell\ell\bar\ell, \mu\mu e}^{\tV\tR\tV\tL\tS*}\big]
+0.2\Re\big[(C_{\ell\ell\bar\ell, \mu\mu e}^{\tS\tL\tS\tL\tS}-C_{\ell\ell\bar\ell, \mu\mu e}^{\tS\tR\tS\tL\tS})C_{\ell\ell\bar\ell, \mu\mu e}^{\tV\tL\tS\tL\tV\tR*}\big]
\nn\\
&-0.2\Re\big(C_{\ell\ell\bar\ell, \mu\mu e}^{\tS\tL\tS\tL\tS}C_{\ell\ell\bar\ell, \mu\mu e}^{\tS\tR\tS\tL\tS*}\big)
-0.1\Re\big(C_{\ell\ell\bar\ell, \mu\mu e}^{\tV\tR\tV\tL\tS}C_{\ell\ell\bar\ell, \mu\mu e}^{\tV\tR\tS\tR\tV\tL*}\big)
-0.005\Re\big(C_{\ell\ell\bar\ell, \mu\mu e}^{\tS\tL\tS\tL\tS}C_{\ell\ell\bar\ell, \mu\mu e}^{\tS\tL\tS\tR\tS*}\big)
\nn\\
&-0.002\Re\big(C_{\ell\ell\bar\ell, \mu\mu e}^{\tV\tR\tV\tL\tS}C_{\ell\ell\bar\ell,  \mu\mu e}^{\tV\tL\tV\tR\tS*}\big) 
+0.001\Re\big[(C_{\ell\ell\bar\ell, \mu\mu  e}^{\tS\tR\tS\tL\tS}-C_{\ell\ell\bar\ell, \mu\mu  e}^{\tS\tL\tS\tL\tS})C_{\ell\ell\bar\ell, \mu\mu e}^{\tV\tR\tV\tL\tS*}\big]
\nn\\
&-0.001\Re\big(C_{\ell\ell\bar\ell, \mu\mu e}^{\tV\tR\tV\tL\tS}C_{\ell\ell\bar\ell, \mu\mu e}^{\tV\tL\tS\tL\tV\tR*}\big)
+6\cdot10^{-4}\Re\big[(C_{\ell\ell\bar\ell, \mu\mu e}^{\tS\tL\tS\tL\tS}-C_{\ell\ell\bar\ell, \mu\mu e}^{\tS\tR\tS\tL\tS})C_{\ell\ell\bar\ell, \mu\mu e}^{\tV\tR\tS\tR\tV\tL*}\big]
\nn\\
&+2\cdot10^{-4}\Re\big(C_{\ell\ell\bar\ell, \mu\mu e}^{\tS\tR\tS\tL\tS}C_{\ell\ell\bar\ell, \mu\mu e}^{\tS\tL\tS\tR\tS*}
+C_{\ell\ell\bar\ell, \mu\mu e}^{\tS\tL\tS\tL\tS}C_{\ell\ell\bar\ell, \mu\mu e}^{\tS\tR\tS\tR\tS*}\big)
+5\cdot10^{-5}\Re\big(C_{\ell\ell\bar\ell, \mu\mu e}^{\tV\tL\tS\tL\tV\tR}C_{\ell\ell\bar\ell, \mu\mu e}^{\tV\tR\tS\tR\tV\tL*}\big)
\nn\\
&+\tL\leftrightarrow\tR,
\\
{\Gamma_{p\to e^+\mu^+e^-} \over 10^{-9}{\rm GeV}^{11}}  
& = 26|C_{\ell\ell\bar\ell,e \mu e}^{\tT\tL\tT\tL\tS}|^2
+2.2\big(|C_{\ell\ell\bar\ell,e \mu e}^{\tV\tR\tV\tL\tS}|^2+|C_{\ell\ell\bar\ell,\mu ee}^{\tV\tR\tV\tL\tS}|^2\big)
+0.5\big(|C_{\ell\ell\bar\ell,e \mu e}^{\tS\tL\tS\tL\tS}|^2
+|C_{\ell\ell\bar\ell,e \mu e}^{\tS\tR\tS\tL\tS}|^2\big)
\nn\\
&+0.2|C_{\ell\ell\bar\ell,e \mu e}^{\tT\tL\tV\tL\tV\tR}|^2
+0.05\big(|C_{\ell\ell\bar\ell,e \mu e}^{\tV\tL\tS\tL\tV\tR}|^2+|C_{\ell\ell\bar\ell, \mu ee}^{\tV\tL\tS\tL\tV\tR}|^2\big)
\nn\\
&+5.1\Re\big(C_{\ell\ell\bar\ell, \mu ee}^{\tV\tR\tV\tL\tS}C_{\ell\ell\bar\ell, e\mu e}^{\tT\tR\tT\tR\tS*}\big)
-0.5\Im\big(C_{\ell\ell\bar\ell, e\mu e}^{\tV\tR\tV\tL\tS}C_{\ell\ell\bar\ell, e\mu e}^{\tT\tL\tV\tL\tV\tR*}\big)
+0.5\Re\big(C_{\ell\ell\bar\ell, e\mu e}^{\tT\tL\tT\tL\tS}C_{\ell\ell\bar\ell, \mu e e}^{\tV\tL\tS\tL\tV\tR*}\big) 
\nn\\
&+0.4\Re\big(C_{\ell\ell\bar\ell, e\mu e}^{\tV\tR\tV\tL\tS}C_{\ell\ell\bar\ell, e\mu  e}^{\tS\tL\tS\tR\tS*}-C_{\ell\ell\bar\ell, \mu ee}^{\tV\tL\tV\tR\tS}C_{\ell\ell\bar\ell, e\mu e}^{\tS\tL\tS\tL\tS*}\big)
+0.1\Re\big(C_{\ell\ell\bar\ell, e\mu e}^{\tS\tL\tS\tL\tS}C_{\ell\ell\bar\ell, \mu ee}^{\tV\tL\tS\tL\tV\tR*}-C_{\ell\ell\bar\ell, e\mu e}^{\tS\tR\tS\tL\tS}C_{\ell\ell\bar\ell,e \mu e}^{\tV\tL\tS\tL\tV\tR*}\big)
\nn\\
&-0.03\Re\big(C_{\ell\ell\bar\ell, e\mu e}^{\tV\tR\tV\tL\tS}C_{\ell\ell\bar\ell, e\mu  e}^{\tT\tR\tT\tR\tS*}\big)
+0.01\Im\big(C_{\ell\ell\bar\ell, e\mu e}^{\tT\tL\tT\tL\tS}C_{\ell\ell\bar\ell, e\mu  e}^{\tT\tL\tV\tL\tV\tR*}\big)
-0.01\Im\big(C_{\ell\ell\bar\ell, e\mu e}^{\tV\tL\tS\tL\tV\tR}C_{\ell\ell\bar\ell, e\mu e}^{\tT\tR\tV\tR\tV\tL*}\big)
\nn\\
&-0.006\Re\big(C_{\ell\ell\bar\ell, e\mu e}^{\tV\tR\tV\tL\tS}C_{\ell\ell\bar\ell, e\mu  e}^{\tT\tL\tT\tL\tS*}\big)
-0.005\Re\big(C_{\ell\ell\bar\ell, e\mu e}^{\tV\tR\tV\tL\tS}C_{\ell\ell\bar\ell, \mu e\mu  }^{\tV\tL\tV\tR\tS*}\big)
+0.005\Im\big(C_{\ell\ell\bar\ell, e\mu e}^{\tT\tL\tT\tL\tS}C_{\ell\ell\bar\ell,  e\mu e}^{\tT\tR\tV\tR\tV\tL*}\big)
\nn\\
&
+0.003\Im\big(C_{\ell\ell\bar\ell, \mu ee}^{\tV\tR\tV\tL\tS}C_{\ell\ell\bar\ell, e\mu e}^{\tT\tL\tV\tL\tV\tR*}\big)
-0.003\Re\big(C_{\ell\ell\bar\ell, e\mu e}^{\tT\tL\tT\tL\tS}C_{\ell\ell\bar\ell, e\mu e}^{\tV\tL\tS\tL\tV\tR*}\big)
\nn\\
&
+0.002\Re\big[(C_{\ell\ell\bar\ell, \mu e e}^{\tV\tR\tV\tL\tS}
-C_{\ell\ell\bar\ell, e\mu e}^{\tS\tL\tS\tL\tS})C_{\ell\ell\bar\ell, e\mu e}^{\tS\tL\tS\tR\tS*}-C_{\ell\ell\bar\ell, e\mu e}^{\tS\tL\tS\tL\tS}C_{\ell\ell\bar\ell, e\mu e}^{\tV\tL\tV\tR\tS*} \big]
\nn\\
&
-0.001\Re\big(C_{\ell\ell\bar\ell, e\mu e}^{\tV\tR\tV\tL\tS}C_{\ell\ell\bar\ell, \mu ee}^{\tV\tL\tS\tL\tV\tR*}
+C_{\ell\ell\bar\ell, \mu ee}^{\tV\tR\tV\tL\tS}C_{\ell\ell\bar\ell, e\mu e}^{\tV\tL\tS\tL\tV\tR*}\big)
+0.001\Im\big(C_{\ell\ell\bar\ell, \mu ee}^{\tV\tR\tV\tL\tS}C_{\ell\ell\bar\ell, e\mu e}^{\tT\tR\tV\tR\tV\tL*}\big)
\nn\\
&-0.001\Re\big(C_{\ell\ell\bar\ell, e\mu e}^{\tT\tL\tT\tL\tS}C_{\ell\ell\bar\ell, e\mu e}^{\tV\tR\tS\tR\tV\tL*}\big)
-0.001\Re\big(C_{\ell\ell\bar\ell, e\mu e}^{\tV\tR\tV\tL\tS}C_{\ell\ell\bar\ell, \mu ee  }^{\tV\tR\tV\tL\tS*}\big)
\nn\\
&+7\cdot10^{-4}\Re\big(C_{\ell\ell\bar\ell, e\mu e}^{\tS\tL\tS\tL\tS}C_{\ell\ell\bar\ell, e\mu e}^{\tV\tL\tS\tL\tV\tR*}
-C_{\ell\ell\bar\ell, e\mu e}^{\tS\tR\tS\tL\tS}C_{\ell\ell\bar\ell, \mu e e}^{\tV\tL\tS\tL\tV\tR*}\big)
\nn\\
&-6\cdot10^{-4}\Re\big(C_{\ell\ell\bar\ell, e\mu e}^{\tV\tR\tV\tL\tS}C_{\ell\ell\bar\ell, \mu ee}^{\tV\tR\tS\tR\tV\tL*}
+C_{\ell\ell\bar\ell, \mu ee}^{\tV\tR\tV\tL\tS}C_{\ell\ell\bar\ell, e\mu e}^{\tV\tR\tS\tR\tV\tL*}\big)
\nn\\
&+5\cdot10^{-4}\Re\big[( C_{\ell\ell\bar\ell, \mu ee}^{\tV\tR\tV\tL\tS}
-C_{\ell\ell\bar\ell, e\mu e}^{\tS\tL\tS\tL\tS})
C_{\ell\ell\bar\ell, e\mu e}^{\tS\tR\tS\tL\tS*}
-C_{\ell\ell\bar\ell, e\mu  e}^{\tS\tL\tS\tL\tS} 
C_{\ell\ell\bar\ell, e\mu e}^{\tV\tR\tV\tL\tS*}\big]
\nn\\
&+3\cdot10^{-4}\Re\big(C_{\ell\ell\bar\ell, e\mu e}^{\tS\tL\tS\tL\tS}C_{\ell\ell\bar\ell, e\mu e}^{\tV\tR\tS\tR\tV\tL*}
-C_{\ell\ell\bar\ell, e\mu e}^{\tS\tR\tS\tL\tS}C_{\ell\ell\bar\ell, \mu e e}^{\tV\tR\tS\tR\tV\tL*}\big)
+2\cdot10^{-4}\Im\big(C_{\ell\ell\bar\ell,\mu e e}^{\tV\tL\tS\tL\tV\tR}C_{\ell\ell\bar\ell, e \mu  e}^{\tT\tL\tV\tL\tV\tR*}\big)
\nn\\
&+10^{-4}\Re\big(C_{\ell\ell\bar\ell, e\mu e}^{\tV\tL\tS\tL\tV\tR}C_{\ell\ell\bar\ell, \mu e e}^{\tV\tL\tS\tL\tV\tR*}\big)
+3\cdot10^{-5}\Re\big(C_{\ell\ell\bar\ell, \mu ee}^{\tV\tR\tV\tL\tS}C_{\ell\ell\bar\ell, e\mu e}^{\tT\tL\tT\tL\tS*}\big)
+3\cdot10^{-5}\Im\big(C_{\ell\ell\bar\ell,\mu e e}^{\tV\tL\tS\tL\tV\tR}C_{\ell\ell\bar\ell, e \mu  e}^{\tT\tR\tV\tR\tV\tL*}\big)
\nn\\
&+2\cdot10^{-5}\Re\big(C_{\ell\ell\bar\ell, e\mu e}^{\tT\tL\tT\tL\tS}C_{\ell\ell\bar\ell, e\mu e}^{\tT\tR\tT\tR\tS*}\big)
+10^{-5}\Re\big(C_{\ell\ell\bar\ell, e\mu e}^{\tV\tL\tS\tL\tV\tR}C_{\ell\ell\bar\ell, \mu e e}^{\tV\tR\tS\tR\tV\tL*}\big)
-6\cdot10^{-6}\Im\big(C_{\ell\ell\bar\ell, e\mu e}^{\tV\tR\tV\tL\tS}C_{\ell\ell\bar\ell, e\mu e}^{\tT\tR\tV\tR\tV\tL*}\big)
\nn\\
&+6\cdot10^{-6}\Re\big(C_{\ell\ell\bar\ell, e\mu e}^{\tT\tL\tT\tL\tS}C_{\ell\ell\bar\ell, \mu e e}^{\tV\tR\tS\tR\tV\tL*}\big)
+3\cdot10^{-6}\Re\big(C_{\ell\ell\bar\ell, e\mu e}^{\tV\tR\tV\tL\tS}C_{\ell\ell\bar\ell, e\mu  e}^{\tS\tR\tS\tL\tS*}-C_{\ell\ell\bar\ell, \mu ee}^{\tV\tR\tV\tL\tS}C_{\ell\ell\bar\ell, e\mu e}^{\tS\tL\tS\tL\tS*}\big)
\nn\\
&+2\cdot10^{-6}\Re\big(C_{\ell\ell\bar\ell, e\mu e}^{\tV\tR\tV\tL\tS}C_{\ell\ell\bar\ell,  e \mu e}^{\tV\tL\tV\tR\tS*}
+C_{\ell\ell\bar\ell, \mu ee}^{\tV\tR\tV\tL\tS}C_{\ell\ell\bar\ell, \mu ee}^{\tV\tL\tV\tR\tS*}\big)
\nn\\
&+2\cdot10^{-6}\Re\big(C_{\ell\ell\bar\ell, e\mu e}^{\tS\tL\tS\tL\tS}C_{\ell\ell\bar\ell, \mu ee}^{\tV\tR\tS\tR\tV\tL*}
-C_{\ell\ell\bar\ell, e\mu e}^{\tS\tR\tS\tL\tS}C_{\ell\ell\bar\ell, e\mu e}^{\tV\tR\tS\tR\tV\tL*}\big)
-10^{-6}\Im\big(C_{\ell\ell\bar\ell, e\mu e}^{\tV\tL\tS\tL\tV\tR}C_{\ell\ell\bar\ell, e\mu  e}^{\tT\tL\tV\tL\tV\tR*}\big)
\nn\\
&+10^{-6}\Re\big(C_{\ell\ell\bar\ell,e \mu  e}^{\tT\tL\tV\tL\tV\tR}C_{\ell\ell\bar\ell,e \mu  e}^{\tT\tR\tV\tR\tV\tL*}\big)
+4\cdot10^{-7}\Re\big(C_{\ell\ell\bar\ell, e\mu e}^{\tS\tL\tS\tL\tS}C_{\ell\ell\bar\ell, e\mu e}^{\tS\tR\tS\tR\tS*}
+C_{\ell\ell\bar\ell, e\mu e}^{\tS\tR\tS\tL\tS}C_{\ell\ell\bar\ell, e\mu e}^{\tS\tL\tS\tR\tS*}\big)
\nn\\
&+2\cdot10^{-7}\Re\big(C_{\ell\ell\bar\ell, e\mu e}^{\tV\tL\tS\tL\tV\tR}C_{\ell\ell\bar\ell, e\mu e}^{\tV\tR\tS\tR\tV\tL*}+C_{\ell\ell\bar\ell,\mu e e}^{\tV\tL\tS\tL\tV\tR}C_{\ell\ell\bar\ell, \mu e e}^{\tV\tR\tS\tR\tV\tL*}\big)
+\tL\leftrightarrow\tR,
\\
{\Gamma_{p\to e^+\mu^+\mu^-}\over 10^{-9}{\rm GeV}^{11}}
& = 24|C_{\ell\ell\bar\ell,e \mu \mu}^{\tT\tL\tT\tL\tS}|^2+2.0\big(|C_{\ell\ell\bar\ell,e \mu \mu}^{\tV\tR\tV\tL\tS}|^2+|C_{\ell\ell\bar\ell,\mu e  \mu}^{\tV\tR\tV\tL\tS}|^2\big)
+0.5\big(|C_{\ell\ell\bar\ell,e \mu \mu}^{\tS\tL\tS\tL\tS}|^2
+|C_{\ell\ell\bar\ell,e \mu \mu}^{\tS\tR\tS\tL\tS}|^2\big)
\nn\\
& +0.2|C_{\ell\ell\bar\ell,e \mu \mu}^{\tT\tL\tV\tL\tV\tR}|^2
+0.05\big(|C_{\ell\ell\bar\ell,e \mu \mu}^{\tV\tL\tS\tL\tV\tR}|^2+|C_{\ell\ell\bar\ell, \mu e\mu}^{\tV\tL\tS\tL\tV\tR}|^2\big)    
\nn\\
&+4.6\Re\big(C_{\ell\ell\bar\ell, \mu e\mu}^{\tV\tR\tV\tL\tS}C_{\ell\ell\bar\ell, e\mu \mu}^{\tT\tR\tT\tR\tS*}\big)
+2.0\Im\big(C_{\ell\ell\bar\ell, e\mu \mu}^{\tT\tL\tT\tL\tS}C_{\ell\ell\bar\ell, e\mu  \mu}^{\tT\tL\tV\tL\tV\tR*}\big)
-1.0\Re\big(C_{\ell\ell\bar\ell, e\mu \mu}^{\tV\tR\tV\tL\tS}C_{\ell\ell\bar\ell, e\mu  \mu}^{\tT\tL\tT\tL\tS*}\big)
\nn\\
&-0.8\Re\big(C_{\ell\ell\bar\ell, e\mu \mu}^{\tV\tR\tV\tL\tS}C_{\ell\ell\bar\ell, \mu e\mu  }^{\tV\tL\tV\tR\tS*}\big)
-0.5\Im\big(C_{\ell\ell\bar\ell, e\mu \mu}^{\tV\tR\tV\tL\tS}C_{\ell\ell\bar\ell, e\mu \mu}^{\tT\tL\tV\tL\tV\tR*}\big)
+0.5\Re\big(C_{\ell\ell\bar\ell, e\mu \mu}^{\tT\tL\tT\tL\tS}C_{\ell\ell\bar\ell, \mu e \mu}^{\tV\tL\tS\tL\tV\tR*}\big)   
\nn\\
&
+0.4\Re\big[(C_{\ell\ell\bar\ell, e\mu \mu}^{\tV\tR\tV\tL\tS}
-C_{\ell\ell\bar\ell, e\mu \mu}^{\tS\tL\tS\tL\tS})
C_{\ell\ell\bar\ell, e\mu  \mu}^{\tS\tL\tS\tR\tS*}
-C_{\ell\ell\bar\ell, \mu e\mu}^{\tV\tL\tV\tR\tS}
C_{\ell\ell\bar\ell, e\mu \mu}^{\tS\tL\tS\tL\tS*}\big]
\nn\\
&-0.2\Re\big(C_{\ell\ell\bar\ell, \mu e\mu}^{\tV\tR\tV\tL\tS}C_{\ell\ell\bar\ell, e\mu \mu}^{\tV\tL\tS\tL\tV\tR*}
+C_{\ell\ell\bar\ell, e\mu \mu}^{\tV\tR\tV\tL\tS}C_{\ell\ell\bar\ell, \mu e \mu}^{\tV\tL\tS\tL\tV\tR*}\big)
+0.2\Im\big(C_{\ell\ell\bar\ell, \mu e\mu}^{\tV\tR\tV\tL\tS}C_{\ell\ell\bar\ell, e\mu \mu}^{\tT\tR\tV\tR\tV\tL*}\big)
\nn\\
&-0.2\Re\big(C_{\ell\ell\bar\ell, e\mu \mu}^{\tT\tL\tT\tL\tS}C_{\ell\ell\bar\ell, e\mu \mu}^{\tV\tR\tS\tR\tV\tL*}\big)   
+0.1\Re\big(C_{\ell\ell\bar\ell, e\mu \mu}^{\tS\tL\tS\tL\tS}C_{\ell\ell\bar\ell, \mu e\mu}^{\tV\tL\tS\tL\tV\tR*}-C_{\ell\ell\bar\ell, e\mu \mu}^{\tS\tR\tS\tL\tS}C_{\ell\ell\bar\ell, e\mu \mu}^{\tV\tL\tS\tL\tV\tR*}\big)
\nn\\
&+0.09\Re\big(C_{\ell\ell\bar\ell, \mu e \mu}^{\tV\tR\tV\tL\tS}C_{\ell\ell\bar\ell, e\mu  \mu}^{\tS\tR\tS\tL\tS*}-C_{\ell\ell\bar\ell, e\mu\mu}^{\tV\tR\tV\tL\tS}C_{\ell\ell\bar\ell, e\mu  \mu}^{\tS\tL\tS\tL\tS*}\big)
\nn\\
&+0.05\Re\big(C_{\ell\ell\bar\ell, e\mu \mu}^{\tS\tL\tS\tL\tS}C_{\ell\ell\bar\ell, e\mu \mu}^{\tV\tR\tS\tR\tV\tL*}-C_{\ell\ell\bar\ell, e\mu \mu}^{\tS\tR\tS\tL\tS}C_{\ell\ell\bar\ell, \mu e \mu}^{\tV\tR\tS\tR\tV\tL*}\big)
+0.04\Im\big(C_{\ell\ell\bar\ell,\mu e \mu}^{\tV\tL\tS\tL\tV\tR}C_{\ell\ell\bar\ell, e \mu\mu}^{\tT\tL\tV\tL\tV\tR*}\big)
\nn\\
& -0.03\Re\big(C_{\ell\ell\bar\ell, e\mu\mu}^{\tV\tR\tV\tL\tS}C_{\ell\ell\bar\ell, e\mu\mu}^{\tT\tR\tT\tR\tS*}\big)
-0.01\Im\big(C_{\ell\ell\bar\ell, e\mu\mu}^{\tV\tL\tS\tL\tV\tR}C_{\ell\ell\bar\ell, e\mu \mu}^{\tT\tR\tV\tR\tV\tL*}\big)
+0.007\Re\big(C_{\ell\ell\bar\ell, e\mu \mu}^{\tV\tL\tS\tL\tV\tR}C_{\ell\ell\bar\ell, \mu e \mu}^{\tV\tR\tS\tR\tV\tL*}\big)
\nn\\
&+0.006\Re\big(C_{\ell\ell\bar\ell, \mu e\mu}^{\tV\tR\tV\tL\tS}C_{\ell\ell\bar\ell, e\mu \mu}^{\tT\tL\tT\tL\tS*}\big)
+0.005\Im\big(C_{\ell\ell\bar\ell, e\mu \mu}^{\tT\tL\tT\tL\tS}C_{\ell\ell\bar\ell,  e\mu \mu}^{\tT\tR\tV\tR\tV\tL*}\big)
+0.004\Re\big(C_{\ell\ell\bar\ell, e\mu \mu}^{\tT\tL\tT\tL\tS}C_{\ell\ell\bar\ell, e\mu \mu}^{\tT\tR\tT\tR\tS*}\big)
\nn\\
&+0.003\Im\big(C_{\ell\ell\bar\ell, \mu e\mu}^{\tV\tR\tV\tL\tS}C_{\ell\ell\bar\ell, e\mu \mu}^{\tT\tL\tV\tL\tV\tR*}\big)
-0.003\Re\big(C_{\ell\ell\bar\ell, e\mu \mu}^{\tT\tL\tT\tL\tS}C_{\ell\ell\bar\ell, e\mu \mu}^{\tV\tL\tS\tL\tV\tR*}\big)
\nn\\
&
+0.002\Re\big(C_{\ell\ell\bar\ell, \mu e\mu}^{\tV\tR\tV\tL\tS}C_{\ell\ell\bar\ell, e\mu\mu}^{\tS\tL\tS\tR\tS*}-C_{\ell\ell\bar\ell, e\mu \mu}^{\tV\tR\tV\tL\tS}C_{\ell\ell\bar\ell, e\mu\mu}^{\tS\tR\tS\tR\tS*}\big)
-0.001\Re\big(C_{\ell\ell\bar\ell, e\mu \mu}^{\tV\tR\tV\tL\tS}C_{\ell\ell\bar\ell, \mu e\mu  }^{\tV\tR\tV\tL\tS*}\big)
\nn\\
&+0.001\Re\big(C_{\ell\ell\bar\ell, e\mu \mu}^{\tT\tL\tT\tL\tS}C_{\ell\ell\bar\ell, \mu e \mu}^{\tV\tR\tS\tR\tV\tL*}\big)
-0.001\Im\big(C_{\ell\ell\bar\ell, e\mu \mu}^{\tV\tR\tV\tL\tS}C_{\ell\ell\bar\ell, e\mu \mu}^{\tT\tR\tV\tR\tV\tL*}\big)
\nn\\
&+7\cdot10^{-4}\Re\big(C_{\ell\ell\bar\ell, e\mu \mu}^{\tS\tL\tS\tL\tS}C_{\ell\ell\bar\ell, e\mu \mu}^{\tV\tL\tS\tL\tV\tR*}-C_{\ell\ell\bar\ell, e\mu \mu}^{\tS\tR\tS\tL\tS}C_{\ell\ell\bar\ell, \mu e \mu}^{\tV\tL\tS\tL\tV\tR*}\big)
\nn\\
&-6\cdot10^{-4}\Re\big(C_{\ell\ell\bar\ell, e\mu \mu}^{\tV\tR\tV\tL\tS}C_{\ell\ell\bar\ell, \mu e\mu}^{\tV\tR\tS\tR\tV\tL*}+C_{\ell\ell\bar\ell, \mu e\mu}^{\tV\tR\tV\tL\tS}C_{\ell\ell\bar\ell, e\mu \mu}^{\tV\tR\tS\tR\tV\tL*}\big)
\nn\\
&+5\cdot10^{-4}\Re\big[(C_{\ell\ell\bar\ell, e\mu \mu}^{\tV\tR\tV\tL\tS}
-C_{\ell\ell\bar\ell, e\mu \mu}^{\tS\tL\tS\tL\tS}) 
C_{\ell\ell\bar\ell, e\mu  \mu}^{\tS\tR\tS\tL\tS*}
-C_{\ell\ell\bar\ell, \mu e\mu}^{\tV\tR\tV\tL\tS} 
C_{\ell\ell\bar\ell, e\mu\mu}^{\tS\tL\tS\tL\tS*}\big]
\nn\\
&+3\cdot10^{-4}\Re\big(C_{\ell\ell\bar\ell, e\mu \mu}^{\tV\tR\tV\tL\tS}C_{\ell\ell\bar\ell,  e \mu \mu}^{\tV\tL\tV\tR\tS*}+C_{\ell\ell\bar\ell, \mu e\mu}^{\tV\tR\tV\tL\tS}C_{\ell\ell\bar\ell, \mu e\mu}^{\tV\tL\tV\tR\tS*}\big)  
\nn\\
&
+3\cdot10^{-4}\Re\big(C_{\ell\ell\bar\ell, e\mu \mu}^{\tS\tL\tS\tL\tS}C_{\ell\ell\bar\ell, \mu e\mu}^{\tV\tR\tS\tR\tV\tL*}-C_{\ell\ell\bar\ell, e\mu \mu}^{\tS\tR\tS\tL\tS}C_{\ell\ell\bar\ell, e\mu \mu}^{\tV\tR\tS\tR\tV\tL*}\big)
-2\cdot10^{-4}\Im\big(C_{\ell\ell\bar\ell, e\mu \mu}^{\tV\tL\tS\tL\tV\tR}C_{\ell\ell\bar\ell, e\mu  \mu}^{\tT\tL\tV\tL\tV\tR*}\big)
\nn\\
&
+10^{-4}\Re\big(C_{\ell\ell\bar\ell, e\mu \mu}^{\tV\tL\tS\tL\tV\tR}C_{\ell\ell\bar\ell, \mu e \mu}^{\tV\tL\tS\tL\tV\tR*}\big)
+10^{-4}\Re\big(C_{\ell\ell\bar\ell,e \mu  \mu}^{\tT\tL\tV\tL\tV\tR}C_{\ell\ell\bar\ell,e \mu  \mu}^{\tT\tR\tV\tR\tV\tL*}\big)
\nn\\
&+8\cdot10^{-5}\Re\big(C_{\ell\ell\bar\ell, e\mu \mu}^{\tS\tL\tS\tL\tS}C_{\ell\ell\bar\ell, e\mu \mu}^{\tS\tR\tS\tR\tS*}
+C_{\ell\ell\bar\ell, e\mu \mu}^{\tS\tR\tS\tL\tS}C_{\ell\ell\bar\ell, e\mu \mu}^{\tS\tL\tS\tR\tS*}\big)
+5\cdot10^{-5}\Im\big(C_{\ell\ell\bar\ell,\mu e \mu}^{\tV\tL\tS\tL\tV\tR}C_{\ell\ell\bar\ell, e \mu  \mu}^{\tT\tR\tV\tR\tV\tL*}\big)
\nn\\
&+4\cdot 10^{-5}\Re\big(C_{\ell\ell\bar\ell, e\mu \mu}^{\tV\tL\tS\tL\tV\tR}C_{\ell\ell\bar\ell, e\mu \mu}^{\tV\tR\tS\tR\tV\tL*}+C_{\ell\ell\bar\ell,\mu e \mu}^{\tV\tL\tS\tL\tV\tR}C_{\ell\ell\bar\ell, \mu e \mu}^{\tV\tR\tS\tR\tV\tL*}\big)
+\tL\leftrightarrow\tR.
\end{align}
\end{subequations}
%%%%%
%%%%%n2llvb
$\bullet\,\,n\to\ell_x^-\ell_y^+\bar{\nu}_z$:
\begin{subequations}	
\label{eq:express_n2llvb}
\begin{align}
{\Gamma_{n\to e^-e^+ \bar{\nu}_z} \over 10^{-9}{\rm GeV}^{11}} & = 
29|C_{\bar\ell \ell \nu,ee z}^{\tT\tL\tT\tL\tS}|^2
+2.4\big(|C_{\bar\ell \ell \nu,ee z}^{\tV\tR\tV\tR\tS}|^2+|C_{\bar\ell \ell \nu,ee z}^{\tV\tL\tV\tR\tS}|^2\big)
+ 0.6\big(|C_{\bar\ell \ell \nu,ee z}^{\tS\tL\tS\tL\tS}|^2+|C_{\bar\ell \ell \nu,ee z}^{\tS\tR\tS\tL\tS}|^2\big)
\nn\\
&
+0.2|C_{\bar\ell \ell \nu,ee z}^{\tT\tR\tV\tR\tV\tL}|^2
+0.06\big(|C_{\bar\ell \ell \nu,ee z}^{\tV\tR\tS\tL\tV\tR}|^2
+|C_{\bar\ell \ell \nu,ee z}^{\tV\tL\tS\tL\tV\tR}|^2\big)
-0.03\Re\big[(C_{\bar\ell \ell \nu,ee z}^{\tV\tR\tV\tR\tS}+C_{\bar\ell \ell \nu,ee z}^{\tV\tL\tV\tR\tS})C_{\bar\ell \ell \nu,ee z}^{\tT\tL\tT\tL\tS*}\big]
\nn\\
&+0.003\Re\big[(C_{\bar\ell \ell \nu,ee z}^{\tV\tL\tS\tL\tV\tR}+C_{\bar\ell \ell \nu,ee z}^{\tV\tR\tS\tL\tV\tR})C_{\bar\ell \ell \nu,ee z}^{\tT\tL\tT\tL\tS*}\big]
+0.003\Im\big[(C_{\bar\ell \ell \nu,ee z}^{\tV\tL\tV\tR\tS}+C_{\bar\ell \ell \nu,ee z}^{\tV\tR\tV\tR\tS})C_{\bar\ell \ell \nu,ee z}^{\tT\tR\tV\tR\tV\tL*}\big]
\nn\\
&
+0.003\Re\big[(C_{\bar\ell \ell \nu,ee z}^{\tS\tL\tS\tL\tS}
-C_{\bar\ell \ell \nu,ee z}^{\tS\tR\tS\tL\tS})
(C_{\bar\ell \ell \nu,ee z}^{\tV\tR\tV\tR\tS*}
-C_{\bar\ell \ell \nu,ee z}^{\tV\tL\tV\tR\tS*})\big]
\nn\\
&
+8\cdot10^{-4}\Re\big[(C_{\bar\ell \ell \nu,ee z}^{\tS\tL\tS\tL\tS}
-C_{\bar\ell \ell \nu,ee z}^{\tS\tR\tS\tL\tS})
(C_{\bar\ell \ell \nu,ee z}^{\tV\tR\tS\tL\tV\tR*}-C_{\bar\ell \ell \nu,ee z}^{\tV\tL\tS\tL\tV\tR*})\big]
-3\cdot10^{-5}\Im\big(C_{\bar\ell \ell \nu,ee z}^{\tT\tL\tT\tL\tS}C_{\bar\ell \ell \nu,ee z}^{\tT\tR\tV\tR\tV\tL*}\big)
\nn\\
&+6\cdot10^{-6}\Re\big(C_{\bar\ell \ell \nu,ee z}^{\tV\tR\tV\tR\tS}C_{\bar\ell \ell \nu,ee z}^{\tV\tL\tV\tR\tS*}\big)
-3\cdot10^{-6}\Re\big(C_{\bar\ell \ell \nu,ee z}^{\tV\tR\tV\tR\tS}C_{\bar\ell \ell \nu,ee z}^{\tV\tL\tS\tL\tV\tR*}
+C_{\bar\ell \ell \nu,ee z}^{\tV\tL\tV\tR\tS}C_{\bar\ell \ell \nu,ee z}^{\tV\tR\tS\tL\tV\tR*}\big)
\nn\\
&-3\cdot10^{-6}\Re\big(C_{\bar\ell \ell \nu,ee z}^{\tS\tL\tS\tL\tS}C_{\bar\ell \ell \nu,ee z}^{\tS\tR\tS\tL\tS*}\big)
-7\cdot10^{-7}\Re\big(C_{\bar\ell \ell \nu,ee z}^{\tV\tR\tS\tL\tV\tR}
C_{\bar\ell \ell \nu,ee z}^{\tV\tL\tS\tL\tV\tR*}\big)
\nn\\
&-9\cdot10^{-9}\Im\big[(C_{\bar\ell \ell \nu,ee z}^{\tV\tL\tS\tL\tV\tR}+C_{\bar\ell \ell \nu,ee z}^{\tV\tR\tS\tL\tV\tR})C_{\bar\ell \ell \nu,ee z}^{\tT\tR\tV\tR\tV\tL*}\big],
\\
{\Gamma_{n\to \mu^-\mu^+ \bar{\nu}_z} \over 10^{-9}{\rm GeV}^{11}} &=
24|C_{\bar\ell \ell \nu,\mu\mu z}^{\tT\tL\tT\tL\tS}|^2
+2.0\big(|C_{\bar\ell \ell \nu,\mu\mu z}^{\tV\tR\tV\tR\tS}|^2+|C_{\bar\ell \ell \nu,\mu\mu z}^{\tV\tL\tV\tR\tS}|^2\big)
+ 0.5\big(|C_{\bar\ell \ell \nu,\mu\mu z}^{\tS\tL\tS\tL\tS}|^2+|C_{\bar\ell \ell \nu,\mu\mu z}^{\tS\tR\tS\tL\tS}|^2\big)
\nn\\
&
+0.2|C_{\bar\ell \ell \nu,\mu\mu z}^{\tT\tR\tV\tR\tV\tL}|^2
+0.05\big(|C_{\bar\ell \ell \nu,\mu\mu z}^{\tV\tR\tS\tL\tV\tR}|^2
+|C_{\bar\ell \ell \nu,\mu\mu z}^{\tV\tL\tS\tL\tV\tR}|^2\big)
\nn\\
&-4.7\Re\big[C_{\bar\ell \ell \nu,\mu\mu z}^{\tT\tL\tT\tL\tS}(C_{\bar\ell \ell \nu,\mu\mu z}^{\tV\tR\tV\tR\tS*}+C_{\bar\ell \ell \nu,\mu\mu z}^{\tV\tL\tV\tR\tS*})\big]
-0.8\Im\big(C_{\bar\ell \ell \nu,\mu\mu z}^{\tT\tL\tT\tL\tS}C_{\bar\ell \ell \nu,\mu\mu z}^{\tT\tR\tV\tR\tV\tL*}\big)
\nn\\ 
&+0.5\Re\big[C_{\bar\ell \ell \nu,\mu\mu z}^{\tT\tL\tT\tL\tS}(C_{\bar\ell \ell \nu,\mu\mu z}^{\tV\tR\tS\tL\tV\tR*}+C_{\bar\ell \ell \nu,\mu\mu z}^{\tV\tL\tS\tL\tV\tR*})\big]
+0.5\Im\big[(C_{\bar\ell \ell \nu,\mu\mu z}^{\tV\tR\tV\tR\tS}+C_{\bar\ell \ell \nu,\mu\mu z}^{\tV\tL\tV\tR\tS})C_{\bar\ell \ell \nu,\mu\mu z}^{\tT\tR\tV\tR\tV\tL*}\big]
\nn\\
&+0.4\Re\big[(C_{\bar\ell \ell \nu,\mu\mu z}^{\tS\tL\tS\tL\tS}
-C_{\bar\ell \ell \nu,\mu\mu z}^{\tS\tR\tS\tL\tS})
(C_{\bar\ell \ell \nu,\mu\mu z}^{\tV\tR\tV\tR\tS*}-C_{\bar\ell \ell \nu,\mu\mu z}^{\tV\tL\tV\tR\tS*})\big]
+0.2\Re\big(C_{\bar\ell \ell \nu,\mu\mu z}^{\tV\tR\tV\tR\tS}C_{\bar\ell \ell \nu,\mu\mu z}^{\tV\tL\tV\tR\tS*}\big)
\nn\\
&
+0.1\Re\big[(C_{\bar\ell \ell \nu,\mu\mu z}^{\tS\tL\tS\tL\tS}
-C_{\bar\ell \ell \nu,\mu\mu z}^{\tS\tR\tS\tL\tS})
(C_{\bar\ell \ell \nu,\mu\mu z}^{\tV\tR\tS\tL\tV\tR*}
-C_{\bar\ell \ell \nu,\mu\mu z}^{\tV\tL\tS\tL\tV\tR*})\big]
\nn\\
&-0.1\Re\big(C_{\bar\ell \ell \nu,\mu\mu z}^{\tV\tR\tV\tR\tS}C_{\bar\ell \ell \nu,\mu\mu z}^{\tV\tL\tS\tL\tV\tR*}+C_{\bar\ell \ell \nu,\mu\mu z}^{\tV\tL\tV\tR\tS}C_{\bar\ell \ell \nu,\mu\mu z}^{\tV\tR\tS\tL\tV\tR*}\big)
-0.09\Re\big(C_{\bar\ell \ell \nu,\mu\mu z}^{\tS\tL\tS\tL\tS}C_{\bar\ell \ell \nu,\mu\mu z}^{\tS\tR\tS\tL\tS*}\big)
\nn\\
&-0.02\Re\big(C_{\bar\ell \ell \nu,\mu\mu z}^{\tV\tR\tS\tL\tV\tR}
C_{\bar\ell \ell \nu,\mu\mu z}^{\tV\tL\tS\tL\tV\tR*}\big)
-0.01\Im\big[(C_{\bar\ell \ell \nu,\mu\mu z}^{\tV\tR\tS\tL\tV\tR}
+C_{\bar\ell \ell \nu,\mu\mu z}^{\tV\tL\tS\tL\tV\tR})C_{\bar\ell \ell \nu,\mu\mu z}^{\tT\tR\tV\tR\tV\tL*}\big],
\\
{\Gamma_{n\to e^-\mu^+ \bar{\nu}_z} \over 10^{-9}{\rm GeV}^{11}} &= 
26|C_{\bar\ell \ell \nu,e\mu z}^{\tT\tL\tT\tL\tS}|^2
+2.2\big(|C_{\bar\ell \ell \nu,e\mu z}^{\tV\tR\tV\tR\tS}|^2+|C_{\bar\ell \ell \nu,e\mu z}^{\tV\tL\tV\tR\tS}|^2\big)
+
0.6\big(|C_{\bar\ell \ell \nu,e\mu z}^{\tS\tL\tS\tL\tS}|^2+|C_{\bar\ell \ell \nu,e\mu z}^{\tS\tR\tS\tL\tS}|^2\big)
\nn\\
&+ 0.2|C_{\bar\ell \ell \nu,e\mu z}^{\tT\tR\tV\tR\tV\tL}|^2
+0.06\big(|C_{\bar\ell \ell \nu,e\mu z}^{\tV\tR\tS\tL\tV\tR}|^2
+|C_{\bar\ell \ell \nu,e\mu z}^{\tV\tL\tS\tL\tV\tR}|^2\big)
\nn\\
&-5.1\Re\big(C_{\bar\ell \ell \nu,e\mu z}^{\tT\tL\tT\tL\tS}C_{\bar\ell \ell \nu,e\mu z}^{\tV\tR\tV\tR\tS*}\big)
+0.5\Im\big(C_{\bar\ell \ell \nu,e\mu z}^{\tV\tL\tV\tR\tS}C_{\bar\ell \ell \nu,e\mu z}^{\tT\tR\tV\tR\tV\tL*}\big)
\nn\\
&+0.5\Re\big(C_{\bar\ell \ell \nu,e\mu z}^{\tT\tL\tT\tL\tS}C_{\bar\ell \ell \nu,e\mu z}^{\tV\tR\tS\tL\tV\tR*}\big)
+0.4\Re\big(C_{\bar\ell \ell \nu,e\mu z}^{\tS\tL\tS\tL\tS}C_{\bar\ell \ell \nu,e\mu z}^{\tV\tR\tV\tR\tS*}
+C_{\bar\ell \ell \nu,e\mu z}^{\tS\tR\tS\tL\tS}C_{\bar\ell \ell \nu,e\mu z}^{\tV\tL\tV\tR\tS*}\big)
\nn\\
&+0.1\Re\big(C_{\bar\ell \ell \nu,e\mu z}^{\tS\tL\tS\tL\tS}C_{\bar\ell \ell \nu,e\mu z}^{\tV\tR\tS\tL\tV\tR*}
+C_{\bar\ell \ell \nu,e\mu z}^{\tS\tR\tS\tL\tS}C_{\bar\ell \ell \nu,e\mu z}^{\tV\tL\tS\tL\tV\tR*}\big)
-0.03\Re\big(C_{\bar\ell \ell \nu,e\mu z}^{\tT\tL\tT\tL\tS}C_{\bar\ell \ell \nu,e\mu z}^{\tV\tL\tV\tR\tS*}\big)
\nn\\
&-0.01\Im\big(C_{\bar\ell \ell \nu,e\mu z}^{\tV\tL\tS\tL\tV\tR}C_{\bar\ell \ell \nu,e\mu z}^{\tT\tR\tV\tR\tV\tL*}\big)
-0.005\Im\big(C_{\bar\ell \ell \nu,e\mu z}^{\tT\tL\tT\tL\tS}C_{\bar\ell \ell \nu,e\mu z}^{\tT\tR\tV\tR\tV\tL*}\big)
+0.003\Re\big(C_{\bar\ell \ell \nu,e\mu z}^{\tT\tL\tT\tL\tS}C_{\bar\ell \ell \nu,e\mu z}^{\tV\tL\tS\tL\tV\tR*}\big)
\nn\\
&
+0.003\Im\big(C_{\bar\ell \ell \nu,e\mu z}^{\tV\tR\tV\tR\tS}C_{\bar\ell \ell \nu,e\mu z}^{\tT\tR\tV\tR\tV\tL*}\big)
-0.002\Re\big(C_{\bar\ell \ell \nu,e\mu z}^{\tS\tL\tS\tL\tS}C_{\bar\ell \ell \nu,e\mu z}^{\tV\tL\tV\tR\tS*}
+C_{\bar\ell \ell \nu,e\mu z}^{\tS\tR\tS\tL\tS}C_{\bar\ell \ell \nu,e\mu z}^{\tV\tR\tV\tR\tS*}\big)
\nn\\
&
+0.001\Re\big(C_{\bar\ell \ell \nu,e\mu z}^{\tV\tR\tV\tR\tS}C_{\bar\ell \ell \nu,e\mu z}^{\tV\tL\tV\tR\tS*}\big)
-7\cdot10^{-4}\Re\big(C_{\bar\ell \ell \nu,e\mu z}^{\tS\tL\tS\tL\tS}C_{\bar\ell \ell \nu,e\mu z}^{\tV\tL\tS\tL\tV\tR*}
+C_{\bar\ell \ell \nu,e\mu z}^{\tS\tR\tS\tL\tS}C_{\bar\ell \ell \nu,e\mu z}^{\tV\tR\tS\tL\tV\tR*}\big)
\nn\\
&-6\cdot10^{-4}\Re\big(C_{\bar\ell \ell \nu,e\mu z}^{\tV\tR\tV\tR\tS}C_{\bar\ell \ell \nu,e\mu z}^{\tV\tL\tS\tL\tV\tR*}
+C_{\bar\ell \ell \nu,e\mu z}^{\tV\tL\tV\tR\tS}C_{\bar\ell \ell \nu,e\mu z}^{\tV\tR\tS\tL\tV\tR*}\big)
-5\cdot10^{-4}\Re\big(C_{\bar\ell \ell \nu,e\mu z}^{\tS\tL\tS\tL\tS}C_{\bar\ell \ell \nu,e\mu z}^{\tS\tR\tS\tL\tS*}\big)
\nn\\
&-10^{-4}\Re\big(C_{\bar\ell \ell \nu,e\mu z}^{\tV\tR\tS\tL\tV\tR}
C_{\bar\ell \ell \nu,e\mu z}^{\tV\tL\tS\tL\tV\tR*}\big)
-3\cdot10^{-5}\Im\big(C_{\bar\ell \ell \nu,e\mu z}^{\tV\tR\tS\tL\tV\tR}C_{\bar\ell \ell \nu,e\mu z}^{\tT\tR\tV\tR\tV\tL*}\big),
\\
{\Gamma_{n\to \mu^-e^+ \bar{\nu}_z} \over 10^{-9}{\rm GeV}^{11}} & = 
26|C_{\bar\ell \ell \nu,\mu e z}^{\tT\tL\tT\tL\tS}|^2
+2.2\big(|C_{\bar\ell \ell \nu,\mu e z}^{\tV\tR\tV\tR\tS}|^2+|C_{\bar\ell \ell \nu,\mu e z}^{\tV\tL\tV\tR\tS}|^2\big)
+
0.6\big(|C_{\bar\ell \ell \nu,\mu e z}^{\tS\tL\tS\tL\tS}|^2+|C_{\bar\ell \ell \nu,\mu e z}^{\tS\tR\tS\tL\tS}|^2\big)
\nn\\
&+ 0.2|C_{\bar\ell \ell \nu,\mu e z}^{\tT\tR\tV\tR\tV\tL}|^2
+0.06\big(|C_{\bar\ell \ell \nu,\mu e z}^{\tV\tR\tS\tL\tV\tR}|^2
+|C_{\bar\ell \ell \nu,\mu e z}^{\tV\tL\tS\tL\tV\tR}|^2\big)
\nn\\
&-5.1\Re\big(C_{\bar\ell \ell \nu,\mu e z}^{\tT\tL\tT\tL\tS}C_{\bar\ell \ell \nu,\mu e z}^{\tV\tL\tV\tR\tS*}\big)
+0.5\Im\big(C_{\bar\ell \ell \nu,\mu e z}^{\tV\tR\tV\tR\tS}C_{\bar\ell \ell \nu,\mu e z}^{\tT\tR\tV\tR\tV\tL*}\big)
\nn\\
&+0.5\Re\big(C_{\bar\ell \ell \nu,\mu e z}^{\tT\tL\tT\tL\tS}C_{\bar\ell \ell \nu,\mu e z}^{\tV\tL\tS\tL\tV\tR*}\big)
-0.4\Re\big(C_{\bar\ell \ell \nu,\mu e z}^{\tS\tL\tS\tL\tS}C_{\bar\ell \ell \nu,\mu e z}^{\tV\tL\tV\tR\tS*}+C_{\bar\ell \ell \nu,\mu e z}^{\tS\tR\tS\tL\tS}C_{\bar\ell \ell \nu,\mu e z}^{\tV\tR\tV\tR\tS*}\big)
\nn\\
&
-0.1\Re\big(C_{\bar\ell \ell \nu,\mu e z}^{\tS\tL\tS\tL\tS}C_{\bar\ell \ell \nu,\mu e z}^{\tV\tL\tS\tL\tV\tR*}
+C_{\bar\ell \ell \nu,\mu e z}^{\tS\tR\tS\tL\tS}C_{\bar\ell \ell \nu,\mu e z}^{\tV\tR\tS\tL\tV\tR*}\big)
-0.03\Re\big(C_{\bar\ell \ell \nu,\mu e z}^{\tT\tL\tT\tL\tS}C_{\bar\ell \ell \nu,\mu e z}^{\tV\tR\tV\tR\tS*}\big)
\nn\\
&-0.01\Im\big(C_{\bar\ell \ell \nu,\mu e z}^{\tV\tR\tS\tL\tV\tR}C_{\bar\ell \ell \nu,\mu e z}^{\tT\tR\tV\tR\tV\tL*}\big)
-0.005\Im\big(C_{\bar\ell \ell \nu,\mu e z}^{\tT\tL\tT\tL\tS}C_{\bar\ell \ell \nu,\mu e z}^{\tT\tR\tV\tR\tV\tL*}\big)
+0.003\Re\big(C_{\bar\ell \ell \nu,\mu e z}^{\tT\tL\tT\tL\tS}C_{\bar\ell \ell \nu,\mu e z}^{\tV\tR\tS\tL\tV\tR*}\big)
\nn\\
&+0.003\Im\big(C_{\bar\ell \ell \nu,\mu e z}^{\tV\tL\tV\tR\tS}C_{\bar\ell \ell \nu,\mu e z}^{\tT\tR\tV\tR\tV\tL*}\big)
+0.002\Re\big(C_{\bar\ell \ell \nu,\mu e z}^{\tS\tR\tS\tL\tS}C_{\bar\ell \ell \nu,\mu e z}^{\tV\tL\tV\tR\tS*}
+C_{\bar\ell \ell \nu,\mu e z}^{\tS\tL\tS\tL\tS}C_{\bar\ell \ell \nu,\mu e z}^{\tV\tR\tV\tR\tS*}\big)
\nn\\
&+0.001\Re\big(C_{\bar\ell \ell \nu,\mu e z}^{\tV\tR\tV\tR\tS}C_{\bar\ell \ell \nu,\mu e z}^{\tV\tL\tV\tR\tS*}\big)
+7\cdot10^{-4}\Re\big(C_{\bar\ell \ell \nu,\mu e z}^{\tS\tL\tS\tL\tS}C_{\bar\ell \ell \nu,\mu e z}^{\tV\tR\tS\tL\tV\tR*}
+C_{\bar\ell \ell \nu,\mu e z}^{\tS\tR\tS\tL\tS}C_{\bar\ell \ell \nu,\mu e z}^{\tV\tL\tS\tL\tV\tR*}\big)
\nn\\
&-6\cdot10^{-4}\Re\big(C_{\bar\ell \ell \nu,\mu e z}^{\tV\tR\tV\tR\tS}C_{\bar\ell \ell \nu,\mu e z}^{\tV\tL\tS\tL\tV\tR*}
+C_{\bar\ell \ell \nu,\mu e z}^{\tV\tL\tV\tR\tS}C_{\bar\ell \ell \nu,\mu e z}^{\tV\tR\tS\tL\tV\tR*}\big)
-5\cdot10^{-4}\Re\big(C_{\bar\ell \ell \nu,\mu e z}^{\tS\tL\tS\tL\tS}C_{\bar\ell \ell \nu,\mu e z}^{\tS\tR\tS\tL\tS*}\big)
\nn\\
&-10^{-4}\Re\big(C_{\bar\ell \ell \nu,\mu e z}^{\tV\tR\tS\tL\tV\tR}
C_{\bar\ell \ell \nu,\mu e z}^{\tV\tL\tS\tL\tV\tR*}\big)
-3\cdot10^{-5}\Im\big(C_{\bar\ell \ell \nu,\mu e z}^{\tV\tL\tS\tL\tV\tR}C_{\bar\ell \ell \nu,\mu e z}^{\tT\tR\tV\tR\tV\tL*}\big).
\label{eq:Gamman2muevb}
\end{align}
\end{subequations}
%%%%%
$\bullet\,\,p \to \nu \bar\nu \ell^+$:
%%%%%p2vvbl
\begin{subequations}
\begin{align}
{\Gamma_{p\to \nu_x\bar{\nu}_y e^+} \over 10^{-9}{\rm GeV}^{11}} & =   
2.4\big(|C_{\bar\nu \nu \ell,xye}^{\tV\tL\tV\tL\tS}|^2+|C_{\bar\nu \nu \ell,xye}^{\tV\tL\tV\tR\tS}|^2\big)
+ 0.06\big(|C_{\bar\nu \nu \ell,xye}^{\tV\tL\tS\tL\tV\tR}|^2+|C_{\bar\nu \nu \ell,xye}^{\tV\tL\tS\tR\tV\tL}|^2\big)
\nn\\
&+0.005\Re\big(C_{\bar\nu \nu \ell,xye}^{\tV\tL\tV\tL\tS}C_{\bar\nu \nu \ell,xye}^{\tV\tL\tV\tR\tS*}\big)
+0.002\Re\big(C_{\bar\nu \nu \ell,xye}^{\tV\tL\tV\tL\tS}C_{\bar\nu \nu \ell,xye}^{\tV\tL\tS\tL\tV\tR*}+C_{\bar\nu \nu \ell,xye}^{\tV\tL\tV\tR\tS}C_{\bar\nu \nu \ell,xye}^{\tV\tL\tS\tR\tV\tL*}\big)
\nn\\
&-4\cdot10^{-9}\Re\big(C_{\bar\nu \nu \ell,xye}^{\tV\tL\tS\tL\tV\tR}C_{\bar\nu \nu \ell,xye}^{\tV\tL\tS\tR\tV\tL*}\big),
\label{eq:Gammap2vvbe}
\\
{\Gamma_{p\to \nu_x\bar{\nu}_y \mu^+} \over 10^{-9}{\rm GeV}^{11}} & =   
2.2\big(|C_{\bar\nu \nu \ell,xy\mu}^{\tV\tL\tV\tL\tS}|^2+|C_{\bar\nu \nu \ell,xy\mu}^{\tV\tL\tV\tR\tS}|^2\big)
+ 0.06\big(|C_{\bar\nu \nu \ell,xy\mu}^{\tV\tL\tS\tL\tV\tR}|^2+|C_{\bar\nu \nu \ell,xy\mu}^{\tV\tL\tS\tR\tV\tL}|^2\big)
\nn\\
&+0.8\Re\big(C_{\bar\nu \nu \ell,xy\mu}^{\tV\tL\tV\tL\tS}C_{\bar\nu \nu \ell,xy\mu}^{\tV\tL\tV\tR\tS*}\big)
+0.3\Re\big(C_{\bar\nu \nu \ell,xy\mu}^{\tV\tL\tV\tL\tS}C_{\bar\nu \nu \ell,xy\mu}^{\tV\tL\tS\tL\tV\tR*}+C_{\bar\nu \nu \ell,xy\mu}^{\tV\tL\tV\tR\tS}C_{\bar\nu \nu \ell,xy\mu}^{\tV\tL\tS\tR\tV\tL*}\big)
\nn\\
& -5\cdot10^{-3}\Re\big(C_{\bar\nu \nu \ell,xy\mu}^{\tV\tL\tS\tL\tV\tR}C_{\bar\nu \nu \ell,xy\mu}^{\tV\tL\tS\tR\tV\tL*}\big).
\end{align}
\end{subequations}
%%%%%
%%%%%n2vbvbv
$\bullet\,\,n \to \bar\nu \bar\nu \nu$:
\begin{subequations}
\begin{align}
{\Gamma_{n\to \bar\nu_x\bar{\nu}_{y\neq x} \nu_z} \over 10^{-9}{\rm GeV}^{11}} & =   
0.6|C_{\nu \nu \bar\nu,xy z}^{\tS\tL\tS\tR\tS}|^2
+ 0.2|C_{\nu \nu \bar\nu,xy z}^{\tT\tL\tV\tL\tV\tR}|^2,
\\
{\Gamma_{n\to \bar\nu_x\bar{\nu}_x \nu_z } \over 10^{-9}{\rm GeV}^{11}} & =   
1.2|C_{\nu \nu \bar\nu,xxz}^{\tS\tL\tS\tR\tS}|^2.
\end{align}
\end{subequations}
%%%%%
%%%%%p2vvl
$\bullet\,\,p\to\nu\nu\ell^+$:
\begin{subequations}
\begin{align}
{\Gamma_{p\to \nu_x\nu_{y\neq x} e^+} \over 10^{-9}{\rm GeV}^{11}} & =
29|C_{\bar\nu \bar\nu \ell,xy e}^{\tT\tR\tT\tR\tS}|^2
+
0.6\big(|C_{\bar\nu \bar\nu \ell,xy e}^{\tS\tR\tS\tR\tS}|^2
+|C_{\bar\nu \bar\nu \ell,xy e}^{\tS\tR\tS\tL\tS}|^2\big)
+0.2|C_{\bar\nu \bar\nu \ell,xy e}^{\tT\tR\tV\tR\tV\tL}|^2
\nn\\
&+0.01\Im\big(C_{\bar\nu \bar\nu \ell,xy e}^{\tT\tR\tT\tR\tS}C_{\bar\nu \bar\nu \ell,xy e}^{\tT\tR\tV\tR\tV\tL*}\big)-0.003\Re\big(C_{\bar\nu \bar\nu \ell,xy e}^{\tS\tR\tS\tR\tS}C_{\bar\nu \bar\nu \ell,xy e}^{\tS\tR\tS\tL\tS*}\big),
\\
{\Gamma_{p\to \nu_x\nu_x e^+} \over 10^{-9}{\rm GeV}^{11}} & =   
1.2\big(|C_{\bar\nu \bar\nu \ell,xx e}^{\tS\tR\tS\tR\tS}|^2
+|C_{\bar\nu \bar\nu \ell,xx e}^{\tS\tR\tS\tL\tS}|^2\big)
-0.005\Re\big(C_{\bar\nu \bar\nu \ell,xx e}^{\tS\tR\tS\tR\tS}C_{\bar\nu \bar\nu \ell,xx e}^{\tS\tR\tS\tL\tS*}\big),
\\
{\Gamma_{p\to \nu_x\nu_{y\neq x} \mu^+} \over 10^{-9}{\rm GeV}^{11}} & = 
26|C_{\bar\nu \bar\nu \ell,xy \mu}^{\tT\tR\tT\tR\tS}|^2
+
0.6\big(|C_{\bar\nu \bar\nu \ell,xy \mu}^{\tS\tR\tS\tR\tS}|^2
+|C_{\bar\nu \bar\nu \ell,xy \mu}^{\tS\tR\tS\tL\tS}|^2\big)
+0.2|C_{\bar\nu \bar\nu \ell,xy \mu}^{\tT\tR\tV\tR\tV\tL}|^2,
\nn\\
&+2.1\Im\big(C_{\bar\nu \bar\nu \ell,xy \mu}^{\tT\tR\tT\tR\tS}C_{\bar\nu \bar\nu \ell,xy \mu}^{\tT\tR\tV\tR\tV\tL*}\big)
-0.4\Re\big(C_{\bar\nu \bar\nu \ell,xy \mu}^{\tS\tR\tS\tR\tS}C_{\bar\nu \bar\nu \ell,xy \mu}^{\tS\tR\tS\tL\tS*}\big),
\\
{\Gamma_{p\to \nu_x\nu_x \mu^+} \over 10^{-9}{\rm GeV}^{11}} & =   
1.1\big(|C_{\bar\nu \bar\nu \ell,xx \mu}^{\tS\tR\tS\tR\tS}|^2
+|C_{\bar\nu \bar\nu \ell,xx \mu}^{\tS\tR\tS\tL\tS}|^2\big)
-0.8\Re\big(C_{\bar\nu \bar\nu \ell,xx \mu}^{\tS\tR\tS\tR\tS}C_{\bar\nu \bar\nu \ell,xx \mu}^{\tS\tR\tS\tL\tS*}\big).
\end{align}
\end{subequations}
%%%%%
%%%%%n2vbvbvb
$\bullet\,\,n\to\bar\nu\bar\nu\bar\nu$:
\begin{subequations}
\begin{align}
{\Gamma_{n\to \bar\nu_e\bar\nu_\mu \bar\nu_\tau} \over 10^{-9}{\rm GeV}^{11}} & =   
0.5\big(|C_{\nu \nu \nu,e\mu\tau}^{\tS\tL\tS\tL\tS}|^2
+|C_{\nu \nu \nu,\mu\tau e}^{\tS\tL\tS\tL\tS}|^2\big)
-0.5\Re\big(C_{\nu \nu \nu,e\mu\tau}^{\tS\tL\tS\tL\tS}C_{\nu \nu \nu,\mu\tau e}^{\tS\tL\tS\tL\tS*}\big),
\\
{\Gamma_{n\to \bar\nu_x\bar\nu_x \bar\nu_{y\neq x}} \over 10^{-9}{\rm GeV}^{11}} & =   
0.7|C_{\nu \nu \nu,xxy}^{\tS\tL\tS\tL\tS}|^2.
\end{align}
\end{subequations}
In the above, the flavor indices $x,y,z$ represent any of $e,\mu,\tau$ unless otherwise specified.  
Similarly, for the other four classes of processes 
$n\to\ell_x^-\ell_y^+\nu_z$
$n\to\nu_x\nu_y\bar\nu_z$,
$p\to\bar\nu_x\bar\nu_y\ell_z^+$,
and $n\to\nu_x\nu_y\nu_z$, their decay widths can be obtained from the corresponding modes above by $\tL\leftrightarrow\tR$ and $\nu\leftrightarrow\bar\nu$ in the WCs.

%%%%%%%%%%%%%%%%%%%%%%%%%%%%%%%%%%%%%%
%%%%%%%%%%%%%%%%%%%%%%%%%%%%%%%%%%%%%%
\bibliography{references.bib}{}
\bibliographystyle{JHEP}

\end{document}